\documentclass[11pt]{article}
\pdfoutput=1
\usepackage{amsmath,amssymb,amsfonts}
\usepackage{a4wide,epsfig,psfrag,scalefnt}
\usepackage[dvipsnames]{xcolor}
\usepackage[hidelinks]{hyperref}
\usepackage{braket}
\usepackage{placeins}
\usepackage{breqn}
\usepackage{slashed}
\usepackage{enumitem}
\usepackage[numbers,sort&compress]{natbib}
\usepackage{caption}
\usepackage{subcaption}
\usepackage{soul}
\usepackage{etoolbox}
\usepackage{listings}

\setlist[itemize]{left=10pt, topsep=2pt, itemsep=2pt, parsep=2pt, partopsep=2pt}

\graphicspath{ {figs/} }

\usepackage[titles]{tocloft}

\allowdisplaybreaks

\newcommand{\ep}{\varepsilon}

\newcommand{\bea}{\begin{eqnarray}}
\newcommand{\eea}{\end{eqnarray}}
\newcommand{\f}[2]{\frac{#1}{#2}}
\newcommand{\lb}{\left(}
\newcommand{\rb}{\right)}
\newcommand{\eps}{\epsilon}

\newcommand{\ri}{\mathrm i}
\newcommand{\rd}{\mathrm d}
\newcommand{\ord}{\mathcal O}
\newcommand{\calS}{\mathcal{S}}
\newcommand{\calI}{\mathcal{I}}
\newcommand{\calO}{\mathcal{O}}
\newcommand{\calT}{\mathcal{T}}
\newcommand{\calU}{\mathcal{U}}
\newcommand{\calF}{\mathcal{F}}
\newcommand{\MB}{\scriptscriptstyle\mathrm{MB}}

\begin{document}

\title{\vskip-3cm{\baselineskip14pt
    \begin{flushleft}
     \normalsize P3H-24-049, TTP24-026
    \end{flushleft}} \vskip2.5cm

\textbf{Massive two-loop four-point Feynman integrals  
at high energies with AsyInt}
\vskip0.3cm
}
 
\author{
  Hantian Zhang
  \\[5pt]
  { \small \it Institut f{\"u}r Theoretische Teilchenphysik,
    Karlsruhe Institute of Technology (KIT),}\\
  {\small\it Wolfgang-Gaede Strasse 1, 76128 Karlsruhe, Germany}
   \\[10pt]
}

\date{}

\maketitle

\thispagestyle{empty}

\begin{abstract}

We present analytic techniques for parametric integrations of massive two-loop four-point Feynman integrals at high energies,
and their implementation in the toolbox \texttt{AsyInt}.
In the high-energy region, 
the Feynman integrals involving external and internal massive particles, such as the top quark, Higgs and vector bosons,
can be asymptotically expanded and directly calculated in the small-mass limit.
With this approach, analytic results for  higher-order terms in the expansion parameter and the dimensional regulator can be obtained with \texttt{AsyInt}.
These results are important ingredients for the two-loop electroweak and QCD corrections for $2 \to 2$ scattering processes in the large transverse momenta region,
which is relevant to both precision collider phenomenology and new physics searches at current and future high-energy colliders.
In this paper, analytic results of representative planar and non-planar Feynman integrals are presented.

\end{abstract}


\newpage

\tableofcontents

\newpage

\section{Introduction}

Since the milestone discovery of the Higgs boson  at the Large Hadron Collider (LHC),
the investigation of Higgs boson properties
has become one of the primary targets of the LHC programme~\cite{ATLAS:2022vkf,CMS:2022dwd}.
This investigation requires the comparisons of theoretical predictions and experimental measurements for Higgs boson production processes at the highest precision.
At the LHC, the high-energy region of $2\to 2$ scattering processes with Higgs boson final states is of particular interest.
In this region, the observables such as Higgs boson transverse momenta~$(p_T)$ spectrum and fiducial cross sections,
offer opportunities to study the Higgs boson's properties under extreme conditions
and probe new physics effects beyond the Standard Model (SM).
%
The investigation in the high-energy region will become particularly relevant in the upcoming high-luminosity phase of LHC experiments and in future high-energy colliders.

This investigation is also challenging from a theoretical perspective,
as it requires highly technical perturbative quantum field theory~(QFT) calculations of higher-order electroweak (EW) and quantum chromodynamic (QCD) corrections.
Given that the Higgs boson interacts with all SM particles, including itself,
the massive particles such as the top quark, Higgs, and gauge bosons will manifest in the virtual loops.
The resolution of massive particles leads to  massive multi-loop Feynman integrals that are significantly more complicated than their massless counterparts.
The complexity of Feynman integrals increases not only with the number of mass scales, but also with the number of massive internal lines in the loops.
To tackle the massive Feynman integrals for $2 \to 2$ processes at the two-loop level, 
various successful methods have been developed in the past decades.
The analytic methods include the differential equations approach~\cite{Gehrmann:1999as,Henn:2013pwa,Bonciani:2016qxi,Adams:2017tga,Delto:2023kqv}, iterated integrals~\cite{Remiddi:1999ew,Panzer:2014caa,Broedel:2017kkb}, 
the Mellin-Barnes method~\cite{Smirnov:2001cm,Smirnov:2004ym,Dubovyk:2022obc} in combination with summation techniques~\cite{Vermaseren:1998uu,Ablinger:2015tua,Blumlein:2021hbq},
the analytic expansion method in high-energy and forward-scattering kinematics~\cite{Bonciani:2018omm,Mishima:2018olh,Bellafronte:2022jmo,Davies:2022ram,Davies:2023vmj},
and also semi-analytic approaches~\cite{Lee:2017qql,Moriello:2019yhu,Hidding:2020ytt,Armadillo:2022ugh,Fael:2021kyg,Fael:2022rgm}.
The numerical methods include the parametric integrations based on sector decomposition~\cite{Binoth:2000ps,Borowka:2016ypz,Borowka:2017idc}, integrand-level subtractions~\cite{Baglio:2018lrj,Anastasiou:2024xvk}, dispersion relations~\cite{Freitas:2022hyp},
and numerical differential equations methods~\cite{Czakon:2008zk}
with auxiliary mass flow~\cite{Liu:2017jxz,Liu:2022chg,Bi:2023bnq}.
These methods have enabled high-precision predictions for many important collider scattering processes.

Among the above methods,
the high-energy expansion  is a well-suited analytic approach for phenomenological studies of scattering processes with large transverse momenta, particularly for $2\to 2$ Higgs boson production processes at the LHC~\cite{Melnikov:2016qoc,Davies:2018qvx,Davies:2020drs,Davies:2022ram}.
%
For example, in the calculations of Higgs boson pair production,
it has been shown that a deep high-energy expansion combined with Pad\'e  approximations can yield precise results across a vast phase space region,
from the high-energy limit down to $p_T \approx 150$~GeV for Yukawa and QCD corrections~\cite{Davies:2022ram,Davies:2023vmj}.
In a similar spirit but from a different perspective,
the high-energy QCD factorisation formulas explored in Refs.~\cite{Penin:2005eh,Mitov:2006xs,Becher:2007cu,Wang:2023qbf}
can capture the leading high-energy behaviour of massive QCD amplitudes.
In order to access most of the phenomenologically relevant regions at the LHC,
the higher-order expansion terms beyond the leading high-energy approximation are necessary.
In this paper, we will present the analytic integration techniques for massive two-loop four-point integrals at high energies, including higher-order terms in this limit.
%
These results serve as crucial boundary conditions for the deep high-energy expansion of
 two-loop EW and QCD corrections for $2 \to 2$ scattering processes at the LHC and future high-energy colliders,
enabling precise predictions over a wide-range of interesting phase space regions.

\begin{figure}[bt]
  \centering
  \begin{tabular}{cccc}
    \scalebox{-1}[1]{\includegraphics[width=.22\textwidth]{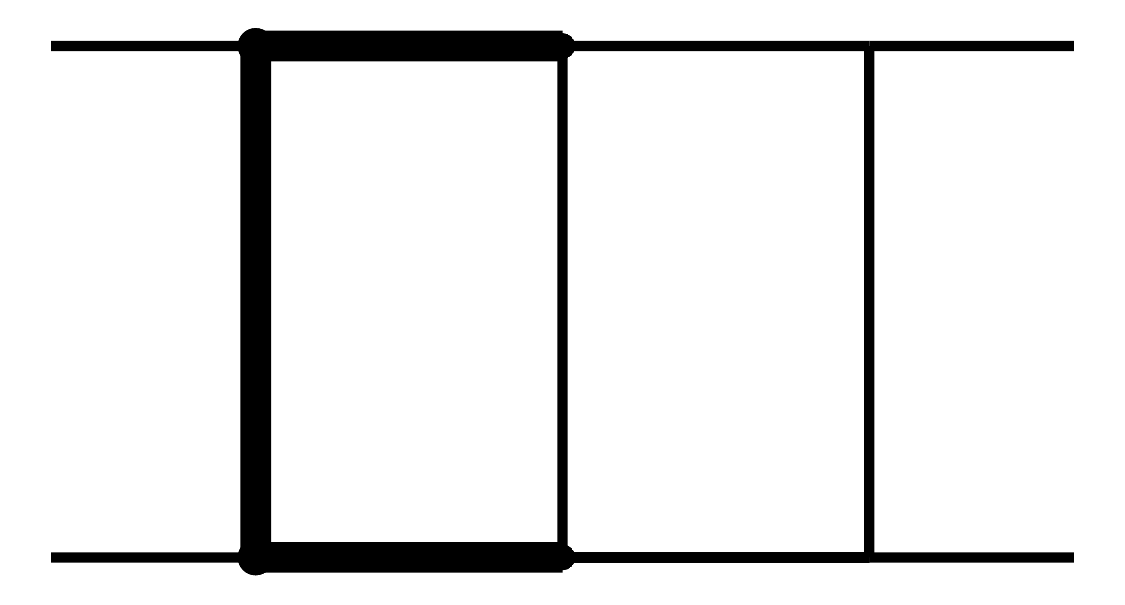}} &
    \scalebox{-1}[1]{\includegraphics[width=.22\textwidth]{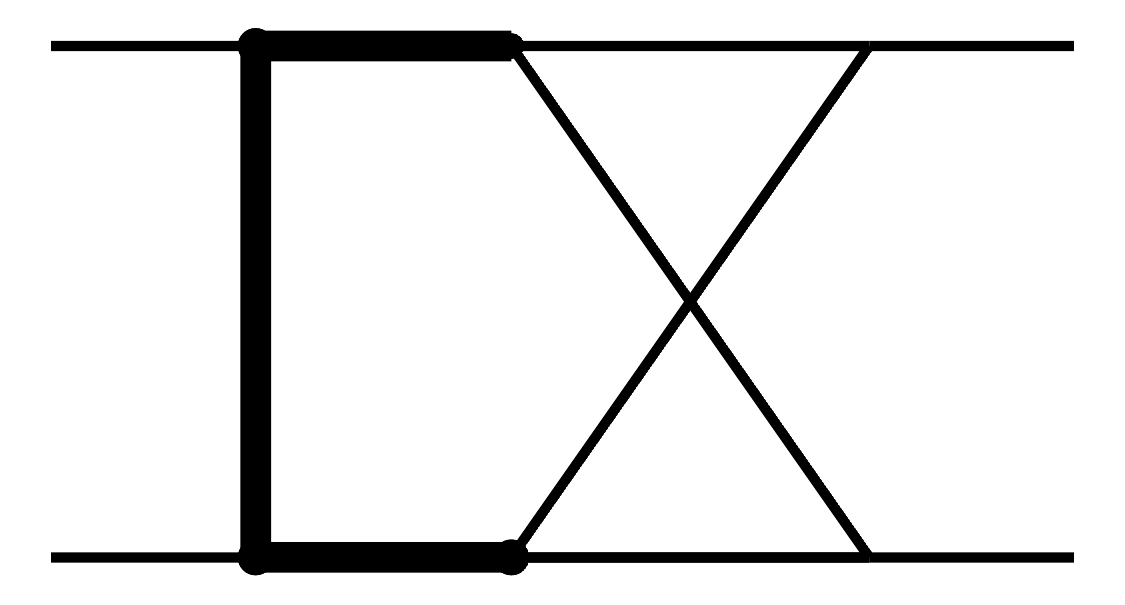}} &
    \scalebox{-1}[1]{\includegraphics[width=.22\textwidth]{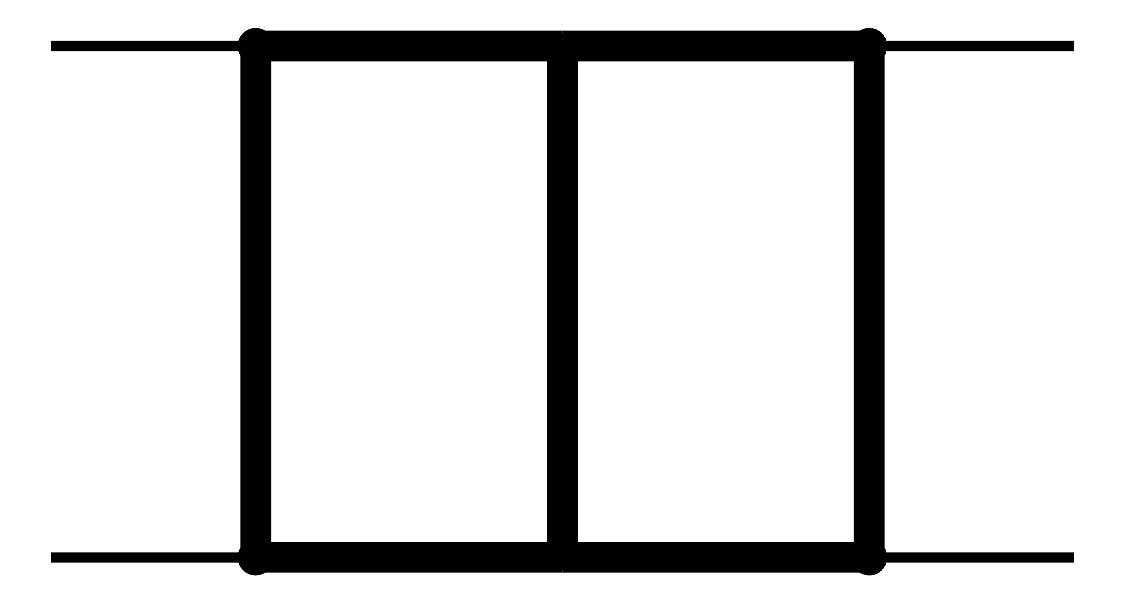}} &
    \includegraphics[width=.22\textwidth]{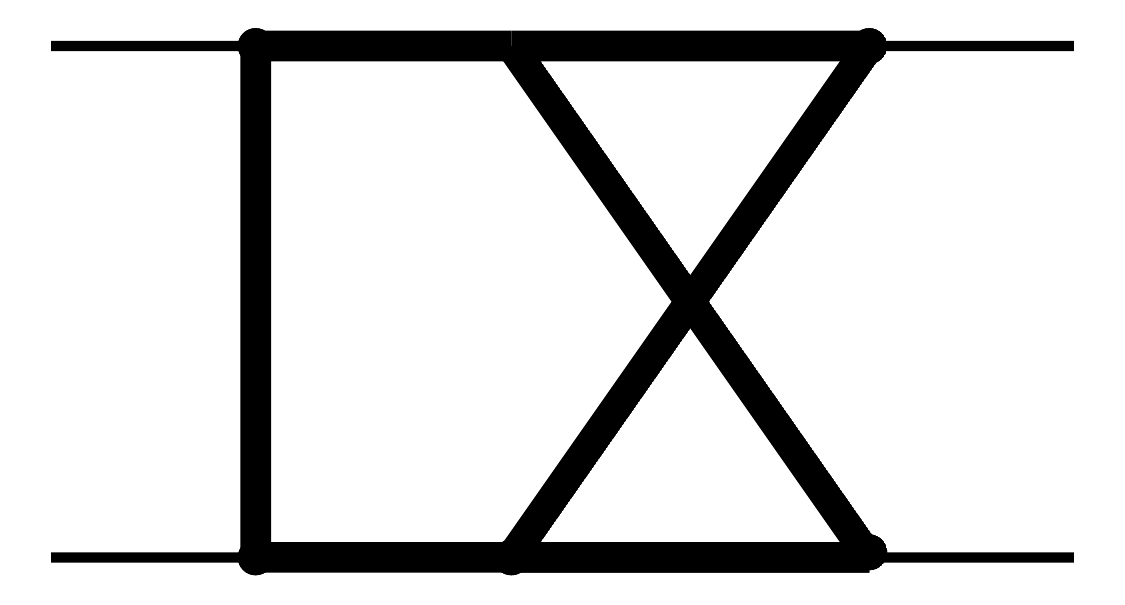} \\
    $\rm PL_1$ & $\rm NPL_1$ & $\rm PL_2$ & $\rm NPL_2$
  \end{tabular}
  \caption{\label{fig:diags}
  Sample planar and non-planar massive two-loop four-point Feynman diagrams with massive propagators represented by thick lines
  and massless ones represented by thin lines.
    }
\end{figure}

In the high-energy region, the kinematic invariants of Feynman diagrams are larger than the mass-square of all SM particles at the EW scale.
Hence, the integrals in most cases can be Taylor expanded in the small-external-mass $(m^{\rm ext} \to 0)$ and equal-internal-mass $( m^{\rm int}  \to m )$ limits~\cite{Davies:2022ram}, resulting in simpler master integrals (MIs) that are analytically tractable.
As examples, sample Feynman diagrams of the representative MIs are shown in Fig.~\ref{fig:diags}.
These MIs are further calculated analytically through
the asymptotic expansion in the small-mass parameter $m$,
which is the main focus of this paper.
In Section~\ref{sec:HErep}, 
we will review the Mellin-Barnes integral representations of higher-order expansion terms at high energies.
In Section~\ref{sec:SolveMB}, we will discuss the analytic techniques for solving the resulting Mellin-Barnes integrals.
The analytic techniques discussed in this paper are implemented in the \texttt{Mathematica} toolbox \texttt{AsyInt}, which is publicly available at:
\[
\texttt{\href{https://gitlab.com/asyint/asyint-public}{https://gitlab.com/asyint/asyint-public}}
\]
The overview of \texttt{AsyInt} is presented in Section~\ref{sec:workflow}, and the commands are summarised in Appendix~\ref{app:func}.
For more details on practical usage,
we also refer to  the working examples in the \texttt{AsyInt} repository.
In Section~\ref{sec:results}, we will discuss calculational steps and present analytic solutions of Feynman integrals shown in Fig.~\ref{fig:diags}.

\section{Integral representations at high energies}
\label{sec:HErep}
In this section, we will review the approach to obtain the Mellin-Barnes (MB) integral representations for massive Feynman integrals at high energies~\cite{Mishima:2018olh,Davies:2022ram}.
The starting point of the whole calculation is the positive definite Symanzik polynomials of a particular Feynman integral.
They can always be obtained in the Euclidean region by using the Euclidean kinematic invariants $S_{ab} = - (q_a + q_b)^2$, where $q_{a}$ and $q_{b}$ denote the external momenta. For four-point integrals, the Euclidean Mandelstam invariants $\{S, T, U\}=\{-s,-t,-u\}$ are used in the calculation.
Note that the final results of four-point integrals are expressed  in terms of positive $\{s,T,U\}$ in the physical region, which requires the analytic continuation from the Euclidean region  by the transformation $S \to s$.

\subsection{Alpha representations and asymptotic expansions}
Let us consider a generic two-loop Feynman integral with  $n$ propagators and $k$ numerators
\bea
\label{eq:feynint}
 \calI_{n,k}  &: =& \int  \prod_{j=1}^{2}  \rd l_j   \, \f{N_1^{\lambda_1} \cdots N_k^{\lambda_k} }{D_1^{1+\delta_1} \cdots D_n^{1+\delta_n}} \, ,
\eea
where the propagator denominators are defined as $ D_i = m_i^2 - p_i^2$ with $m_i \in \{ 0,m\} $ and $p_i$ being the momentum.
The numerators can be either irreducible scalar products or denominators with negative powers.
The integration measure is defined as
\bea
\label{eq:measure}
\int \rd l_j &:=& \f{\mu^{2 \, \eps} \, e^{\gamma_{_{\rm E}} \eps} }{\ri \, \pi^{d/2}} \, \int  \rd^d l_j \; \quad \mbox{with } \; d = 4 - 2\, \eps \,,
\eea
where $\gamma_{_{\rm E}}$ is the Euler-Mascheroni constant and $\eps$ is the dimensional regulator.
The $\delta_i$ are real numbers that serve as the additional regulators to the propagator denominator $D_i$ and also the shifts of propagator denominator powers.\footnote{For example,  the shift $\delta_i \to \delta_i + a$ yields an integral with higher-power denominator $D_i^{a+1+\delta_i}$.}
The Feynman integral in Eq.~\eqref{eq:feynint} admits an $n$-line Feynman diagram, whose Symanzik polynomials are denoted by $\mathcal{U}$ and $\mathcal{F}$ with $\{N_1, \dots, N_k\} = 1$.
The $k$-numerator extension to its integral definition is a necessary completion in order to perform the integration-by-parts (IBP) reduction at two loops.
The Symanzik polynomials for this extended $k$-numerator  Feynman integral are denoted by  $\mathcal{\tilde U}$ and $\mathcal{\tilde F}$.
Note that $\mathcal{U}$ and $\mathcal{F}$ depend on $n$ alpha parameters, while $\mathcal{\tilde U}$ and $\mathcal{\tilde F}$ depend on $(n+k)$ alpha parameters. 

Now the so-called alpha representation (or Schwinger-parametric form) for this Feynman integral can be written as
\bea
\label{eq:alphaNum}
 \calI_{n,k}  
 & = &
 \int_0^\infty \rd^n \alpha^\delta \,
\lb \lb \prod_{t=1}^{k}  \, \f{(-\partial)^{|\lambda_t|}}{\partial \alpha_{n+t}^{|\lambda_t|}} \rb \,  \mathcal{\tilde U}^{-d/2} \, e^{\cal -{\tilde F}/{\tilde U}} \rb \bigg|_{\alpha_{n+1}=\cdots=\alpha_{n+k}=0}  \,,
\eea
where the integration measure is
\bea
\int_0^{\infty} \rd^n \alpha^\delta &:=& \int_0^\infty \prod_{i=1}^n \rd \alpha_i \, \f{\alpha_i^{\delta_i}}{\Gamma(1+\delta_i)} \;.
\eea
This alpha representation is obtained by treating the numerator on the same footing as the propagator \cite{Borowka:2015mxa}
\bea
\label{eq:Prop2Alpha}
\f{1}{(D_i)^\lambda} &=& \begin{cases}
\displaystyle \; \f{1}{\Gamma(\lambda)} \int_0^\infty \rd \alpha_i \, \alpha_i^{\lambda-1} \, e^{-D_i \, \alpha_i} \;  & \text{for}\; \lambda >0  \\[6mm]
\displaystyle \; (-1)^{|\lambda|} \, \f{\partial^{|\lambda|}}{\partial \alpha_i^{|\lambda|}} \, e^{-D_i \, \alpha_i}\, \Big|_{\alpha_i=0} \;  & \text{for}\;   \lambda <0 \;.
\end{cases} 
\eea
By taking the derivative in Eq.~\eqref{eq:alphaNum}, the following alpha representation can be obtained
\bea
\label{eq:alphaNum2}
 \calI_{n,k}  
 &=&   \int_0^\infty\rd^n \alpha^\delta  \,
  \left[\, \mathcal{U}^{-d/2} \, e^{\cal -F/U} \, \right] \, \hat{\calO}_k \big( \{S_{ab},m^2 \}, \{\alpha_n\} \big) \;,
\eea
which only depends on  $\mathcal{U}, \mathcal{F}$ polynomials, $n$ alpha parameters $ \{\alpha_n\} $ and invariants $ \{S_{ab},m^2 \}$.
As a consequence of taking derivatives, the additional rational function  $\hat{\calO}_k$ in terms of alpha parameters and invariants is generated.
Note that for integrals without numerators, the rational function reduces to unity ($\hat{\calO}_0 = 1$).

In the next step, one can readily apply the high-energy asymptotic expansions on the alpha representation of the integral.
First, one needs to find all relevant regions by the method of regions~\cite{Beneke:1997zp} with the following high-energy hierarchy
\bea
\label{eq:scaling}
m^2 /S_{ab} \, \sim \, \rho \, \ll \, 1\,.
\eea
With the positive definite $\mathcal{U}, \mathcal{F}$ polynomials as inputs, \texttt{AsyInt} relies on \texttt{asy2.1.m}~\cite{Pak:2010pt} to obtain all the high-energy regions,
including one hard region and various asymptotic regions.
Here we use the terminology of asymptotic regions to denote all kinds of soft, ultrasoft, collinear regions etc.
In the hard region, the scaling is given by
\bea
\mbox{hard scaling:} & m^2 /S_{ab} \, \sim \, \rho, \quad  \{ \alpha_n \} \sim 1.
\eea
In the asymptotic regions, certain alpha parameters can have the scalings of the order $\rho$ or $\rho^{1/2}$.
Suppose that the set of alpha parameters which have $\rho$-scaling is given by $\{\alpha_s\}$, and the remaining set of alpha parameters is given by $ \{ \alpha_{n/s} \}$,
then the scaling of an asymptotic region is denoted by
\bea
\label{eq:asyscale}
\mbox{asymptotic scaling:} & m^2 /S_{ab} \, \sim \, \rho, \quad \{\alpha_s\} \sim \mbox{$\rho$ or $\sqrt{\rho}$} \,, \quad  \{ \alpha_{n/s} \}  \sim 1.
\eea

In the hard region,
since the only scaling parameter is $m^2 \sim \rho$, its high-energy expansion can be obtained by a simple Taylor expansion as
\bea
\label{eq:hard}
\mathcal{I}_{n,k}^{(\mathrm{hard})} &=& 
 \sum_{j=0}^{\infty} \f{\big(\rho \, m^2\big)^j}{j!} \, \f{\partial^j}{\partial (m^2)^j}\mathcal{I}_{n,k}\bigg|_{m^2=0} \;.
\eea
The integrals on the r.h.s.~can be reduced to the massless MIs, which are well studied in the literature~\cite{Smirnov:1999gc,Tausk:1999vh,Henn:2013pwa,Henn:2020lye}.

The calculation of the asymptotic region integrals is more involved. 
Suppose there are $R$ asymptotic regions in total,
and each region $r$ has its own asymptotic scaling in Eq.~\eqref{eq:asyscale}.
Each scaling can yield the following expansions on $\mathcal{U}, \mathcal{F}$ polynomials
\bea
\mathcal{U}_r \;=\; \sum_{j=0}^{\infty} \, \sqrt{\rho}^{\, j} \, \mathcal{U}_r^{(j)}, 
&&
\mathcal{F}_r \;=\; \sum_{j=0}^{\infty} \, \sqrt{\rho}^{\, j} \, \mathcal{F}_r^{(j)},
\eea
and also the expansion of the $\hat{\calO}_k$ rational function
\bea
\hat{\calO}_{k,r} \;=\; \sum_{j=0}^{\infty} \, \sqrt{\rho}^{\, j-j_{\min} } \, \hat{\calO}_{k,r}^{(j)},
\eea
where $\sqrt{\rho}^{\, -j_{\min}}$ denotes the leading order expansion term of the rational function.
Inserting the above expansions into Eq.~\eqref{eq:alphaNum2} and re-expanding around $\sqrt{\rho}$ while ignoring any overall factor $\sqrt{\rho}^{\, -j_{\min}}$,
the alpha representation in each region $r$ at high energies can be schematically written as
\bea
\label{eq:alphaHE}
\calI_{n,k}^{(r)} &=&
 \int_0^\infty\rd^n \alpha^\delta  \, \frac{\sqrt{\rho}^{\,j} }{j!} \Bigg[ \frac{\partial^j}{\partial (\sqrt{\rho})^{\, j} }
  \left(\, \mathcal{U}_r^{-d/2} \, e^{- \mathcal{F}_r/ \mathcal{U}_r} \, \hat{\calO}_{k,r}  \right) \Bigg] \Bigg|_{\rho = 0} \nonumber \\
&=& 
\int_0^\infty\rd^n \alpha^\delta  \,\bigg[
 \Big( \mathcal{U}_r^{(0)} \Big)^{-\f{d}{2}} \, e^{-\mathcal{F}_r^{(0)}/\mathcal{U}_r^{(0)}}  \bigg] \, 
 \Bigg(  
   \sum_{j=0}^{\infty} \sqrt{\rho}^{\, j} \hat{\calS}_{k,r}^{(j)}
 \Bigg)\,.
\eea
The rational function $\hat{\calS}_{k,r}^{(j)}$ depends on $ \{\alpha_n\} $ and $ \{S_{ab},m^2,d \}$,
and  can be regarded as the shift operators of the so-called template integrals, which can be obtained by the integration of integrands in the square bracket of Eq.~\eqref{eq:alphaHE}. More details will be discussed in the next subsection. Note that $\hat{\calS}_{k,r}^{(0)} = \hat{\calO}_{k,r}^{(0)}$, and we have $\hat{\calS}_{0,r}^{(0)} = 1$ if there is no numerator.

\subsection{Mellin-Barnes representations and template integrals}
\label{sec:tempint}
In this section, we will review the template integral approach~\cite{Mishima:2018olh} based on Mellin-Barnes (MB) integral representations.
The template integral $\calT_{n}^{(r)}$ in region $r$ is the leading-order approximation in the high energy expansion of Eq.~\eqref{eq:alphaHE} 
\bea
\label{eq:template}
\calT_{n}^{(r)} (\{ \delta_i \}, \eps) &:=&
\int_0^\infty\rd^n \alpha^\delta  \,\bigg[
 \Big( \mathcal{U}_r^{(0)} \Big)^{-\f{d}{2}} \, e^{-\mathcal{F}_r^{(0)}/\mathcal{U}_r^{(0)}}  \bigg] \,,
\eea
which only depends on the propagators of the integral definition.
The MB representation of the template integral can be obtained by means of the alpha-parametric integrations
\bea
\label{eq:inttype1}
\int_{0}^{\infty} \rd \alpha \, \alpha^a \, e^{-A \alpha} &=& \Gamma (1+a) \, A^{-1-a} \,, \\
\int_0^\infty \rd \alpha \, \alpha^a \, (A+B\alpha) &=& A^{1+a+b} \, B^{-1-a} \, \frac{\Gamma(1+a) \, \Gamma(-1-a-b)}{\Gamma(-b)}
\label{eq:inttype2}
\eea
and the applications of the Mellin transformation into the complex plane
\bea
\label{eq:mellin}
(X+Y)^\lambda &=&  \int_{\mathrm{Re}(z)- \ri \infty}^{\mathrm{Re}(z)+ \ri \infty} \f{\rd z}{2 \pi \ri} \, \f{\Gamma(-\lambda+z) \, \Gamma(-z)}{\Gamma(-\lambda)} \, X^{z} \, Y^{\lambda-z} \;.
\eea
By employing the rescaling transformation $\alpha_i \to \alpha_i \alpha_j $, 
one can obtain (multi-dimensional) MB representations.
Note that the dimensionality of MB representations of $\calT_{n}^{(r)}$
can be reduced by well-chosen variable transformations and integration strategies.
Alternatively, one can also use automated tools like \texttt{AMBRE}~\cite{Gluza:2007rt} or \texttt{MBcreate.m}~\cite{Belitsky:2022gba} to obtain the MB representations.

The higher-order expansion terms in Eq.~\eqref{eq:alphaHE} can be obtained by promoting the rational functions $ \hat{\calO}_{k,r}^{(0)}$ and $\hat{\calS}_{k,r}^{(j)} $
to shift operators that shift the $\{\delta_i\}$ and $\eps$ indices of the template integrals. 
Schematically, these functions take the form
\bea
\label{eq:shiftOP1}
\hat{\calS}_{k,r}^{(j)}  &=& 
\sum_{\sigma} \Big[ \{ S_{ab}, \, m^2, d \}\text{-monomial} \Big]_{\sigma} \times 
 \frac{  \Big[ \{\alpha_i\}\text{-polynomial} \Big]_{\sigma}}{\lb \mathcal{U}_r^{(0)} \rb^{\eta_\sigma}}  \, ,
\eea
where $\sigma$ runs over all possibilities for taking derivatives in Eq.~\eqref{eq:alphaHE}, and $\eta_\sigma$ are non-negative integers.
The function $ \hat{\calO}_{k,r}^{(0)}$ takes the same form.
The action of one particular term in these functions applied to the template integral is given by
\bea
\label{eq:ShiftRule}
\hat{\calS}_{k,r}^{(j)} \, \circ \, \calT_{n}^{(r)} (\{ \delta_i \}, \eps) &\supset& 
 \{ S_{ab}, \, m^2, d \}\text{-monomial} \times 
 \frac{ \prod_{i=1}^{n}  \alpha_i^{\beta_i} }{\big( \mathcal{U}_r^{(0)} \big)^{\eta}}  \,  \circ \, \calT_{n}^{(r)} (\{ \delta_i \}, \eps) \nonumber \\
 &=& 
  \{ S_{ab}, \, m^2, d \}\text{-monomial} \times 
 \lb \prod_{i=1}^{n} \mathcal{P}_{1+\delta_i}^{\beta_i} \rb  \,  \calT_{n}^{(r)} (\{ \delta_i + \beta_i \}, \eps - \eta)\,,
\eea
where $\beta_i$ and $\eta$ are non-negative integers, 
and $\mathcal{P}_{1+\delta_i}^{\beta_i} = \Gamma(1+\delta_i + \beta_i) / \Gamma(1+\delta_i)$ 
is the Pochhammer function.
Now the high-energy expansion in Eq.~\eqref{eq:alphaHE}
can be cast into a form in terms of shift operators and template integrals, and the summation over all $R$ asymptotic regions gives 
\bea
\label{eq:templateHE}
\mathcal{I}_{n,k}^{(\mathrm{asy})} &=& 
\sum_{r=1}^R
\Bigg[ \, \sum_{j=0}^{\infty}\, \sqrt{\rho}^{\, j} \, \hat{\calS}_{k,r}^{(j)} \, \circ \, \calT_{n}^{(r)} (\{ \delta_i \}, \eps) \, \Bigg]  \;.
\eea

After applying the shift operators,
we need to resolve the $\delta_i$- and $\epsilon$-singularities in the Gamma functions,
and perform series expansions in the limits $\eps \to 0$ and $ \delta_i \to 0$.
In order to explore the cancellations of $\delta_i$-poles,
these singularities must be resolved in the same sequence of $\{ \delta_1, \dots, \delta_n, \eps \}$ for all regions.
Note that the $\delta$-singularities are present in each individual region, but their sum across all asymptotic regions is free from $\delta$-singularities.
To achieve this, \texttt{AsyInt} chooses the integration lines parallel to the imaginary axis, e.g. $\mathrm{Re}(z) = \{ -1/7, -1/11, \dots\}$ for each integration variable $z=z_1, z_2, \dots$,
and employs  \texttt{MB.m}~\cite{Czakon:2005rk} to find the integration contour by assigning real values to $\{\delta_i,\eps\}$
such that all left poles in Gamma functions $\Gamma(\dots + z)$ are on the left side of integration contour,
while all right poles in Gamma functions $\Gamma(\dots - z)$ are on the right side of integration contour.
The subsequent analytic continuation is performed by \texttt{MB.m} to move $\{\delta_i,\eps\}$ across the integration contour,
in order to perform series expansions in the limits $\delta_i \to 0$ and $\eps \to 0$ .
Note that after the analytic continuation and expansion, merging poles can appear where the locations of left poles and right poles in Gamma functions coincide.
In this approach, one can obtain the power series of Eq.~\eqref{eq:templateHE} up to a certain order $\mathcal{O}\lb m^{j_\max} \rb$ and $\mathcal{O}\lb \eps^{v_\max} \rb$ in MB representations expressed in terms of Gamma functions $\Gamma( z)$ 
and PolyGamma functions $\psi^{(i)}(z) = \dfrac{d^{i+1}}{dz^{i+1}} \ln \Big[ \Gamma(z) \Big]$.
Schematically, the final MB representation can be written as the series
\bea
\label{eq:MBHE}
\mathcal{I}_{n,k}^{(\mathrm{asy})} &=& 
\sum_{u,j,v} \, \eps^{u} \, m^{j} \, \log(m^2)^{v}  \, \sum_{i_{\MB}}  
 \left[ \;\, \int \limits_{-\ri \infty}^{+ \ri \infty}   \prod_l  \frac{\rd z_l}{2\pi \ri}  \;  f_{i_{\MB}}^{\, ujv} \Big( \Gamma, \psi^{(i)} ; z_1, z_2, \dots \Big) \right] \,,
\eea
where each of the MB integrals inside the square brackets can contain either  $z_l$-independent Gamma and PolyGamma functions or various $z_l$-integrations in the multi-dimensional complex space.
The real values of integration lines $\mathrm{Re}(z_l)$ are omitted to simplify the notation in the following.

\section{Solving irreducible Mellin-Barnes integrals}
\label{sec:SolveMB}
In this section, we will describe the algorithmic approaches to solve a large number of MB integrals appearing in Eq.~\eqref{eq:MBHE}
to higher-order expansion terms in $m^2$. 
%
%
Usually, the first step is to reduce the dimensionality of the MB integrals whenever possible through various corollaries of Barnes' lemma~\cite{Smirnov:2004ym,Mishima:2018olh,Belitsky:2022gba}.
Note that \texttt{AsyInt} is not designed for the MB integral reduction problem but aims to solve the irreducible MB integrals,
which are the ``master integrals'' in this context.
Here, irreducible MB integrals are broadly defined as those that either cannot be reduced or for which a reduction method is unknown.
\texttt{AsyInt} offers three approaches to address the irreducible MB integrals: the analytic summation, the numerical reconstruction,
and the analytic \texttt{Expand\&Fit} method.
We will discuss their applicabilities to different types of MB integrals in the following subsections.\footnote{There are other approaches for solving particular types of MB integrals, such as the series-representation solutions~\cite{Ananthanarayan:2020fhl} and integral-representation solutions in terms of  transcendental functions~\cite{Passarino:2024ugq}.
For pedagogical reviews of MB methods, please refer to the textbooks~\cite{Smirnov:2004ym,Dubovyk:2022obc}
}

\subsection{Analytic summation}
\label{sec:AnaSum}
The analytic summation method 
is the core approach in solving the MB integrals by summing over residues based on the Cauchy theorem.
Given a one-dimensional MB integral with a straight-line integration contour $\mathrm{Re}(z) < 0$ and no right pole in the interval $(\mathrm{Re}(z),0)$, 
we can close the integration contour with the right semi-circle
\bea
 \int \limits_{- \ri \infty}^{ +\ri \infty} \frac{\rd z}{2 \pi \ri} \, f \Big( \Gamma, \psi^{(i)} ; z \Big) \,=\,
 -\sum_{k=0}^{\infty} \mathrm{Res}_{z=k} \Big[ f \Big( \Gamma, \psi^{(i)} ; z \Big) \Big] - \int \limits_{\mathrm{arc}}  \frac{\rd z}{2 \pi \ri} \, f \Big( \Gamma, \psi^{(i)} ; z \Big) \,,
\eea
or with the left semi-circle.
This approach relies on two assumptions: (1) the residue series in the right or left half-plane is convergent; (2) the arc contribution on the left or right semi-circle is vanishing.
In cases where these two assumptions are satisfied,
\texttt{AsyInt} can extract the sum over infinitely many residues of MB integrals and pass them to 
\texttt{HarmonicSums.m}~\cite{HarmonicSums}, \texttt{Sigma.m}~\cite{Sigma} and \texttt{EvaluateMultiSums.m}~\cite{EvaluateMultiSums} for analytic summation.
Often, this approach 
can successfully solve  scaleless and one-scale MB integrals up to two dimensions
within the capabilities of the modern summation techniques.
However, 
there are in general also more complicated multi-dimensional MB integrals appearing in Eq.~\eqref{eq:MBHE},
such as those with non-vanishing arc contributions on the semi-circles,\footnote{
For more details on the discussion of non-vanishing arc contributions, please refer to Ref.~\cite{Davies:2022ram}.
}
and those involving two kinematic scales.
For those complicated MB integrals, one can resort to the \texttt{Expand\&Fit} method that will be discussed in the Section~\ref{sec:expfit}.
Note that the results obtained from the analytic summation can provide useful insights into the structures of transcendental functions and rational functions that will appear, which are important ingredients for the \texttt{Expand\&Fit} method.

\subsection{Numerical reconstruction}
\label{sec:NumRec}
The numerical reconstruction method is a widely used approach, particularly suitable for solving multi-dimensional scaleless MB integrals 
when the basis of constants is known a priori.
The strategy is to first reduce their dimensionality by making changes of variables and
applying corollaries of Barnes' lemma~\cite{Smirnov:2004ym,Mishima:2018olh,Belitsky:2022gba}.
%
%
Note that except  for the first and second Barnes lemma that are automated in \texttt{MB.m},
the dimensionality reduction step is usually conducted on a case-by-case basis.
\texttt{AsyInt} currently does not offer MB reduction routines.
If the MB integrals can be reduced to one-dimensional ones, then in most cases their high-precision numerical evaluations  can be obtained by using  \texttt{MB.m} to more than a thousand digits of precision.
%
The \texttt{PSLQ} algorithm~\cite{pslq} can be applied to reconstruct the analytic expressions in terms of special constants from the high-precision numerical values,
provided that the basis of the constants is sufficiently complete for the problem.
Based on the author's experience, the required constants for massive two-loop four-point integrals 
up to transcendental-weight five can be constructed from the basis
\begin{align}
\label{eq:constlist}
\Big\{
&1,\, \sqrt{3}, \, 
\log (2), \, \log (3), \, \pi, \,
\psi ^{(1)}\big(1/3 \big), \, c_Z \,,
\zeta (3) ,  \, \text{Li}_4\left({1}/{2} \right) ,\,
 \zeta(5)
\Big\} 
\end{align}
where $c_Z$ is a new constant that admits a series representation
\bea
\label{eq:cz}
c_Z &=&\int_{0}^{\infty}  \, \frac{\rd \alpha_1 \, \rd \alpha_2}{\sqrt{\alpha _1 \, \alpha _2 \, \big( \alpha _1+\alpha _2+1\big) \, \big( \alpha _2 \alpha _1+\alpha _1+\alpha _2\big) } }  \nonumber \\[4pt]
& =& \sum_{k=0}^{\infty}\frac{2\, \Gamma \left(k+\frac{1}{2}\right)^4 \left[\psi ^{(0)}(k+1)-2 \psi ^{(0)}\left(k+\frac{1}{2}\right)+\psi ^{(0)}(2 k+1)\right]}{\pi (k!)^2 \, \Gamma (2 k+1)} \, .
\eea
The new constant $c_Z$ appears in the fully-massive non-planar two-loop four-point Feynman integral.
Note that the constants 
 $\Big\{ \, \mathrm{Im} \left[\text{Li}_3\left(\frac{\pm \ri}{\sqrt{3}}\right)\right], \mathrm{Im} \left[\text{Li}_3\left( \frac{\ri \sqrt{3}+1}{4}\right)\right]  \Big\}$
 may also appear in the intermediate steps of the calculation.

\subsection{Analytic \texttt{Expand\&Fit} method}
\label{sec:expfit}

In this section, we will present  the analytic \texttt{Expand\&Fit} method for addressing the complicated MB integrals with one or two scales appearing in Eq.~\eqref{eq:MBHE}.
The previously discussed approaches serve as the building blocks for this method.
The basic idea is to expand these MB integrals in certain kinematic limits to high orders,
and then perform a fitting procedure using an ansatz of linear combinations of transcendental and rational functions.
%
Note that this approach has been used for solving one-scale one-dimensional MB integrals in Ref.~\cite{Mishima:2018olh}.\footnote{
In general, the expansion-and-fitting is a well-known mathematical method that has been used in various contexts. The early application of this idea can also be found in Ref.~\cite{Harlander:2005rq}.
}
In this paper,
the \texttt{Expand\&Fit} method is devised to solve the following types of more complicated MB integrals:
\bea
\label{eq:I1}
I_1 &=&  \int \limits_{- \ri \infty}^{+ \ri \infty} \frac{\rd z_1}{2 \pi \ri} \frac{\rd z_2}{2 \pi \ri} \; x^{z_1} \, f_1 \Big( \Gamma, \psi^{(i)} ; z_1, z_2 \Big) \,, \\[4pt]
I_2 &=&  \int \limits_{- \ri \infty}^{ + \ri \infty} \frac{\rd z_1}{2 \pi \ri} \frac{\rd z_2}{2 \pi \ri} \; x^{z_1} \, y^{z_2} \, f_2 \Big( \Gamma, \psi^{(i)} ; z_1, z_2 \Big)\,,
\label{eq:I2}
\eea
where $I_1$ contains one scale $x$ and possibly nested non-vanishing arc contributions in the $z_2$ integration,
and $I_2$ contains two scales $x$ and $y$, making the calculations much more involved.
In practical calculations of two-loop four-point integrals,
one usually has $x = T/S$ and $y=U/S$. 
Note that there are no specific restrictions or assumptions that $f_1$ and $f_2$ must satisfy.
The \texttt{Expand} step can always be performed given sufficient computational resources,
while the \texttt{Fit} step only requires that the function space is known a priori.

In the applications that we have considered so far, the difficult integrals of $I_2$-type appear only in the non-planar integrals,
as the positive definiteness of  $\mathcal{U}$ and $ \mathcal{F}$ polynomials requires $S,T$ and $U$ to be independent.
This requirement has two reasons:
(1) the expansion-by-regions method in the parametric space requires the positive definite  $ \mathcal{F}$ polynomial to work correctly;
(2)  analytic integrations with positive definite  $ \mathcal{F}$ polynomials are simpler than those with $ \mathcal{F}$ polynomials involving indefinite signs, as the latter can develop singularities.
Given that
$U$ cannot be completely eliminated in terms of $S$ and $T$ without introducing a negative sign in the $ \mathcal{F}$ polynomial,
it can only be partially eliminated by imposing the kinematic relation.
For example, with the kinematic relation $S+T+U=0$,
and since $\mathcal{F}$ is linear in $S,T,U$ and $m^2$,
we can subtract zero-terms $\big[ (\dots)(S+T+U)\big]$ inside the $ \mathcal{F}$ polynomial
while maintaining its positive definiteness in a minimal way.
This polynomial minimisation step is implemented in \texttt{AsyInt}.
Moreover, rewriting $U$ in terms of $S$ and $T$ will introduce additional Mellin transformations that increase the dimensionality of MB representations.
Thus, it cannot simplify the problem with the MB approach.

Note that the simplification of eliminating $U$-dependence will only manifest in the final physical-region results with positive $s$ and $T$.
To arrive at this point, we need to first obtain the full functional dependence of $U$ in the Euclidean region and perform the analytic continuation afterwards.
Therefore, the \texttt{Expand\&Fit} method is well-suited for $I_2$-type integrals, as the two-scale problem in the Euclidean region 
eventually simplifies to a one-scale problem in the physical region.
More details will be discussed in Section~\ref{sec:expfit2scale}.

\subsubsection{\texttt{Expand\&Fit} for one-scale MB integrals}
\label{sec:expfit1scale}
The \texttt{Expand\&Fit} method can be used to solve the $I_1$-type MB integrals as in Eq.~\eqref{eq:I1},
which can contain non-vanishing arc contributions on the semi-circle of the $z_2$-plane.
To keep the technical discussion concise, let us assume that only integer-valued poles appear in $f_1$.
%

In the \texttt{Expand} step, the $z_1$ right-residues are extracted
such that the expansion in the $x \to 0$ limit can be taken.
There are two types of  right-residues for $z_1$
\begin{align}
\label{eq:resZ1}
\begin{cases}
z_1 = 0, \, 1,\,  2,\, \dots \\[2mm]
z_1 = g(z_2), \, g(z_2)+1, \, g(z_2) +2, \, \dots
\end{cases} \,,
\end{align}
where the second-type residue arises from Gamma functions of the type $\Gamma(\dots \pm  z_2 -z_1)$. 
Note that when evaluating at the real values of the $z_1$- and $z_2$-integration contours  (straight lines),
the function $g(z_2) = \dots \pm z_2$ gives the minimal real value such that  $\mathrm{Re}\big(g(z_2)) >  \mathrm{Re}(z_1) > \mathrm{Re}\big(g(z_2)) - 1,$ 
Now taking the first type of residues in Eq.~\eqref{eq:resZ1} yields
\bea
\label{eq:resType1}
I_1^{(1)} &=&  - \int \limits_{- \ri \infty}^{+ \ri \infty}  \frac{d z_2}{2 \pi \ri} \, \sum_{k=0}^{\infty}\mathrm{Res}_{z_1=k} \, x^{z_1} \, f_1 \Big( \Gamma, \psi^{(i)}; z_1, z_2 \Big) \nonumber \\[2pt]
&=& - \sum_{k=0}^{\infty} \sum_{n=0}^{N} \, x^{k} \log(x)^{n} \,  \int \limits_{- \ri \infty}^{+ \ri \infty}  \frac{d z_2}{2 \pi \ri} \, \hat{f}_{1,(k,n)}^{(1)} \Big( \Gamma, \psi^{(i)}; k_1, z_2 \Big) \,,
\eea
where $ \hat{f}_{1,(k,n)}^{(1)}$ and the power-log series in $x$ are the resultant residue functions from taking $k$-th right-residue in $z_1$,
and the negative sign arises from closing the integration contour in the right half of the plane. 
The remaining $z_2$-integrals in Eq.~\eqref{eq:resType1} need to be computed to obtain an analytic finite series for $I_1^{(1)}$.
In order to account for the arc contributions in these $z_2$-integrals,
\texttt{AsyInt} uses the aforementioned numerical reconstruction method with the \texttt{PSLQ} algorithm.

Taking the second type of residues in Eq.~\eqref{eq:resZ1} gives
\bea
\label{eq:resType2}
I_1^{(2)} 
&=& - \sum_{k_1=0}^{\infty} \,  \int \limits_{- \ri \infty}^{+ \ri \infty}  \frac{\rd z_2}{2 \pi \ri} \, x^{g(z_2)+ k_1}  \,  \hat{f}_1^{(2)}\ \Big( \Gamma, \psi^{(i)} ; g(z_2)+ k_1, z_2 \Big)  \nonumber \\
&=& 
\sum_{k= 0}^{k_\max} \sum_{n=0}^{N} \, x^{k} \log(x)^n \, c_{(k,n)}^{(2)} \Big( \Gamma, \psi^{(i)} \Big)
\eea
where in the first line the $z_2$ left- or right-residues need to be taken according to the criteria $g(z_2) |_{z_2 \to \pm \infty} >0$.
Note that in this case, the arc contributions inside the $z_2$-integrals vanish, regulated by the scale $x$ assuming $1>x>0$.
In the second line, the expressions are re-organised into a finite  series, 
and $c_{(k,n)}^{(2)}$ are coefficients expressed in terms of Gamma and PolyGamma functions.
From Eqs.~\eqref{eq:resType1} and \eqref{eq:resType2}, we obtain a  power-log series of the form
\bea
\label{eq:series1}
 I_1 &\overset{ x \to 0}{=}&  I_1^{(1)} + I_1^{(2)} \; = \; \sum_{k=0}^{k_\max} \sum_{n=0}^{N} \,  c_{(k,n)} \, x^k \, \log(x)^{n} \,,
\eea
which we truncate at a finite order $k_\max$.

In a second step we apply a \texttt{Fit} procedure to reconstruct the full functional dependence of $I_1$.
For this purpose, we make an ansatz for which we also expand in the $x \to 0$ limit.
The comparison with Eq.~\eqref{eq:series1} determines the unknown coefficient constants in the ansatz.
%
The ansatz can be constructed from a basis of rational functions in combination with another basis of transcendental functions.
For example, the following weight-5 function bases typically appear in the high-energy massive two-loop four-point integrals calculations 
to $\ord(\eps^1)$
\bea
f_{\rm{RF}} &=& \Big\{ 1 \,, \, \frac{1}{x}\, ,\, \frac{1}{1+x} \,, \, \frac{1}{x^2} \,, \, \frac{1}{(1+x)^2} \,, \, \dots  \Big\}  \,, \nonumber \\
f_{\mathrm{HPL}} &= & \Big\{ H_{0}(x) \,, \, H_{-1}(x) \, ,\, H_{0,-1}(x) \, ,\, \dots \, ,\, H_{0,0,0,0,-1}(x) \Big\} \,,
\eea
in the Euclidean region with $x = T/S$ (or $x = U/S$).
The Harmonic PolyLogarithms (HPLs) are defined by
\bea
H_{k_1,\dots,k_n}(x) &=&\int_0^x \frac{dt_1}{\left| k_1\right| -\mathrm{sgn}\left(k_1\right) \, t_1 }  H_{k_2,\dots,k_n} ( t_1)  \,,
\eea
with $H(x) = 1$ and $H_0(x) = \log(x)$.
The corresponding bases in the physical region can be obtained by the transformation $x \to x' = T/s = -x$.
Note that  fitting higher-order expansion terms in $m$ typically requires a larger basis $f_{\mathrm{RF}}$ with higher-inverse-power rational functions,
while fitting higher-order terms in $\eps$ requires a larger basis $f_{\mathrm{HPL}}$ with higher-weight transcendental functions. 
Now the ansatz can take the form 
\bea
\label{eq:ansatz1}
I_1^{(\mathrm{ans})} &= & \sum_{i,j} \, a_{(i,j)} \, f_{\mathrm{RF}}^{(i)} (x) \, f_{\mathrm{HPL}}^{(j)} (x) \,,
\eea
where $a_{(i,j)}$ denote  the unknown  coefficient constants,
and the indices $i,j$ denote the $i$-th and $j$-th elements of the list $ f_{\mathrm{RF}}$ and $ f_{\mathrm{HPL}}$, respectively.
The series expansion of the ansatz gives
\bea
\label{eq:ansexp1}
I_1^{(\mathrm{ans})} & \overset{x\to 0}{=} &   
\sum_{k=0}^{k_\max} \sum_{n=0}^{N} \, f_{(k,n)}(a_{(i,j)}) \,  x^{k} \, \log(x)^n  \,,
\eea
where the function $f_{(k,n)}$ is expressed in terms of $a_{(i,j)}$ and other constants. 
The value of $k_\max$ depends on the number of unknown coefficients in the ansatz.
Comparing Eq.~\eqref{eq:ansexp1} and Eq.~\eqref{eq:series1} yields relations that form a linear system of equations
\bea
\label{eq:relation1}
f_{(k,n)}(a_{(i,j)}) &=& c_{(k,n)}\,.
\eea
Now the final step of the \texttt{Fit} procedure is to solve the linear system of equations 
to determine the coefficients $a_{(i,j)}$ for the ansatz in Eq.~\eqref{eq:ansatz1}.

Note that for practical calculations, it is advantageous to apply this \texttt{Fit} procedure to a suitable combination of MB integrals in Eq.~\eqref{eq:MBHE} at a particular order,
including the computed ones.
The combination of MB integrals can effectively exploit the cancellation of spurious rational functions in the final expressions,
thereby reducing the size of ansatz in Eq.~\eqref{eq:ansatz1} needed for the \texttt{Fit} procedure.
The \texttt{Expand\&Fit} method also works for more general cases where the transcendental functions are beyond HPLs.
The only requirement is that  the function space must be known a priori.

\subsubsection{\texttt{Expand\&Fit} for two-scale MB integrals}
\label{sec:expfit2scale}
The \texttt{Expand\&Fit} method for the $I_2$-type integrals as in Eq.~\eqref{eq:I2} is more involved.
As mentioned before, this type of integral typically appears in the non-planar two-loop four-point topologies,
in order to keep the  $\mathcal{F}$ polynomial positive definite.
In the following discussions, we have $x = T/S$ and $y=U/S$.

In the \texttt{Expand} step, the $z_1$ right-residues are first extracted in the same way as in Eq.~\eqref{eq:resZ1}.
Taking the first type of $z_1$ residues gives
\bea
\label{eq:res2Type1}
I_2^{(1)} &=& 
- \sum_{k=0}^{\infty} \sum_{n=0}^{N} \, x^{k} \log(x)^{n} \,  \int \limits_{- \ri \infty}^{+ \ri \infty}  \frac{d z_2}{2 \pi \ri} \, y^{z_2} \, \hat{f}_{2,(k,n)}^{(1)} \Big( \Gamma, \psi^{(i)}; k_1, z_2 \Big) \nonumber \\
&=&
\sum_{k=0}^{\infty} \sum_{n=0}^{N} \, x^{k} \log(x)^{n} \, F^{(1)}_{(k,n)}\big(\{H\};y \big) \,,
\eea
where in the second line the $z_2$-integration is computed analytically by the summation method.
The exact function $F^{(1)}_{(k,n)}\big(\{H\};y \big)$ is expressed in terms of HPLs with argument $y$.
Taking the second type of $z_1$ right-residues gives
\bea
\label{eq:res2Type2}
I_2^{(2)} 
&=& - \sum_{k_1=0}^{\infty} \,  \int \limits_{- \ri \infty}^{+ \ri \infty}  \frac{\rd z_2}{2 \pi \ri} \, x^{g(z_2)+ k_1}  \, y^{z_2} \, \hat{f}_2^{(2)}\ \Big( \Gamma, \psi^{(i)} ; g(z_2)+ k_1, z_2 \Big)  \nonumber \\
&=& 
\sum_{k_1= 0}^{\infty} \sum_{k_2= 0}^{\infty} \sum_{n_1=0}^{N_1} \sum_{n_2=0}^{N_2}  \,  c_{(k_1,n_1,k_2,n_2)}^{(2)} \, x^{k_1} \log(x)^{n_1} \, y^{k_2} \log(y)^{n_2} \,.
\eea
At this stage, the full functional dependence of $y$ in the Euclidean region has been obtained,
and we can perform the analytic continuation into the physical region
to take advantage of the simplification in that region.
Since all masses are expanded,
and the $m$-dependence is factorised,
we can use the relation $U = -S - T$ and $S = -s$,
 and perform the transformation $ S - \ri \varepsilon \to s + \ri \varepsilon$ where $\ri \varepsilon$ is the causal prescription.
 This corresponds to the transformations
 \bea
 x \;=\; \frac{T}{-(s+\ri \varepsilon)} \;\to\; x' \; = \;  \frac{T}{s + \ri \varepsilon}  & \; \mbox{and}\; &
 y \;=\;  \frac{s-T}{-(s+\ri \varepsilon)} \;\to\; y' \; = \; \frac{s-T}{s + \ri \varepsilon} \,,
 \eea
and the branch cuts through the analytic continuation
\bea
\log(x) \,=\, \log(x') + \ri \pi  \;\quad \mbox{and} \quad\; \log(y) \,=\, \log(y') + \ri \pi \,.
\eea
The subsequent transformations on the HPLs are needed such that all HPLs are expressed with argument $x'$. 
By default, \texttt{AsyInt} relies on \texttt{HarmonicSums.m} to perform transformations of HPLs.
One can also use other tools 
for this purpose.
Once the transformations are done, the expressions can be re-expanded in the $x' \to 0$ limit 
and truncated to a finite power-log series
\bea
\label{eq:series2}
 I_2 &\overset{ x' \to 0}{=}&  I_2^{(1)} + I_2^{(2)} \Big|_{x,y \to x'} \; = \; \sum_{k=0}^{k_\max} \sum_{n=0}^{N} \,  c'_{(k,n)} \, x'{}^{k} \, \log(x')^{n} \,,
\eea
with complex-valued coefficient constants $c'_{(k,n)}$.

The \texttt{Fit} step is similar to the case in Section~\ref{sec:expfit1scale}, but with a different function basis in the physical region.
Now an ansatz 
\bea
\label{eq:ansatz2}
I_2^{(\mathrm{ans})} &= & \sum_{i,j} \, a'_{(i,j)} \, f_{\mathrm{RF}}^{(i)} (x') \, f_{\mathrm{HPL}}^{(j)} (x') \,,
\eea
can be constructed from the following bases
\bea
f_{\rm{RF}} &=& \Big\{ 1 \,, \, \frac{1}{x'}\, ,\, \frac{1}{1-x'} \,, \, \frac{1}{x'{}^2} \,, \, \frac{1}{(1-x')^2} \,, \, \dots  \Big\}  \,, \nonumber \\
f_{\mathrm{HPL}} &= & \Big\{ H_{0}(x') \,, \, H_{1}(x') \, ,\, H_{0,1}(x') \, ,\, H_{0,0,1}(x')  \,,\, \dots \, \Big\} \,.
\eea
By expanding the ansatz in the $x' \to 0$ limit and comparing to Eq.~\eqref{eq:series2},
we can determines the coefficient constants $a'_{(i,j)}$ in Eq.~\eqref{eq:ansatz2}.

\section{Overview of \texttt{AsyInt}}
\label{sec:workflow}
In this section, we will overview the program structure of \texttt{AsyInt} 
by presenting the workflow, main commands, and a simple one-loop example.
With \texttt{AsyInt},  users can perform analytic calculations of two-loop four-point integrals with a few simple commands.
More details of two-loop calculations will discussed in Section~\ref{sec:results}.

\subsection{Workflow and main commands}
\begin{figure}[tb]
  \centering
   \includegraphics[width=0.95\textwidth]{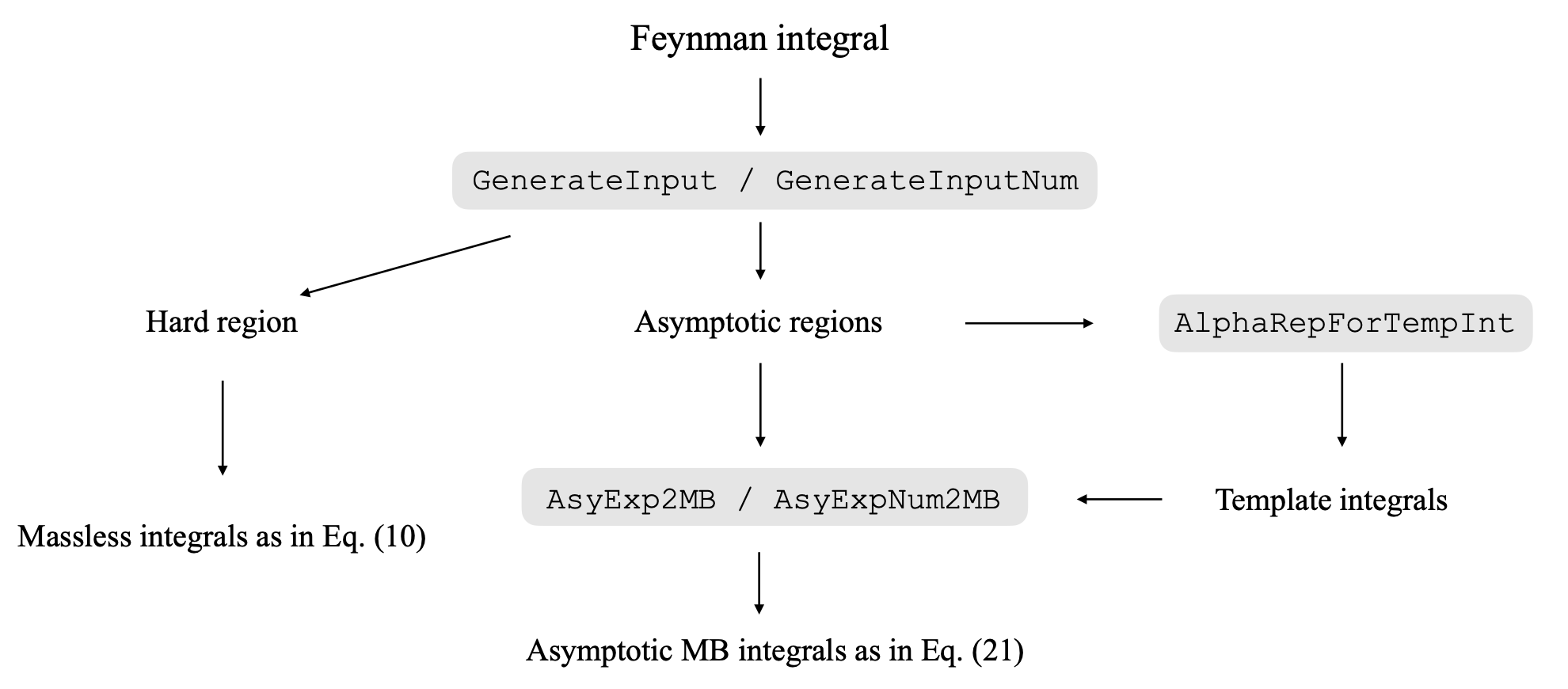} 
  \caption{\label{fig:workflow1}
  Workflow of \texttt{AsyInt} toolkit I: generate integrals.
    }
\end{figure}

\begin{figure}[tb]
  \centering
   \includegraphics[width=.9\textwidth]{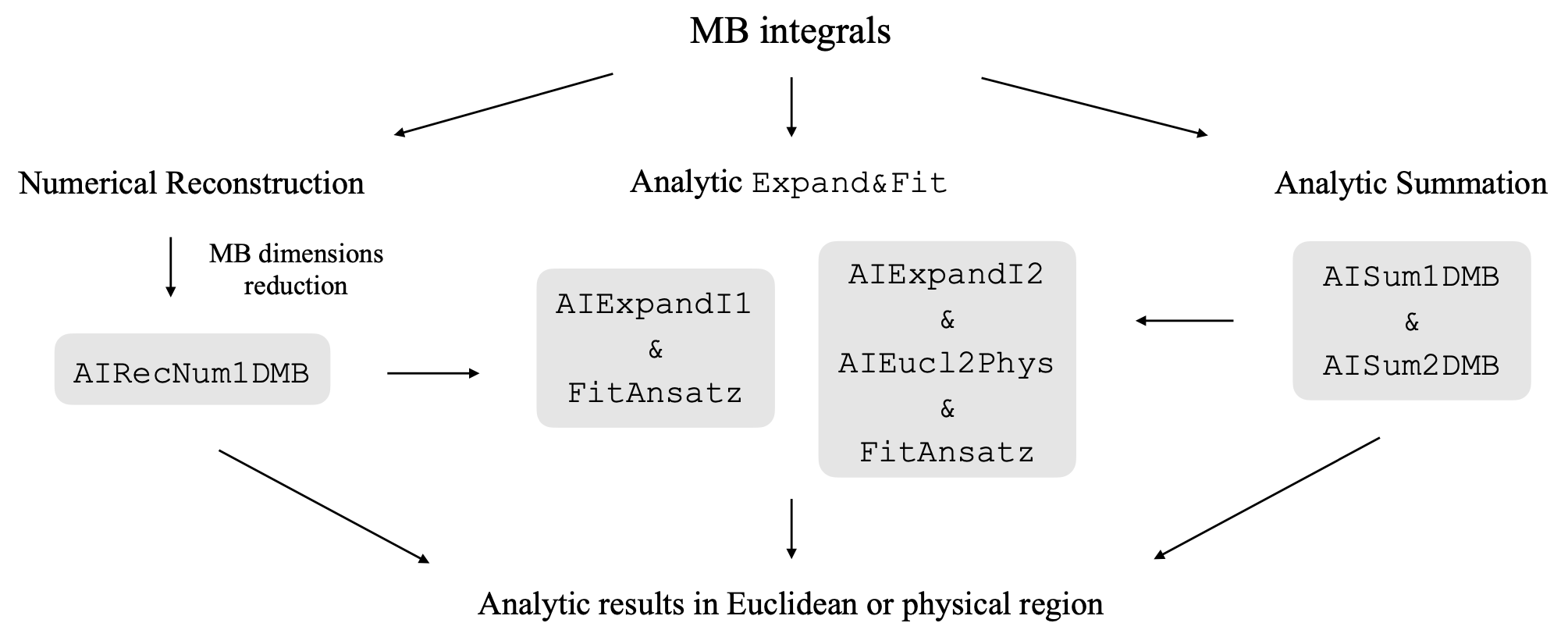} 
  \caption{\label{fig:workflow2}
  Workflow of \texttt{AsyInt} toolkit II: solve integrals.
    }
\end{figure}

\texttt{AsyInt} contains two toolkits: the first is for generating MB integrals, and the second is for solving the MB integrals.
Their workflows are shown in Figs.~\ref{fig:workflow1} and \ref{fig:workflow2}, respectively.
The \texttt{AsyInt} main commands (or interfaces) are depicted inside grey blocks, 
and their inputs and outputs are summarised in Appendix~\ref{app:func}.
%
%

In toolkit I, \texttt{AsyInt} relies on \texttt{asy2.1.m} for finding the regions and \texttt{MB.m} for resolving singularities.
By specifying the small expansion parameter, 
the command
\[
\texttt{GenerateInput}  \, \mbox{ or }\,
\texttt{GenerateInputNum}
\]
generates the Symanzik polynomials and returns relevant regions with corresponding scalings,
for the Feynman integrals without or with numerators, respectively.
For the hard region, the massless master integrals up to weight-6 are provided in the \texttt{AsyInt} repository.
With these master integrals, users can straightforwardly obtain the expansion of hard regions integrals, as shown in Eq.~\eqref{eq:hard}. 
For the asymptotic regions,
the command
\[ \texttt{AlphaRepForTempInt}\]
generates the alpha representation for the template integrals.
Note that the MB representations of template integrals are not unique.
It depends on the users' choice on the change of variables and Mellin transformations.
\texttt{AsyInt} does not offer an automated routine for integrating alpha parameters,
but it does provide helper functions such as \texttt{IntTypeA}, \texttt{IntTypeB} and \texttt{MBsplit} commands, which enable users to perform integrations.
With the scalings and template integrals,
the command 
\[
\texttt{AsyExp2MB}  \, \mbox{ or } \, \texttt{AsyExpNum2MB} 
\]
automatically performs the steps discussed in Section~\ref{sec:tempint} 
and generates MB integrals to the desired order in the expansion parameter and the dimensional regulator $\eps$,
for Feynman integrals without and with numerators, respectively.
Integrals with higher-power denominators can also be generated by providing the \texttt{DotShift} optional input.

In toolkit II, \texttt{AsyInt} relies on \texttt{MB.m} for numerical integrations and \texttt{HarmonicSums.m}, \texttt{Sigma.m} and \texttt{EvaluateMultiSums.m}
for analytic summations of residues.
The analytic summation approach for scaleless and one-scale MB integrals are implemented in the commands
\[
\texttt{AISum1DMB} \, \mbox{ and } \, \texttt{AISum2DMB}
\]
for one- and two-dimensional integrals, respectively.
However,  due to the potential presence of non-vanishing arc contributions,
it is advised to always verify the results against numerical evaluations using \texttt{MB.m}.
If the numerical validation fails, users can resort to the \texttt{Expand\&Fit} method.
For scaleless MB integrals, the numerical reconstruction method is implemented in the command
\[
\texttt{AIRecNum1DMB}
\]
with a user-defined constant list, which can be built from the basis in Eq.~\eqref{eq:constlist}.
Note that as mentioned before, \texttt{AsyInt} is not designed to solve the MB dimensionality reduction problem.
Therefore, users are required to perform the MB dimensionality reduction themselves if necessary.

The above commands \texttt{AISum1DMB} and \texttt{AIRecNum1DMB} further serve as the building blocks
for the analytic \texttt{Expand\&Fit} method.
For $I_1$-type integrals as in Eq.\eqref{eq:I1},
the command
\[
\texttt{AIExpandI1}
\]
performs the expansion in the $x\to0$ limit and employs the \texttt{AIRecNum1DMB} command to evaluate the nested integration.
The output of the \texttt{AIExpandI1} command can be further processed by the fitting command
\[
\texttt{FitAnsatz}
\]
with a user-defined ansatz as in Eq.~\eqref{eq:ansatz1}.
For $I_2$-type integrals as in Eq.\eqref{eq:I2},
the command 
\[
\texttt{AIExpandI2}
\]
performs the expansion in the $x\to0$ limit and employs the \texttt{AISum1DMB} command to evaluate the nested integration with $y$-dependence.
Note that obtaining higher-order expansion terms for $I_2$-type integrals with analytic $y$-dependence can be computationally very expensive.
Therefore, users can switch to the lower-level command
\[
\texttt{AIExpand2DMB}
\]
in combination with the \texttt{AISum1DMB} command for parallel computations on a computing cluster.
The command
\[
\texttt{AIEucl2Phys}
\]
further processes the output of the \texttt{AIExpandI2} command for the analytic continuation from the Euclidean region to the physical region. 
Now the output is rewritten in terms $x'=-x$ and can be re-expanded in the $x'\to 0$ limit.
The re-expanded results are passed to the command \texttt{FitAnsatz} to perform the fitting procedure with an ansatz as in Eq.~\eqref{eq:ansatz2}.

\subsection{A simple example: one-loop box integral}
\begin{figure}[tb]
  \centering
    \scalebox{-1}[1]{\includegraphics[width=.24\textwidth]{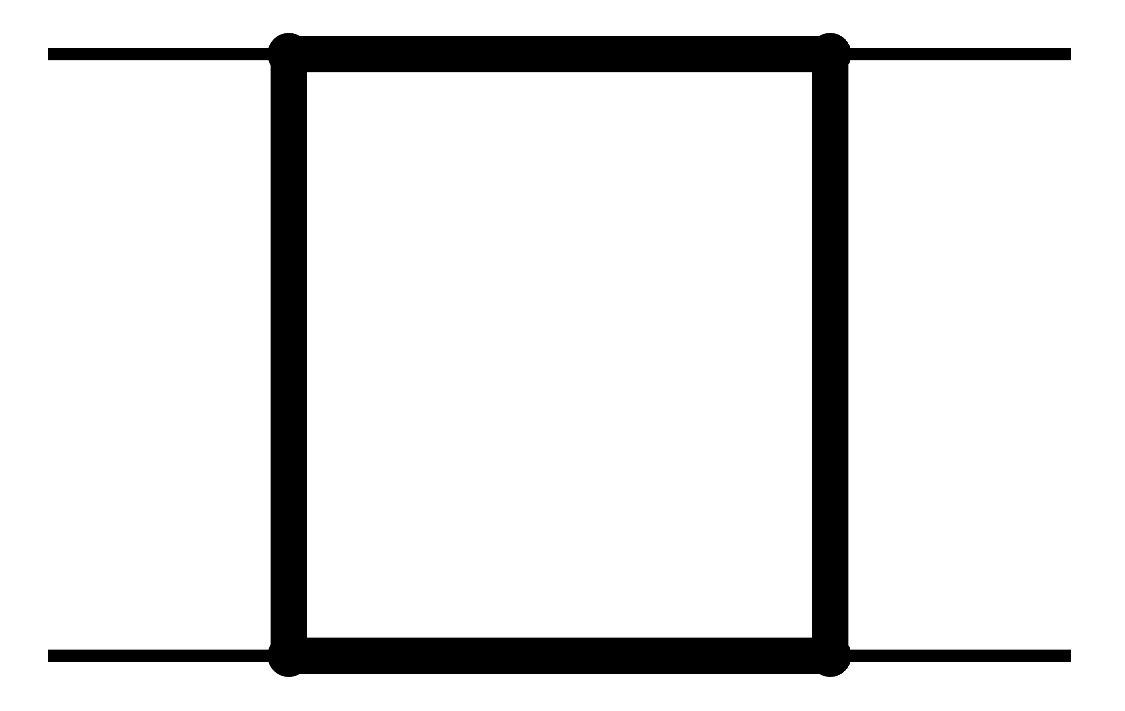} }
  \caption{\label{fig:box1l}
   One-loop box diagram.
    }
\end{figure}
In this subsection, we provide a simple example of a massive one-loop box integral, shown in Fig.~\ref{fig:box1l}, to demonstrate the basic usage of \texttt{AsyInt}.
With \texttt{AsyInt}, we can compute this integral to $\ord(\eps^2)$ and $\ord(m^{8})$ efficiently.
We will keep this subsection brief and provide more detailed discussions of calculational steps for the two-loop cases in Section~\ref{sec:results}.

The propagators of this integral are
\bea
\Big\{m^2-l^2,m^2-(l+q_2)^2,m^2-(l-q_1-q_3)^2,m^2-(l-q_1)^2\Big\}
\eea
where the external kinematics are described by
\bea
q_1^2 = q_2^2 = q_3^2= 0 \, , \quad (q_1+q_2)^2=s\, ,  \quad (q_1+q_3)^2=t\, , \quad (q_2+q_3)^2=u\,.
\eea
The Euclidean invariants are $S=-s, \, T=-t, \, U=-u$.

First, we use the  \texttt{GenerateInput} command to generate
the Symanzik polynomials 
\bea
\mathcal{U} &=& \alpha_1 + \alpha_2 + \alpha_3 + \alpha_4 \,, \\[2pt]
\mathcal{F} &=& m^2 \, (\alpha_1+\alpha_2+\alpha_3+\alpha_4)^2+ S \,  \alpha_1 \alpha_3+ T\, \alpha_2 \alpha_4
\eea
and also all regions with corresponding scalings under the hierarchy $S,T,U \sim 1$ and $m^2 \sim \rho \ll 1$.
They are listed in Table~\ref{tab:box1lregion}.
\begin{table}[b]
\centering
\begin{tabular}{ccccc}
\text{Region} & $\alpha_1$ & $\alpha_2$ & $\alpha_3$ & $\alpha_4$ \\ \hline
$R_1$ & 0 & 0 & 0 & 0 \\
$R_2$ & 0 & 0 & 1 & 1 \\
$R_3$ & 0 & 1 & 1 & 0 \\
$R_4$ & 1 & 0 & 0 & 1 \\
$R_5$ & 1 & 1 & 0 & 0 \\
\end{tabular}
\caption{Scalings of alpha parameters for the one-loop box integral, i.e. $\alpha_i \sim \rho^p$ where $p$ is the power in the table. $R_1$ is the hard region  and $R_2, \dots, R_4$ are the asymptotic regions.}
\label{tab:box1lregion}
\end{table}
Since the one-loop example is simple,
it is not necessary to compute the hard region integral separately.
We can treat all five regions on the same footing using the template integrals approach.

We then use the \texttt{AlphaRepForTempInt} command to generate the alpha representations,
and the MB representations are obtained by using the integration commands: \texttt{IntTypeA}, \texttt{IntTypeB} and \texttt{MBsplit}.
The following abbreviations are adopted for later convenience
\bea
\Gamma[x_1, \dots, x_n] \; := \; \prod_{i=1}^n \Gamma(x_i) \;, \quad \; \delta_{i_1 \dots i_n} \; := \; \sum_{k=1}^n \delta_{i_k} \;. 
\eea
The MB representations are
\bea
\calT_{\rm box1l}^{(1)}  \!\! &=& \!\! 
\int \limits_{- \ri \infty}^{+ \ri \infty}  \frac{\rd z_1}{2 \pi \ri} 
\frac{1}{S^{\delta _{1234}+\epsilon +2}} \lb \frac{T}{S} \rb^{z_1}
\frac{\Gamma\left[ -\delta _{413}-z_1-\epsilon -1,\delta _{1234}+z_1+\epsilon +2,-z_1 \right] }{\Gamma\left[\delta _3+1,\delta _4+1,-\delta _{1234}-2 \epsilon ,\delta _1+1,\delta _2+1\right]}
 \nonumber \\[2pt]
&&\hspace{-1.2cm}  \times \, \Gamma\big[ \delta _1+z_1+1,\delta _3+z_1+1,-\delta _{312}-z_1-\epsilon -1 ]
 \,,  
\\[4pt]
\calT_{\rm box1l}^{(2)}  \!\! &=& \!\! 
\frac{ T^{-\delta _3-1} m^{2 \left(-\delta _{12}-\epsilon \right)} }{S^{\delta _4+1} }
\frac{ \Gamma\left[\delta _{12}+\epsilon ,\delta _1-\delta _3,\delta _2-\delta _4\right]}{\Gamma\left[\delta _{12}-\delta _{34},\delta _1+1,\delta _2+1\right]}
\,,
\\[4pt]
\calT_{\rm box1l}^{(3)}  \!\! &=& \!\! 
\frac{ T^{-\delta _3-1} m^{2 \left(-\delta _{14}-\epsilon \right)} }{S^{\delta _2+1}}
\frac{  \Gamma\left[\delta _{14}+\epsilon ,\delta _1-\delta _3,\delta _4-\delta _2\right]}{\Gamma\left[\delta _{14}-\delta _{23},\delta _1+1,\delta _4+1\right]}
\,,
\\[4pt]
\calT_{\rm box1l}^{(4)}  \!\! &=& \!\! 
\frac{ T^{-\delta _1-1} m^{2 \left(-\delta _{23}-\epsilon \right)} }{S^{\delta _4+1} }
\frac{ \Gamma\left[ \delta _{23}+\epsilon ,\delta _3-\delta _1,\delta _2-\delta _4\right]}{\Gamma\left[\delta _{23}-\delta _{14},\delta _2+1,\delta _3+1\right]}
\,,
\\[4pt]
\calT_{\rm box1l}^{(5)}  \!\! &=& \!\! 
\frac{ T^{-\delta _1-1} m^{2 \left(-\delta _{34}-\epsilon \right)} }{S^{\delta _2+1}}
\frac{ \Gamma\left[\delta _{34}+\epsilon ,\delta _3-\delta _1,\delta _4-\delta _2\right]}{\Gamma\left[\delta _{34}-\delta _{12},\delta _3+1,\delta _4+1\right]}
\,.
\eea
From the template integrals, we use the \texttt{AsyExp2MB} command to generate MB integrals up to $\ord(\eps^2)$ and $\ord(m^8)$.
We use the \texttt{SortPatternList} command to select all one-dimensional MB integrals.
In the next step, we use the \texttt{AISum1DMB} command to solve  these MB integrals order-by-order in $\eps$.
In the last step, we use the \texttt{AIEucl2Phys} command to perform the analytic continuation to the physical region.
The results to $\ord(\eps^2)$ and $\ord(m^8)$ are expressed as
\bea
\calI_{\rm box1l} [1,1,1,1] & = &
\lb \frac{\mu^2}{s} \rb^{2 \epsilon} \sum_{i=0}^{2} \sum_{j = 0}^{4} \,  \eps^{i} \, \lb m^2\rb^{j}  \, f_{(i,j)}\big(s,T,m^2 \big)\,,
\eea
with the coefficient functions
\bea
f_{(0,0)} &=&
\frac{1}{s T} \bigg[
        \pi ^2
        +2 i \pi  H_0\big(
                \hat{T}\big)
        -2 H_0\big(
                \hat{m}^2\big)^2
        +2 H_0\big(
                \hat{m}^2
        \big)
\big(-i \pi +H_0\big(
                        \hat{T}\big)\big)
\bigg] \,, 
\\[4pt]
f_{(1,0)} &=&
\frac{1}{s T} \bigg[
        \frac{i \pi ^3}{3}
        +\big(
                -\frac{4 \pi ^2}{3}
                -2 i \pi  H_1\big(
                        \hat{T}\big)
                +2 H_{0,1}\big(
                        \hat{T}\big)
        \big) H_0\big(
                \hat{T}\big)
        -\frac{1}{3} H_0\big(
                \hat{T}\big)^3
        \nonumber\\
        &&+\frac{2}{3} \pi ^2 H_0\big(
                \hat{m}^2\big)
        +\frac{4}{3} H_0\big(
                \hat{m}^2\big)^3
        +2 i \pi  H_{0,1}\big(
                \hat{T}\big)
        -2 H_{0,0,1}\big(
                \hat{T}\big)
        \nonumber\\
        &&+i H_0\big(
                \hat{m}^2\big)^2 \big(
                \pi +i H_0\big(
                        \hat{T}\big)\big)
        -i H_0\big(
                \hat{T}\big)^2 \big(
                \pi -i H_1\big(
                        \hat{T}\big)\big)
        +14 \zeta (3)
\bigg] \,,
\\[4pt]
f_{(2,0)} &=&
\frac{1}{s T} \bigg[
        \frac{7 \pi ^4}{180}
        +\big(
                -\frac{i \pi ^3}{6}
                +i \pi  H_1\big(
                        \hat{T}\big)^2
                +2 H_{0,1,1}\big(
                        \hat{T}\big)
                +H_1\big(
                        \hat{T}
                \big)
\big(\pi ^2-2 H_{0,1}\big(
                                \hat{T}\big)\big)
                \nonumber\\
                &&-6 \zeta (3)
        \big) H_0\big(
                \hat{T}\big)
        +\big(
                \frac{\pi ^2}{2}
                +i \pi  H_1\big(
                        \hat{T}\big)
                +\frac{1}{2} H_1\big(
                        \hat{T}\big)^2
                -H_{0,1}\big(
                        \hat{T}\big)
        \big) H_0\big(
                \hat{T}\big)^2
        \nonumber\\
        &&+\frac{1}{6} H_0\big(
                \hat{T}\big)^4
        -\frac{1}{2} \pi ^2 H_0\big(
                \hat{m}^2\big)^2
        -\frac{1}{2} H_0\big(
                \hat{m}^2\big)^4
        +\big(
                \frac{i \pi ^3}{3}
                -2 i \pi  H_{0,1}\big(
                        \hat{T}\big)
                \nonumber\\
                &&+2 H_{0,0,1}\big(
                        \hat{T}\big)
                -2 \zeta (3)
        \big) H_1\big(
                \hat{T}\big)
        -\pi ^2 H_{0,1}\big(
                \hat{T}\big)
        -2 i \pi  H_{0,0,1}\big(
                \hat{T}\big)
        \nonumber\\
        &&+2 i \pi  H_{0,1,1}\big(
                \hat{T}\big)
        +2 H_{0,0,0,1}\big(
                \hat{T}\big)
        -2 H_{0,0,1,1}\big(
                \hat{T}\big)
        \nonumber\\
        &&+\frac{1}{3} H_0\big(
                \hat{m}^2\big)^3 \big(
                -i \pi +H_0\big(
                        \hat{T}\big)\big)
        +\frac{1}{3} H_0\big(
                \hat{T}\big)^3 \big(
                i \pi +2 H_1\big(
                        \hat{T}\big)\big)
        \nonumber\\
        &&+H_0\big(
                \hat{m}
                ^2
        \big)
\big(-\frac{i \pi ^3}{6}+\frac{1}{6} \pi ^2 H_0\big(
                        \hat{T}
                \big)
-4 \zeta (3)\big)
        +4 i \pi  \zeta (3)
\bigg] \,,
\eea
where $\hat{T} = T/s$, $\hat{m}^2 = m^2/s$.
The higher-order terms to $\ord(m^8)$ can be found in the ancillary file~\cite{ttplink}.
These results are cross checked against numerical evaluations with \texttt{AMFlow}~\cite{Liu:2022chg} on various phase space points in the high-energy limit.
For example, with the phase space point $\sqrt{s} = 2$~TeV, $p_T = \sqrt{u\,t/s} = 400$~GeV and $m=80$~GeV, the results to   $\ord(m^8)$ agree with \texttt{AMFlow} to 5 digits.
The accuracy can be easily improved by computing high-order expansion terms in $m$.
For the deeper expansion, we recommend the differential equations approach as discussed in Refs.~\cite{Davies:2022ram, Davies:2018qvx}.
%

\section{Analytic results of massive two-loop integrals}
\label{sec:results}
In this section, we will discuss the calculational steps and present the analytic results
for two-loop $\mathrm{PL_1}$, $\mathrm{NPL_1}$ and $\mathrm{NPL_2}$ integrals shown in Fig.~\ref{fig:diags}.
Note that the fully-massive planar integral $\mathrm{PL_2}$ (with and without numerators)  also calculated by \texttt{AsyInt} has been discussed in Ref.~\cite{Davies:2022ram}.
%

\subsection{Planar integral $\mathrm{PL_1}$ with numerators}
%
As a first two-loop example, we will discuss the three-massive-line planar integral $\rm PL_1$ with numerators in more detail.
\begin{figure}[b]
  \centering
    \scalebox{-1}[1]{\includegraphics[width=.28\textwidth]{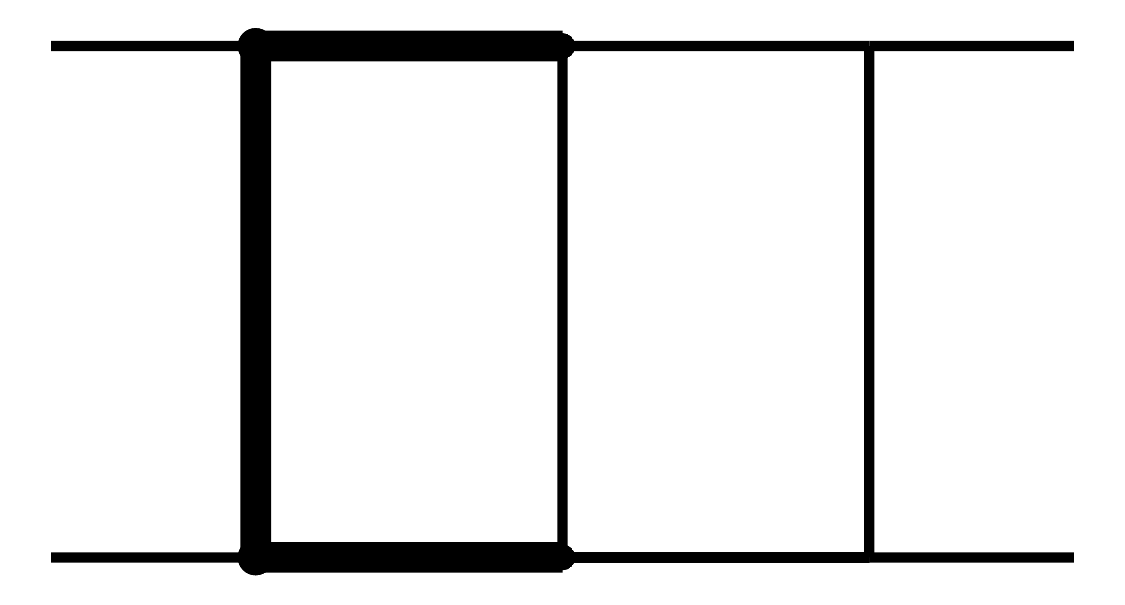} }
  \caption{\label{fig:pl1}
   Two-loop $\mathrm{PL_1}$ diagram.
    }
\end{figure}
The $\mathrm{PL_1}$ integral is defined through the propagators
\bea
& \Big\{ 
 m^2-l_2^2, \; -l_l^2,\; m^2-\left(l_2+q_1+q_2\right){}^2,\; m^2-\left(l_2-q_3\right){}^2,\; -\left(l_l+q_1\right){}^2,\; -\left(l_l-q_2\right){}^2, \nonumber \\
&-\left(-l_2+l_l-q_2\right){}^2,\; -\left(l_2+l_l\right){}^2,\; -\left(l_l+q_3\right){}^2
\Big \}
\eea
where the last two propagators denote the irreducible numerators.
The kinematics of external lines are given by
\bea
\label{eq:kin}
q_1^2 = q_2^2 = q_3^2= 0 \, , \quad (q_1+q_2)^2=s\, ,  \quad (q_1+q_3)^2=t\, , \quad (q_2+q_3)^2=u\,,
\eea
with the relation $s+t+u =0 $. 
In the direct integration approach with \texttt{AsyInt}, we start with the Euclidean kinematics $\{S,T,U\} = -\{ s,t,u\}$ 
by assuming $S, T$ and $U$ to be positive.
Note that in the physical region, our ``Minkowski" kinematics is $s,T,U > 0$.
For the present discussion of $\rm PL_1$ integral with numerators, 
the extended Symanzik polynomials associated with the nine propagators are 
\bea
\tilde{\mathcal{U}} &=& \left(\alpha _1+\alpha _3+\alpha _4\right) \left(\alpha _2+\alpha _5+\alpha _6\right)+\left(\alpha _1+\alpha _2+\alpha _3+\alpha _4+\alpha _5+\alpha _6\right) \alpha _7 \nonumber \\
&& +\left(\alpha _1+\alpha _2+\alpha _3+\alpha _4+\alpha _5+\alpha _6+4 \alpha _7\right) \alpha _8+\left(\alpha _1+\alpha _3+\alpha _4+\alpha _7+\alpha _8\right) \alpha _9 \,,
 \nonumber \\[4pt]
\tilde{\mathcal{F}} &=& \left(\alpha _1+\alpha _3+\alpha _4\right) \Big[\alpha _3 \alpha _5+\alpha _4 \alpha _5+\alpha _7 \alpha _5+\alpha _8 \alpha _5+\alpha _3 \alpha _6+\alpha _4 \alpha _6+\alpha _3 \alpha _7+\alpha _4 \alpha _7+\alpha _6 \alpha _7 \nonumber \\
&&+\alpha _3 \alpha _8+\alpha _4 \alpha _8+\alpha _6 \alpha _8+4 \alpha _7 \alpha _8+\alpha _2 \left(\alpha _3+\alpha _4+\alpha _7+\alpha _8\right)+\left(\alpha _3+\alpha _4+\alpha _7+\alpha _8\right) \alpha _9 \nonumber \\
&& +\alpha _1 \left(\alpha _2+\alpha _5+\alpha _6+\alpha _7+\alpha _8+\alpha _9\right)\Big] \, m^2 
 +\Big[\alpha _5 \big(2 \alpha _7 \alpha _8+\alpha _6 \left(\alpha _7+\alpha _8\right)+\left(\alpha _7+\alpha _8\right) \alpha _9 \nonumber \\
 &&+\alpha _4 \left(\alpha _6+\alpha _8+\alpha _9\right)\big)+\alpha _1 \left(\alpha _2 \alpha _3+\left(\alpha _5+\alpha _6+\alpha _7+\alpha _8+\alpha _9\right) \alpha _3+\alpha _5 \left(\alpha _6+\alpha _7+\alpha _9\right)\right) \nonumber \\
 &&+\alpha _3 \left(\alpha _2 \alpha _8+\left(\alpha _5+\alpha _7\right) \left(\alpha _6+2 \alpha _8+\alpha _9\right)\right)\Big] \, S
 +\Big[\alpha _2 \alpha _4 \alpha _7+\big(\left(\alpha _5+\alpha _6\right) \alpha _7 \nonumber \\
 &&+\alpha _1 \left(\alpha _5+\alpha _6+\alpha _7\right)+\alpha _3 \left(\alpha _5+\alpha _6+\alpha _7\right)+\left(\alpha _5+\alpha _6+2 \alpha _7\right) \alpha _8\big) \alpha _9 \nonumber \\
 && +\alpha _4 \left(\alpha _5+\alpha _6+2 \alpha _7\right) \left(\alpha _8+\alpha _9\right)\Big] \, U \,,
 \eea
 where the  relation $S+T+U =0 $ has been used.
The  Symanzik polynomials associated with the seven-propagator Feynman diagram can be obtained by
\bea
\calU \;=\; \tilde{\calU}\Big|_{\alpha_8 = \alpha_9 = 0}\quad \text{and} \quad \calF = \tilde{\calF}\Big|_{\alpha_8 = \alpha_9 = 0}\,.
\eea

In the following, we will concentrate on the calculation of the one-numerator (8-th propagator) top-sector integral $\calI_{\rm PL_1}[1,1,1,1,1,1,1,-1,0] $  to  $\ord(m^0)$,
whose complexity is similar to the integral without numerators $\calI_{\rm PL_1}[1,1,1,1,1,1,1,0,0]$  to  $\ord(m^2)$.
The alpha representation of this one-numerator integral can be obtained by taking derivative w.r.t. $\alpha_8$ as described in Eq.~\eqref{eq:alphaNum} and~\eqref{eq:alphaNum2} such that
\bea
\label{eq:pl1alpha}
\calI_{\rm PL_1}[1,1,1,1,1,1,1,-1,0]   \, =\,
 \int_0^\infty\rd^7 \alpha^\delta  \,
  \left[\, \mathcal{U}^{-d/2} \, e^{\cal -F/U} \, \right] \, \hat{\calO}_1 \Big( \{S,U,m^2 \}, \{\alpha_1,\dots,\alpha_7\} \Big) \;, 
\eea
where $\hat{\calO}_1$ is the resulting rational function from the derivative operation. 
At this stage, one observes that the alpha representation of this integral  only depends on seven alpha parameters,
hence the high-energy asymptotic expansion directly applies to the $\calU, \calF$ polynomials and the $\hat{\calO}_1$ function, which are associated to the seven-propagator Feynman  diagram.
\texttt{AsyInt} relies on \texttt{asy2.1.m} to find all asymptotic regions by imposing the kinematic scaling $S,U \sim 1$ and $m^2 \sim \rho \ll 1$. The resulting scalings of the seven alpha parameters in all regions are listed in Table~\ref{tab:pl1region}.
\begin{table}[bt]
\centering
\begin{tabular}{cccccccc}
\text{Region} & $\alpha_1$ & $\alpha_2$ & $\alpha_3$ & $\alpha_4$ & $\alpha_5$ & $\alpha_6$ & $\alpha_7$ \\ \hline
$R_1$ & 0 & 0 & 0 & 0 & 0 & 0 & 0 \\
$R_2$ & 0 & 0 & 1 & 1 & 1 & 0 & 0 \\
$R_3$ & 0 & 1 & 1 & 0 & 1 & 0 & 0 \\
$R_4$ & 0 & 1 & 1 & 0 & 1 & 1 & 1 \\
$R_5$ & 1 & 0 & 0 & 1 & 0 & 1 & 0 \\
$R_6$ & 1 & 0 & 1 & 2 & 1 & 1 & 0 \\
$R_7$ & 1 & 1 & 0 & 0 & 0 & 1 & 0 \\
$R_8$ & 1 & 1 & 0 & 0 & 1 & 1 & 1 \\
\end{tabular}
\caption{Scalings of alpha parameters for the $\rm PL_1$ integral, i.e. $\alpha_i \sim \rho^p$ where $p$ is the power in the table. $R_1$ is the hard region  and $R_2, \dots, R_8$ are the asymptotic regions.}
\label{tab:pl1region}
\end{table}
Note that the above Symanzik polynomials, the relevant regions and scalings can be obtained by using the \texttt{GenerateInputNum} command.

Among these regions, the hard region $R_1$ yields  massless integrals, which can be conveniently calculated by the canonical differential equation approach with \texttt{Canonica}~\cite{Meyer:2017joq} in combination with IBP reductions by \texttt{LiteRed}~\cite{Lee:2012cn}, using the Taylor expansion shown in~Eq.~\eqref{eq:hard}.

For the asymptotic regions $R_2,\dots,R_8$, \texttt{AsyInt} employs the template-integral approach as described in Section~\ref{sec:tempint}.
By applying the scalings of the alpha parameters and invariants to $\calU, \calF$ and $\hat{\calO}_1$ in Eq.~\eqref{eq:pl1alpha},
one can re-expand the alpha representation in $\sqrt{\rho}$ and obtain the expanded alpha representation for each asymptotic region $r$ as in
Eq.~\eqref{eq:alphaHE} with additional shift-operators $\hat\calS_{1,r}$.
Then the template integral $\calT_{\rm PL_1}^{(r)}$ for each region can be extracted as in Eq.~\eqref{eq:template},
such that the sum of asymptotic regions can be written as in Eq.~\eqref{eq:templateHE} by
\bea
\label{eq:tempHEpl1}
\calI_{\rm PL_1}^{\rm (asy)} [1,1,1,1,1,1,1,-1,0] &=&
\sum_{r=2}^8
\Bigg[ \, \sum_{j=0}^{\infty}\, \sqrt{\rho}^{\, j} \, \hat{\calS}_{1,r}^{(j)} \, \circ \, \calT_{\rm PL_1}^{(r)} (\{ \delta_i \}, \eps) \, \Bigg]  \;.
\eea
Note that the template integrals $\calT_{\rm PL_1}^{(r)}$ describe the leading-order high-energy behaviour
of the original Feynman diagram with the first seven propagators,
and the shift operators $\hat{\calS}_{1,r}$ encodes both the numerator and higher-order high-energy expansion effects.
As the template integrals in different regions have different leading-order scaling of $m$,
the  $\sqrt{\rho}$-scaling  in Eq.~\eqref{eq:tempHEpl1} does not have a one-to-one correspondence with the $m$-scaling in the final results.

The alpha representations of template integrals are generated by the \texttt{AlphaRepForTempInt} command.
Their MB representations  can be obtained  through using the commands: \texttt{IntTypeA}, \texttt{IntTypeB} and \texttt{MBsplit}.
%
We obtain the  template integrals
\bea
\calT_{\rm PL_1}^{(2)}  \!\! &=& \!\! 
\int \limits_{- \ri \infty}^{+ \ri \infty}  \frac{\rd z_1}{2 \pi \ri}
 \frac{U^{-\delta _4-1}  m^{-2 \delta _{1267}-4 \epsilon }}{ S^{\delta _{35}+2} }
\frac{ \Gamma\left[-z_1,\delta _2-\delta _4+z_1,-\delta _{25}-\epsilon ,-\delta _{26}-\epsilon, \delta _{1267}+2 \epsilon   \right]}{\Gamma\left[-\delta _{534}-\epsilon -1,\delta _1+1,-\delta _2-\epsilon +1,\delta _2+1 \right]} \nonumber \\
&&\hspace{-1.2cm}  \times \frac{\Gamma\big[-\delta _{27}+\delta _4-z_1-\epsilon +1,-\delta _{534}+z_1-\epsilon -1,\delta _{726}-\delta _4+z_1+\epsilon \big]}{\Gamma \big[ \delta _6+1,\delta _7+1,-\delta _{45}+z_1-\epsilon  \big] } \,,  
\\[4pt]
\calT_{\rm PL_1}^{(3)}  \!\! &=& \!\! 
\int \limits_{- \ri \infty}^{+ \ri \infty}  \frac{\rd z_1}{2 \pi \ri} \frac{\rd z_2}{2 \pi \ri}
\frac{U^{-\delta _2-1} m^{-2 \delta _{1467}-4 \epsilon }}{  S^{\delta _{35}+2} }
\frac{ \Gamma\left[-\delta _2+\delta _4+z_1+z_2,-\delta _7+z_2-\epsilon +1, -z_1,-z_2 \right]}{\Gamma\left[-\delta _{523}-\epsilon -1,\delta _1+1,-\delta _2+z_1+z_2-\epsilon +1 \right]} \nonumber \\
&& \hspace{-1.2cm} \times \frac{\Gamma\big[ -\delta _{23}+z_1+z_2-\epsilon ,-\delta _{26}+z_1-\epsilon ,\delta _{67}+\epsilon ,-\delta _{523}+z_1-\epsilon -1,\delta _{1267}-z_1-z_2+2 \epsilon \big]}{\Gamma \big[ \delta _4+1,\delta _6+1,\delta _7+1,-\delta _{23}+z_1-\epsilon \big]} \,,
 \\[4pt]
\calT_{\rm PL_1}^{(4)}  \!\! &=& \!\! 
 \int \limits_{- \ri \infty}^{+ \ri \infty}  \frac{\rd z_1}{2 \pi \ri} \frac{\rd z_2}{2 \pi \ri}
 \frac{ m^{-2 \delta _{14}-2 \epsilon } }{S^{\delta _{23567}+\epsilon +3}} \lb \frac{U}{S} \rb^{z_1}
\frac{ \Gamma\left[\delta _4+z_1+1,\delta _7+z_1+z_2+1,\delta _{14}+\epsilon \right]}{\Gamma\left[\delta _1+1,\delta _2+1,\delta _4+1,\delta _5+1,\delta _6+1,\delta _7+1 \right]} \nonumber \\
&& \hspace{-1.2cm} \times \frac{\Gamma \big[ -\delta _{725}-z_1-z_2-\epsilon -1,-\delta _{726}-z_1-\epsilon -1,\delta _{2567}+z_1+z_2+\epsilon +2,-z_1,-z_2 \big]}{\Gamma \big[ \delta _{14}-\delta _3+z_1+z_2+1,-\delta _{2567}-2 \epsilon \big]} \nonumber \\
&& \hspace{-1.2cm}  \times \, \Gamma \big[\delta _1-\delta _3+z_2 , \delta _2+z_1+1 \big] \,,
 \\[4pt]
\calT_{\rm PL_1}^{(5)} \!\! &=& \!\!
 \int \limits_{- \ri \infty}^{+ \ri \infty}  \frac{\rd z_1}{2 \pi \ri}
 \frac{U^{-\delta _4-1}  m^{-2 \delta _{2357}-4 \epsilon}}{S^{\delta _{16}+2}}
 \frac{ \Gamma\left[-z_1,\delta _2-\delta _4+z_1,-\delta _{25}-\epsilon ,-\delta _{26}-\epsilon \right]}{\Gamma\left[-\delta _{614}-\epsilon -1,-\delta _2-\epsilon +1,\delta _2+1,\delta _3+1,\delta _5+1 \right]} \nonumber \\
 && \hspace{-1.2cm}  \times \frac{\Gamma \big[-\delta _{27}+\delta _4-z_1-\epsilon +1,-\delta _{614}+z_1-\epsilon -1,\delta _{725}-\delta _4+z_1+\epsilon ,\delta _{2357}+2 \epsilon  \big]}{\Gamma \big[ \delta _7+1,-\delta _{46}+z_1-\epsilon \big]} \,,
 \\[4pt]
\calT_{\rm PL_1}^{(6)}  \!\! &=& \!\! 
\frac{U^{-\delta _4-1} m^{-2 \delta _{1356}-4 \delta _{27}-8 \epsilon } }{ S^{-\delta _{27}-2 \epsilon +2}  }
\frac{\Gamma\left[-\delta _{26}-\epsilon ,\delta _{1267}+2 \epsilon ,\delta _{2357}+2 \epsilon ,-\delta _7-\epsilon +1,-\delta _{25}-\epsilon \right]}{\Gamma\left[ \delta _2+1,\delta _3+1,\delta _7+1,\delta _1+1,-\delta _2-\epsilon +1\right] } \,,
 \\[4pt]
\calT_{\rm PL_1}^{(7)}  \!\! &=& \!\!
 \int \limits_{- \ri \infty}^{+ \ri \infty}  \frac{\rd z_1}{2 \pi \ri} \frac{\rd z_2}{2 \pi \ri}
 \frac{U^{-\delta _2-1} m^{-2 \delta _{3457}-4 \epsilon }}{ S^{\delta _{16}+2} }
\frac{ \Gamma\left[ -\delta _2+\delta _4+z_1+z_2,-\delta _7+z_2-\epsilon +1 ,-z_1,-z_2  \right]}{\Gamma\left[-\delta _{612}-\epsilon -1,-\delta _2+z_1+z_2-\epsilon +1,\delta _3+1,\delta _4+1 \right]} \nonumber \\
&& \hspace{-1.2cm}  \times \frac{\Gamma \big[ -\delta _{12}+z_1+z_2-\epsilon ,-\delta _{25}+z_1-\epsilon ,\delta _{57}+\epsilon ,-\delta _{612}+z_1-\epsilon -1,\delta _{2357}-z_1-z_2+2 \epsilon \big] }{\Gamma\big[ \delta _5+1,\delta _7+1,-\delta _{12}+z_1-\epsilon  \big]} \,,
  \\[4pt]
\calT_{\rm PL_1}^{(8)}  \!\! &=& \!\!
 \int \limits_{- \ri \infty}^{+ \ri \infty}  \frac{\rd z_1}{2 \pi \ri} \frac{\rd z_2}{2 \pi \ri}
 \frac{ m^{-2 \delta _{34}-2 \epsilon }  }{ S^{\delta_{12567}+\epsilon +3} } \lb \frac{U}{S} \rb^{z_1}
 \frac{\Gamma\left[\delta _5+z_2+1,\delta _{34}+\epsilon ,-\delta _{625}-z_2-\epsilon -1 \right]}{\Gamma\left[\delta _2+1,\delta _3+1,\delta _4+1,\delta _5+1,\delta _6+1,\delta _7+1 \right]}   \nonumber \\
 && \hspace{-1.2cm}  \times \frac{\Gamma \big[ -\delta _{725}-z_1-\epsilon -1,\delta _{2567}+z_1+z_2+\epsilon +2,-\delta _{56712}+\delta _3-z_1-z_2-\epsilon -2 , -z_1,-z_2 \big]}{ \Gamma \big[ -\delta _{2567}-2 \epsilon ,\delta _{34}-\delta _{56712}-z_2-\epsilon -1 \big] }
\nonumber \\
&& \hspace{-1.2cm}  \times  \, \Gamma\big[ \delta _2+z_1+1,\delta _4+z_1+1  \big] \,.
\eea
Since  the MB representations of template integrals are not unique,
it is crucial to derive the minimal MB representations with the lowest possible MB dimensionality.
With the MB representations derived, the action of shift operators on the template integrals in Eq.~\eqref{eq:tempHEpl1} can be determined according to the shifting rules in Eq.~\eqref{eq:ShiftRule},
yielding the integral representations of the high-energy expansion to the desired order.
Then \texttt{AsyInt} employs \texttt{MB.m} to perform the the analytic continuation to separate left poles and right poles in the Gamma functions
with pre-determined straight lines as the integration contour, e.g. $\mathrm{Re}(z_1) = -1/7$ and $\mathrm{Re}(z_2) = -1/11$.
The analytically continued integrals are then expanded in the sequence of $\{ \delta_1, \dots, \delta_7, \epsilon \}$  to the order $\ord(\delta_i^0)$ and $\ord(\epsilon^{\epsilon_\max})$.
Note that each region can have $\delta$-singularities, while they cancel in the sum of all asymptotic regions in Eq.~\eqref{eq:tempHEpl1}.
Truncating the resulting integrals to the order $\ord(m^0)$ and $\ord(\epsilon^1)$, we obtain 5730 one-dimensional MB integrals and 3536 two-dimensional MB integrals, among which there are 3497 two-dimensional ones of the $I_1$-type as in Eq.~\eqref{eq:I1}.
This procedure is performed by the \texttt{AsyExpNum2MB} command.

The next step is solving the resulting MB integrals. The one-dimensional and simple two-dimensional ones can be solved by the analytic summation and numerical reconstruction methods with \texttt{AsyInt} commands: \texttt{AISum1DMB}, \texttt{AISum2DMB} and \texttt{AIRecNum1DMB}.
The more difficult two-dimensional ones of $I_1$-type need the \texttt{Expand\&Fit} method as discussed in Section~\ref{sec:expfit1scale}.
A priori, it is unclear how many two-dimensional MB integrals contain non-vanishing nested arc contributions.
This usually only becomes evident during the numerical validation stage.
Hence, in practice, we can either pass all $I_1$-type integrals to the \texttt{Expand\&Fit} procedure,
or proceed with \texttt{AISum2DMB} command and validate the results numerically.
If the validation fails, one resorts to the \texttt{Expand\&Fit} procedure.
In this procedure, the \texttt{AIExpandI1} command turns these two-dimensional MB integrals into a power-log series 
\bea
\label{eq:exppl1}
I^{\rm(exp)} &=& \sum_{k=0}^{k_\max} \sum_{n=0}^{N} \,  c_{(k,n)} \, y^k \, \log(y)^{n} \
\eea
 in the $y = U/S \to 0$ limit.
 The value of $k_\max$ depends on the order of $\eps$ and $m^2$.
 For this example up to  $\ord(m^0)$ and $\ord(\epsilon^1)$, it is sufficient to have $k_\max = 120$.
The coefficients $c_{(k,n)}$ are correctly computed by the \texttt{AIExpandI1} command,
which also accounts for non-vanishing arc contributions.
Note that computing higher-order $c_{(k,n)}$ terms for $k>100$ can be computationally expensive.
It can require one- or two-thousand-digit precision for numerical evaluations of MB integrals,
such that the \texttt{PSLQ} algorithm can reconstruct the analytic ratios of two large integer numbers.
%
Since the planar integral can be solved directly in the Euclidean region, we can set $S=1$ and write down an ansatz from the following basis rational functions and HPLs
\bea
f_{\rm RF} &=& \Big\{ 1, \, \frac{1}{y}, y \Big\}  \,,\nonumber \\[4pt]
f_{\rm HPL} &=& \Big\{H_{-1}(y),H_{0,-1}(y),H_{0,-1,-1}(y),H_{0,0,-1}(y),H_{0,-1,-1,-1}(y),H_{0,0,-1,-1}(y), \nonumber \\
&& H_{0,0,0,-1}(y),H_{0,-1,-1,-1,-1}(y),H_{0,-1,0,-1,-1}(y),H_{0,0,-1,-1,-1}(y),\nonumber \\
&&H_{0,0,-1,0,-1}(y),  H_{0,0,0,-1,-1}(y),H_{0,0,0,0,-1}(y)
\Big\} \,.
\eea
It is not necessary to include $H_0(y) = \log (y)$ in the basis $f_{\rm HPL} $, since $\log(y)$ is explicitly present in Eq.~\eqref{eq:exppl1}.
As discussed in Section~\ref{sec:expfit1scale}, the size of rational function basis depends on the expansion order in $m$.
Hence, we have a more compact basis at $\ord(m^0)$.
%
Now the ansatz to transcendental weight-5 can be constructed as
\bea
I^{\rm (ans)} &=&   \sum_{i,j} \, a_{(i,j)} \, f_{\mathrm{RF}}^{(i)} (y) \, f_{\mathrm{HPL}}^{(j)} (y) \,,
\eea
which contains 96 undetermined coefficients $a_{(i,j)}$.
By using the \texttt{FitAnsatz} command, we can require
\bea
\lim_{y \to 0}  I^{\rm (ans)}  &=&  I^{\rm(exp)}
\eea
to the order $\ord(y^{120})$ and solve the resulting linear system of equations.
Now the fitting procedure is complete and the coefficients $a_{(i,j)}$ are determined.
Strictly speaking, the first 96 equations are sufficient to fix all $a_{(i,j)}$,
but we prefer to use all 121 equations, with the last 25 equations serving as cross checks.
%
%
%

At the end, the asymptotic-region results are combined with hard-region results, 
then the reconstruction of $S$-dependence and the analytic continuation to the physical region in terms of $s$ and $T$ is performed
 by the \texttt{AIEucl2Phys} command.
The final analytic results for this integral to $\ord(m^0)$ and $\ord(\eps^1)$  are expressed as
\bea
\calI_{\rm PL_1} [1,1,1,1,1,1,1,-1,0] & = &
\lb \frac{\mu^2}{s} \rb^{2 \epsilon} \sum_{i=-2}^{1} \,  \eps^{i}  \, f_{(i)}\big(s,T,U,m^2 \big)\,,
\eea
with coefficient functions
\bea
f_{(-2)}
&=&
\frac{1}{s \, U} \bigg[
        -\frac{\pi ^2 }{6}
        -\frac{1}{2}  H_0\big(
                \hat{m}^2\big)^2
        - H_0\big(
                \hat{m}^2\big) H_1\big(
                \hat{T}\big)
        -\frac{1}{2}  H_1\big(
                \hat{T}\big)^2
\bigg] \,, \\[4pt]
f_{(-1)}
&=&
\frac{1}{s^2 U} \bigg[
        -\frac{1}{6} i \pi ^3 (s
        +2 T
        )
        +\big(
                -\frac{\pi ^2 T}{3}
                +2 i \pi  U H_1\big(
                        \hat{T}\big)
                -2 s H_1\big(
                        \hat{T}\big)^2
        \big) H_0\big(
                \hat{m}^2\big)
        \nonumber\\
        &&+\frac{1}{3} U H_0\big(
                \hat{m}^2\big)^3
        +\pi ^2 \big(
                -\frac{5 s}{3}
                +T
        \big) H_1\big(
                \hat{T}\big)
        -T H_0\big(
                \hat{m}^2\big)^2 H_1\big(
                \hat{T}\big)
        \nonumber\\
        &&+i \pi  U H_1\big(
                \hat{T}\big)^2
        +\frac{1}{3} (-4 s
        +T
        ) H_1\big(
                \hat{T}\big)^3
        +i \pi  (5 s
        -2 T
        ) H_{0,1}\big(
                \hat{T}\big)
        \nonumber\\
        &&+(-5 s
        +2 T
        ) H_{0,1,1}\big(
                \hat{T}\big)
        +(3 s
        +4 T
        ) \zeta (3)
\bigg] \,, 
\\[4pt]
f_{(0)}
&=&
\frac{1}{s^2 U} \bigg[
        \frac{ \pi ^4}{180} (64 s
        +111 T
        )
        +\big(
                \frac{i \pi ^3}{2}  (-T
                +U
                )
                +\frac{\pi ^2}{3}  (-5 s
                +7 T
                ) H_1\big(
                        \hat{T}\big)
                +2 i \pi  U H_1\big(
                        \hat{T}\big)^2
                \nonumber\\
                &&+\frac{1}{6} (-9 s
                +4 T
                ) H_1\big(
                        \hat{T}\big)^3
                +i \pi  (7 s
                -4 T
                ) H_{0,1}\big(
                        \hat{T}\big)
                +(-7 s
                +4 T
                ) H_{0,1,1}\big(
                        \hat{T}\big)
                \nonumber\\
                &&+(19 s
                -14 T
                ) \zeta (3)
        \big) H_0\big(
                \hat{m}^2\big)
        +\big(
                \frac{1}{3} \pi ^2 (s
                +T
                )
                +\frac{3}{4} s H_1\big(
                        \hat{T}\big)^2
        \big) H_0\big(
                \hat{m}^2\big)^2
                \nonumber\\
                &&+\big( \frac{i \pi  U}{3}
                +\frac{1}{6} (3 s
                +4 T
                ) H_1\big(
                        \hat{T}\big)
        \big) H_0\big(
                \hat{m}^2\big)^3
        +\frac{1}{24} (s
        +6 T
        ) H_0\big(
                \hat{m}^2\big)^4
                \nonumber\\
                && +\big( \frac{1}{6} i \pi ^3 (-19 s
                +10 T
                )
                +i \pi  (19 s
                -10 T
                ) H_{0,1}\big(
                        \hat{T}\big)
                +(-19 s
                +10 T
                ) H_{0,1,1}\big(
                        \hat{T}\big)
                \nonumber\\
                &&+(13 s
                +6 T
                ) \zeta (3)
        \big) H_1\big(
                \hat{T}\big)
        +\frac{1}{6} \pi ^2 (-15 s
        +13 T
        ) H_1\big(
                \hat{T}\big)^2
        +\frac{4}{3} i \pi  U H_1\big(
                \hat{T}\big)^3
        \nonumber\\
        &&+\big(
                -\frac{11 s}{8}
                +\frac{2 T}{3}
        \big) H_1\big(
                \hat{T}\big)^4
        +\frac{1}{6} \pi ^2 (-57 s
        +26 T
        ) H_{0,1}\big(
                \hat{T}\big)
        +i \pi  (23 s
        -10 T
        ) H_{0,0,1}\big(
                \hat{T}\big)
        \nonumber\\
        &&+3 i \pi  (-7 s
        +4 T
        ) H_{0,1,1}\big(
                \hat{T}\big)
        +(-23 s
        +10 T
        ) H_{0,0,1,1}\big(
                \hat{T}\big)
        +(23 s
        -14 T
        ) H_{0,1,1,1}\big(
                \hat{T}\big)
        \nonumber\\
        &&+4 i \pi  (3 s
        -4 T
        ) \zeta (3)
\bigg] \,, 
\\[4pt]
f_{(1)}
&=&
\frac{1}{s^2 U} \bigg[
        \frac{1}{360} i \pi ^5 (-17 s
        +194 T
        )
        +\big(
                \frac{1}{360} \pi ^4 (-183 s
                +136 T
                )
                +\big(
                        \frac{1}{6} i \pi ^3 (-11 s
                        +8 T
                        )
                        \nonumber\\
                        &&+i \pi  (11 s
                        -8 T
                        ) H_{0,1}\big(
                                \hat{T}\big)
                        +(-11 s
                        +8 T
                        ) H_{0,1,1}\big(
                                \hat{T}\big)
                        +\frac{1}{3} (11 s
                        -6 T
                        ) \zeta (3)
                \big) H_1\big(
                        \hat{T}\big)
                \nonumber\\
                &&+\frac{1}{6} \pi ^2 (-11 s
                +10 T
                ) H_1\big(
                        \hat{T}\big)^2
                +i \pi  U H_1\big(
                        \hat{T}\big)^3
                +\frac{1}{4} (-3 s
                +2 T
                ) H_1\big(
                        \hat{T}\big)^4
                \nonumber\\
                &&+\frac{1}{6} \pi ^2 (-31 s
                +20 T
                ) H_{0,1}\big(
                        \hat{T}\big)
                +i \pi  (13 s
                -8 T
                ) H_{0,0,1}\big(
                        \hat{T}\big)
                \nonumber\\
                &&+i \pi  (-13 s
                +10 T
                ) H_{0,1,1}\big(
                        \hat{T}\big)
                +(-13 s
                +8 T
                ) H_{0,0,1,1}\big(
                        \hat{T}\big)
                +3 (5 s
                -4 T
                ) H_{0,1,1,1}\big(
                        \hat{T}\big)
                \nonumber\\
                &&+2 i \pi  (-T
                +U
                ) \zeta (3)
        \big) H_0\big(
                \hat{m}^2\big)
        +\big(
                \frac{1}{12} i \pi ^3 (-11 s
                +14 T
                )
                -\frac{1}{6} \pi ^2 (s
                +4 T
                ) H_1\big(
                        \hat{T}\big)
                \nonumber\\
                &&-\frac{1}{2} i \pi  U H_1\big(
                        \hat{T}\big)^2
                +\frac{1}{6} (3 s
                -T
                ) H_1\big(
                        \hat{T}\big)^3
                +i \pi  \big(
                        -\frac{5 s}{2}
                        +T
                \big) H_{0,1}\big(
                        \hat{T}\big)
                \nonumber\\
                &&+\big(
                        \frac{5 s}{2}
                        -T
                \big) H_{0,1,1}\big(
                        \hat{T}\big)
                +\big(
                        -\frac{103 s}{6}
                        +13 T
                \big) \zeta (3)
        \big) H_0\big(
                \hat{m}^2\big)^2
        +\big(
                \frac{1}{18} \pi ^2 (-15 s
                +2 T
                )
                \nonumber\\
                &&-\frac{1}{3} i \pi  U H_1\big(
                        \hat{T}\big)
                -\frac{1}{6} s H_1\big(
                        \hat{T}\big)^2
        \big) H_0\big(
                \hat{m}^2\big)^3
        +\big(
                -\frac{1}{3} i \pi  U
                +\frac{1}{12} (-2 s
        -T
        ) H_0\big(
                \hat{m}^2\big)^5
                \nonumber\\
                &&+\frac{1}{12} (-5 s
                -3 T
                ) H_1\big(
                        \hat{T}\big)
        \big) H_0\big(
                \hat{m}^2\big)^4
                  +\big( \frac{1}{360} \pi ^4 (1153 s
                -500 T
                )
                \nonumber\\
                && +\frac{\pi ^2}{6}  (-159 s
                +86 T
                ) H_{0,1}\big(
                        \hat{T}\big)
+i \pi  (61 s
                -30 T
                ) H_{0,0,1}\big(
                        \hat{T}\big)
                +i \pi  (-61 s
                +40 T
                ) H_{0,1,1}\big(
                        \hat{T}\big)
                \nonumber\\
                &&+(-61 s
                +30 T
                ) H_{0,0,1,1}\big(
                        \hat{T}\big)
                +(67 s
                -46 T
                ) H_{0,1,1,1}\big(
                        \hat{T}\big)
                +10 i \pi  (-2 s
                +T
                ) \zeta (3)
        \big) H_1\big(
                \hat{T}\big)
        \nonumber\\
        &&+\big(
                \frac{1}{12} i \pi ^3 (-55 s
                +34 T
                )
                +i \pi  \big(
                        \frac{55 s}{2}
                        -17 T
                \big) H_{0,1}\big(
                        \hat{T}\big)
                +\big(
                        -\frac{55 s}{2}
                        +17 T
                \big) H_{0,1,1}
                \big(
                        \hat{T}\big)
                \nonumber\\
                &&+\big(
                        \frac{167 s}{6}
                        -8 T
                \big) \zeta (3)
        \big) H_1\big(
                \hat{T}\big)^2
        +\pi ^2 \big(
                -\frac{41 s}{18}
                +2 T
        \big) H_1\big(
                \hat{T}\big)^3
        +\frac{11}{12} i \pi  U H_1\big(
                \hat{T}\big)^4
        \nonumber\\
        &&+\frac{1}{60} (-53 s
        +33 T
        ) H_1\big(
                \hat{T}\big)^5
        +\big(
                \frac{1}{6} i \pi ^3 (-71 s
                +34 T
                )
                +4 (3 s
                -4 T
                ) \zeta (3)
        \big) H_{0,1}\big(
                \hat{T}\big)
        \nonumber\\
        &&+2 i \pi  U H_{0,1}\big(
                \hat{T}\big)^2
        +\frac{1}{6} \pi ^2 (-183 s
        +86 T
        ) H_{0,0,1}\big(
                \hat{T}\big)
                +\frac{\pi ^2}{6}  (179 s
        -100 T
        ) H_{0,1,1}\big(
                \hat{T}\big)
        \nonumber\\
        &&
        +5 i \pi  (13 s
        -6 T
        ) H_{0,0,0,1}\big(
                \hat{T}\big)
+i \pi  (
        40 T-71 s
        ) H_{0,0,1,1}\big(
                \hat{T}\big)
        +i \pi  (61 s
        -40 T
        ) H_{0,1,1,1}\big(
                \hat{T}\big)
        \nonumber\\
        &&+(-65 s
        +30 T
        ) H_{0,0,0,1,1}\big(
                \hat{T}\big)
        +(65 s
        -34 T
        ) H_{0,0,1,1,1}\big(
                \hat{T}\big)
        -4 U H_{0,1,0,1,1}\big(
                \hat{T}\big)
        \nonumber\\
        &&+(-67 s
        +46 T
        ) H_{0,1,1,1,1}\big(
                \hat{T}\big)
        +\frac{1}{9} \pi ^2 (-146 s
        +195 T
        ) \zeta (3)
        +(51 s
        +4 T
        ) \zeta (5)
\bigg]\,,
\eea
where $\hat{T} = T/s$, $\hat{m}^2 = m^2 /s$ and $U = s- T$.
Note that the presence of $U$ is simply to shorten the expressions, 
and all HPLs are expressed with argument $\hat{T}$.
In the ancillary file~\cite{ttplink}, the analytic results for another integral $\calI_{\rm PL_1}[1,1,1,1,1,1,1,0,-1]$
are also provided.

These results are the boundary conditions for determining the deep high-energy expansion, e.g. up to $\ord(m^{120})$.
%
For details on this aspect, please refer to Refs.~\cite{Davies:2018qvx,Davies:2022ram}. 
The analytic results are cross checked against numerical evaluations with \texttt{AMFlow} on various phase space points in the high-energy limit.
For example, the results to $\ord(m^0)$ agree with \texttt{AMFlow} at the percent level  with $\sqrt{s} = 2$~TeV, $p_T = \sqrt{u\,t/s} = 400$~GeV and $m=80$~GeV.
The high-energy expansion to $\ord(m^{30})$ agrees with \texttt{AMFlow} to more than 35 digits.
%

\subsection{Non-planar integral $\mathrm{NPL_1}$}
As a second two-loop example, we will sketch the calculational steps for the three-massive-line non-planar integral $\rm NPL_1$, and present the analytic results to  $\ord(\eps^1)$ and $\ord(m^2)$, i.e., four expansion terms in $m$.
\begin{figure}[b]
  \centering
    \scalebox{-1}[1]{\includegraphics[width=.28\textwidth]{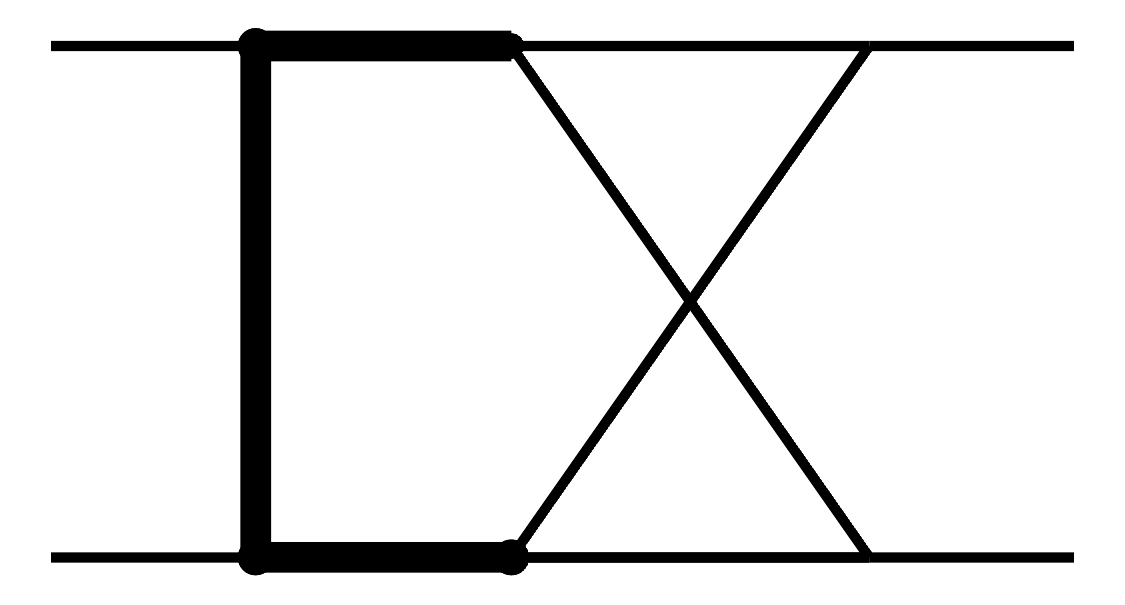} }
  \caption{\label{fig:npl1}
   Two-loop $\mathrm{NPL_1}$ diagram.
    }
\end{figure}
The $\rm NPL_1$ integral has the following propagators
\bea
&\Big\{m^2-l_1^2, \; -l_2^2, \; m^2-\left(-l_1+q_1+q_2\right){}^2, \;  m^2-\left(l_1-q_1-q_2-q_3\right){}^2, \;  -\left(l_2+q_1\right){}^2, \nonumber \\
&-\left(l_1-l_2-q_1-q_2\right){}^2, \;  -\left(l_1-l_2-q_1\right){}^2 \Big\} \,,
\eea
where the kinematics of the external lines are the same as Eq.~\eqref{eq:kin} and the numerators are not shown.
In this example, we will discuss the seven-line top-sector integral $\calI_{\rm NPL_1}[1,1,1,1,1,1,1]$.
By using the \texttt{GenerateInput} command, the Symanzik polynomials for this integral are
\bea
\mathcal{U} &=& \alpha _3 \alpha _5+\alpha _4 \alpha _5+\alpha _6 \alpha _5+\alpha _3 \alpha _6+\alpha _4 \alpha _6+\left(\alpha _3+\alpha _4+\alpha _5\right) \alpha _7\nonumber \\
&& +\alpha _2 \left(\alpha _3+\alpha _4+\alpha _6+\alpha _7\right) +\alpha _1 \left(\alpha _2+\alpha _5+\alpha _6+\alpha _7\right) \,, \nonumber \\[4pt]
\mathcal{F}  &=&
\Big[\alpha _3 \alpha _5 \alpha _7+\alpha _1 \left(\alpha _2 \left(\alpha _3+\alpha _6\right)+\alpha _3 \left(\alpha _5+\alpha _6+\alpha _7\right)\right)\Big] \, S+ \big( \alpha _4 \alpha _5 \alpha _6 \big) \, T+\big( \alpha _2 \alpha _4 \alpha _7 \big) \, U \nonumber \\
&& +\left(\alpha _1+\alpha _3+\alpha _4\right) \Big[\alpha _3 \alpha _5+\alpha _4 \alpha _5+\alpha _6 \alpha _5+\alpha _3 \alpha _6+\alpha _4 \alpha _6+\left(\alpha _3+\alpha _4+\alpha _5\right) \alpha _7 \nonumber \\
&&+\alpha _2 \left(\alpha _3+\alpha _4+\alpha _6+\alpha _7\right)+\alpha _1 \left(\alpha _2+\alpha _5+\alpha _6+\alpha _7\right)\Big] \, m^2 \,,
\eea
where the relation $S+T+U =0 $ is used to obtain the minimal positive definite $\mathcal{F}$ polynomial.
By imposing $S,T,U \sim 1$ and $m^2 \sim \rho$, the scalings of alpha-parameters in all regions are listed in Table~\ref{tab:npl1region}.
\begin{table}[bt]
\centering
\begin{tabular}{cccccccc}
\text{Region} & $\alpha_1$ & $\alpha_2$ & $\alpha_3$ & $\alpha_4$ & $\alpha_5$ & $\alpha_6$ & $\alpha_7$ \\ \hline
$R_1$ & 0 & 0 & 0 & 0 & 0 & 0 & 0 \\
$R_2$ & 0 & 0 & 1 & 1 & 0 & 1 & 0 \\
$R_3$ & 0 & 1/2 & 1 & 1/2& 0 & 1/2 & 0 \\
$R_4$ & 0 & 1 & 1 & 0 & 0 & 1 & 0 \\
$R_5$ & 0 & 1 & 1 & 0 & 1 & 1 & 1 \\
$R_6$ & 0 & 1 & 1 & 1 & 0 & 0 & 0 \\
$R_7$ & 1 & 0 & 0 & 0 & 1 & 0 & 1 \\
$R_8$ & 1 & 0 & 0 & 1/2 & 1/2 & 0 & 1/2 \\
$R_9$ & 1 & 0 & 0 & 1 & 0 & 0 & 1 \\
$R_{10}$ & 1 & 0 & 0 & 1 & 1 & 0 & 0 \\
$R_{11}$ & 1 & 0 & 1 & 2 & 1 & 1 & 0 \\
$R_{12}$ & 1 & 1 & 0 & 0 & 1 & 1 & 1 \\
$R_{13}$ & 1 & 1 & 1 & 2 & 0 & 0 & 1 \\
\end{tabular}
\caption{Scalings of alpha parameters for the $\rm NPL_1$ integral, i.e. $\alpha_i \sim \rho^p$ where $p$ is the power in the table. 
$R_1$ is the hard region and $R_2, \dots, R_{13}$ are the asymptotic regions.}
\label{tab:npl1region}
\end{table}

For the hard region $R_1$, the massless non-planar master integrals  can be calculated by the corresponding canonical differential equations.
Its high-energy expansion to $\ord(m^2)$ can be  obtained by using the \texttt{Dinv} command from \texttt{LiteRed} and performing IBP reductions.

For the asymptotic regions $R_2,\dots,R_{13}$, the alpha representations of template integrals are obtain by using \texttt{AlphaRepForTempInt} command. The MB representations of these template integrals are
\bea
\calT^{(2)}_{\rm NPL_1}
\!\! &=& \!\!
 \int \limits_{- \ri \infty}^{+ \ri \infty}  \frac{\rd z_1}{2 \pi \ri} 
 \frac{U^{-\delta _4-1}  m^{-2 \delta _{1257}-4 \epsilon }}{S^{\delta _{36}+2}}
 \frac{ \Gamma\left[\delta _{1257}+2 \epsilon ,-z_1,-\delta _{25}-\epsilon ,-\delta _{27}+\delta _4-z_1-\epsilon +1 \right]}{\Gamma\left[-\delta _{46}+z_1-\epsilon ,-\delta _{634}-\epsilon -1,\delta _1+1,\delta _2+1\right]} \nonumber \\
 && \hspace{-1.2cm} \times \frac{\Gamma \big[ -\delta _{4,6}+\delta _2+z_1-1,-\delta _{6,3,4}+z_1-\epsilon -1,\delta _{7,2,5}-\delta _4+z_1+\epsilon \big]}{\Gamma\big[ \delta _5+1,\delta _7+1 \big]} \,,
  \\[4pt]
 \calT^{(3)}_{\rm NPL_1}
\!\! &=& \!\!
\frac{ T^{-\frac{\delta _{46}}{2}+\frac{\delta _2}{2}-\frac{1}{2}} U^{-\frac{\delta _{24}}{2}+\frac{\delta _6}{2}-\frac{1}{2}} m^{-\delta _{624}-2 \delta _{715}-4 \epsilon -1}}{ S^{\frac{\delta _{26}}{2}+\delta _3-\frac{\delta _4}{2}+\frac{3}{2}} }
\frac{ \Gamma\Big[ \frac{1}{2} \left(\delta _{624}+2 \delta _{715}+4 \epsilon +1\right), -\delta _{25}-\epsilon \Big]}{2 \, \Gamma\left[\frac{1}{2} \left(-\delta _{624}-2 \epsilon +1\right),\delta _1+1,\delta _2+1\right]} \nonumber \\
&& \hspace{-1.2cm} \times \frac{\Gamma \Big[ \frac{1}{2} \left(\delta _{24}-\delta _6+1\right) ,\frac{1}{2} \left(\delta _{26}-\delta _4+1\right),\frac{1}{2} \left(\delta _{26}-\delta _4+2 \delta _5+1\right)+\delta _7+\epsilon ,\frac{1}{2} \left(\delta _{46}-\delta _2+1\right) \Big]}{\Gamma \Big[ \delta _4+1,\delta _5+1,\delta _6+1,\delta _7+1 \Big]} \nonumber \\
&& \hspace{-1.2cm} \times \, \Gamma(-\delta _{67}-\epsilon) \,,
  \\[4pt]
 \calT^{(4)}_{\rm NPL_1}
\!\! &=& \!\!
 \int \limits_{- \ri \infty}^{+ \ri \infty}  \frac{\rd z_1}{2 \pi \ri} 
 \frac{ T^{-\delta _6-1} U^{-\delta _2-1} m^{-2 \delta _{1457}-4 \epsilon }}{ S^{\delta _3+1} }
 \frac{ \Gamma\left[-z_1,-\delta _{25}-\epsilon ,-\delta _{26}+\delta _4+z_1-1,\delta _{57}+\epsilon \right]}{\Gamma\left[-\delta _{2567}-2 \epsilon ,\delta _1+1,\delta _4+1,\delta _5+1,\delta _7+1\right]} \nonumber \\
 &&\hspace{-1.2cm} \times \frac{\Gamma \big[ -\delta _{67}-\epsilon ,-\delta _{623}+z_1-\epsilon -1,-\delta _{2567}+z_1-2 \epsilon ,\delta _{56712}-z_1+2 \epsilon +1 \big]}{ \Gamma\big[ -\delta _{26}+z_1-\epsilon ,-\delta _{623}-\epsilon -1 \big] } \,,
 \\[4pt]
 \calT^{(5)}_{\rm NPL_1}
\!\! &=& \!\!
 \int \limits_{- \ri \infty}^{+ \ri \infty}  \frac{\rd z_1}{2 \pi \ri}    \frac{\rd z_2}{2 \pi \ri} 
 \frac{ m^{-2 \delta _{1,4}-2 \epsilon } }{  S^{\delta _{56723}+\epsilon +3} } \lb  \frac{T}{S} \rb^{z_1}  \lb  \frac{U}{S} \rb^{z_2}
 \frac{ \Gamma\left[\delta _7+z_2+1,\delta _{1,4}+\epsilon \right]}{\Gamma\left[\delta _1+1,\delta _2+1,\delta _4+1,\delta _5+1,\delta _6+1 \right]} \nonumber \\
 &&\hspace{-1.2cm} \times \frac{\Gamma \big[ -\delta _{725}-z_2-\epsilon -1,-\delta _{756}-z_1-\epsilon -1,\delta _{2567}+z_1+z_2+\epsilon +2 \big]}{\Gamma \big[ \delta _7+1,-\delta _{2567}-2 \epsilon ,\delta _{1,4}-\delta _{56723}-\epsilon -1\big]} \nonumber \\
 &&\hspace{-1.2cm} \times \, \Gamma \big[ -\delta _{56723}+\delta _1-z_1-z_2-\epsilon -2,-z_1,-z_2,\delta _4+z_1+z_2+1,\delta _5+z_1+1 \big] \,,
  \\[4pt]
 \calT^{(6)}_{\rm NPL_1}
\!\! &=& \!\!
 \int \limits_{- \ri \infty}^{+ \ri \infty}  \frac{\rd z_1}{2 \pi \ri}  
 \frac{T^{-\delta _4-1} m^{-2 \delta _{1567}-4 \epsilon } }{ S^{\delta _{23}+2}  }
 \frac{\Gamma\left[ \delta _{1567}+2 \epsilon ,-z_1,-\delta _{24}+\delta _6+z_1-1 \right]}{\Gamma\left[-\delta _{24}+z_1-\epsilon ,-\delta _{423}-\epsilon -1,\delta _1+1,\delta _5+1\right]}
 \nonumber \\
 &&\hspace{-1.2cm} \times \frac{\Gamma \big[ -\delta _{56}+\delta _4-z_1-\epsilon +1,-\delta _{67}-\epsilon ,-\delta _{423}+z_1-\epsilon -1,\delta _{756}-\delta _4+z_1+\epsilon \big]}{\Gamma \big[ \delta _6+1,\delta _7+1 \big]} \,,
   \\[4pt]
 \calT^{(7)}_{\rm NPL_1}
\!\! &=& \!\!
 \int \limits_{- \ri \infty}^{+ \ri \infty}  \frac{\rd z_1}{2 \pi \ri}  
 \frac{ T^{-\delta _5-1} U^{-\delta _7-1} m^{-2 \delta _{2346}-4 \epsilon }}{S^{\delta _1+1}}
 \frac{ \Gamma\left[-z_1,-\delta _{25}-\epsilon ,\delta _{26}+\epsilon,-\delta _{67}-\epsilon \right]}{\Gamma\left[-\delta _{2567}-2 \epsilon ,\delta _2+1,\delta _3+1,\delta _4+1,\delta _6+1 \right]}\nonumber \\
 &&\hspace{-1.2cm} \times \frac{\Gamma \big[ -\delta _{57}+\delta _4+z_1-1 ,-\delta _{715}+z_1-\epsilon -1,-\delta _{2567}+z_1-2 \epsilon ,\delta _{56723}-z_1+2 \epsilon +1 \big]}{\Gamma \big[ -\delta _{57}+z_1-\epsilon ,-\delta _{715}-\epsilon -1 \big]} \,,
    \\[4pt]
 \calT^{(8)}_{\rm NPL_1}
\!\! &=& \!\!
\frac{ T^{-\frac{\delta _{45}}{2}+\frac{\delta _7}{2}-\frac{1}{2}} U^{-\frac{\delta _{47}}{2}+\frac{\delta _5}{2}-\frac{1}{2}} m^{-2 \delta _{623}-\delta _{745}-4 \epsilon -1} }{ S^{\frac{\delta _{57}}{2}+\delta _1-\frac{\delta _4}{2}+\frac{3}{2}} }
\frac{\Gamma\left[\delta _{623}+\frac{1}{2} \delta _{745}+2 \epsilon +\frac{1}{2},-\delta _{25}-\epsilon  \right]}{2 \Gamma\left[-\frac{1}{2} \delta _{745}-\epsilon +\frac{1}{2},\delta _2+1,\delta _3+1,\delta _4+1 \right]} \nonumber \\
 &&\hspace{-1.2cm} \times \frac{\Gamma \big[ \frac{\delta _{45}}{2}-\frac{\delta _7}{2}+\frac{1}{2},\frac{\delta _{47}}{2}-\frac{\delta _5}{2}+\frac{1}{2},\frac{\delta _{57}}{2}-\frac{\delta _4}{2}+\frac{1}{2},\delta _{26}+\frac{\delta _{57}}{2}-\frac{\delta _4}{2}+\epsilon +\frac{1}{2},-\delta _{67}-\epsilon \big]}{\Gamma \big[ \delta _5+1,\delta _6+1,\delta _7+1 \big]} \,,
     \\[4pt]
 \calT^{(9)}_{\rm NPL_1}
\!\! &=& \!\!
 \int \limits_{- \ri \infty}^{+ \ri \infty}  \frac{\rd z_1}{2 \pi \ri}  
 \frac{T^{-\delta _4-1}  m^{-2 \delta _{2356}-4 \epsilon }}{ S^{\delta _{17}+2} }
 \frac{ \Gamma\left[\delta _{2356}+2 \epsilon ,-z_1,-\delta _{25}-\epsilon ,-\delta _{47}+\delta _5+z_1-1 \right]}{\Gamma\left[-\delta _{47}+z_1-\epsilon ,-\delta _{714}-\epsilon -1,\delta _2+1,\delta _3+1\right]}\nonumber \\
 &&\hspace{-1.2cm} \times \frac{\Gamma \big[ -\delta _{56}+\delta _4-z_1-\epsilon +1,\delta _{625}-\delta _4+z_1+\epsilon ,-\delta _{714}+z_1-\epsilon -1 \big]}{\Gamma \big[ \delta _5+1,\delta _6+1 \big]} \,,
 \\[4pt]
 \calT^{(10)}_{\rm NPL_1}
\!\! &=& \!\!
 \int \limits_{- \ri \infty}^{+ \ri \infty}  \frac{\rd z_1}{2 \pi \ri}  
 \frac{U^{-\delta _4-1}  m^{-2 \delta _{2367}-4 \epsilon }}{S^{\delta _{15}+2}}
 \frac{ \Gamma\left[\delta _{2367}+2 \epsilon ,-z_1,-\delta _{27}+\delta _4-z_1-\epsilon +1 \right]}{\Gamma\left[-\delta _{45}+z_1-\epsilon ,-\delta _{514}-\epsilon -1,\delta _2+1,\delta _3+1\right]}\nonumber \\
 &&\hspace{-1.2cm} \times \frac{\Gamma \big[ -\delta _{45}+\delta _7+z_1-1,-\delta _{67}-\epsilon ,-\delta _{514}+z_1-\epsilon -1,\delta _{726}-\delta _4+z_1+\epsilon \big]}{\Gamma \big[ \delta _6+1,\delta _7+1 \big]} \,,
  \\[4pt]
 \calT^{(11)}_{\rm NPL_1}
\!\! &=& \!\!
\frac{U^{-\delta _4-1}  m^{-4 \delta _{27}-2 \delta _{1356}-8 \epsilon } }{ S^{-\delta _{27}-2 \epsilon +2} }
\frac{\Gamma\left[-\delta _{25}-\epsilon ,-\delta _{67}-\epsilon ,\delta _{1257}+2 \epsilon ,\delta _{2367}+2 \epsilon \right]}{\Gamma\left[\delta _1+1,\delta _2+1,\delta _3+1,\delta _7+1\right]} \,,
  \\[4pt]
 \calT^{(12)}_{\rm NPL_1}
\!\! &=& \!\!
 \int \limits_{- \ri \infty}^{+ \ri \infty}  \frac{\rd z_1}{2 \pi \ri}  \frac{\rd z_2}{2 \pi \ri}  
 \frac{ m^{-2 \delta _{34}-2 \epsilon } }{ S^{\delta _{56712}+\epsilon +3} } \lb \frac{T}{S}\rb^{z_1}  \lb \frac{U}{S}\rb^{z_2}
 \frac{ \Gamma\left[\delta _6+z_1+1,\delta _{34}+\epsilon ,-\delta _{625}-z_1-\epsilon -1 \right]}{\Gamma\left[\delta _2+1,\delta _3+1,\delta _4+1,\delta _5+1,\delta _6+1\right]} \nonumber \\
 &&\hspace{-1.2cm} \times \frac{\Gamma \big[ -\delta _{726}-z_2-\epsilon -1,\delta _{2567}+z_1+z_2+\epsilon +2,-\delta _{56712}+\delta _3-z_1-z_2-\epsilon -2 \big]}{\Gamma \big[ \delta _7+1, -\delta _{2567}-2 \epsilon ,\delta _{34}-\delta _{56712}-\epsilon -1 \big]} \nonumber \\
 &&\hspace{-1.2cm} \times \, \Gamma \big[ -z_1,-z_2,\delta _2+z_2+1,\delta _4+z_1+z_2+1 \big] \,,
   \\[4pt]
 \calT^{(13)}_{\rm NPL_1}
\!\! &=& \!\!
\frac{T^{-\delta _4-1} m^{-4 \delta _{56}-2 \delta _{1237}-8 \epsilon } }{  S^{-\delta _{56}-2 \epsilon +2} }
\frac{\Gamma\left[-\delta _{25}-\epsilon ,-\delta _{67}-\epsilon ,\delta _{1567}+2 \epsilon ,\delta _{2356}+2 \epsilon \right]}{\Gamma\left[\delta _1+1,\delta _3+1,\delta _5+1,\delta _6+1\right]} \,.
\eea
In case of $R_3$ and $R_8$, some of the $\alpha_i$ scale as $\sqrt{\rho}$.
Here it is convenient to apply the following variable transformations
\bea
\alpha_i \to \sqrt{ \frac{\beta_1 \beta_2}{\beta_3}} \,, \quad \alpha_j \to \sqrt{ \frac{\beta_1 \beta_3}{\beta_2}} \,, \quad \alpha_k \to \sqrt{ \frac{\beta_2 \beta_3}{\beta_1}} \,,
\eea
in the corresponding $\mathcal{U}$ and $\mathcal{F}$ polynomials and take into account the Jacobian determinant  $\lb2\sqrt{\beta_1 \beta_2 \beta_3} \, \rb^{-1}$.

With these  template integrals, 
the \texttt{AsyExp2MB} command can generate the high-energy expansion using shift operators
and perform the analytic continuation and the series expansion in the sequence of $\{\delta_1, \dots, \delta_7, \epsilon\}$.
After resolving the $\delta$- and $\epsilon$-singularities, the cancellation of $\delta$-singularities is observed in the sum of all asymptotic regions.
%
%
Truncating the expanded MB integrals to the order $\ord(\eps^1)$ and $\ord(m^2)$,
we obtain 11650 MB integrals, among which there are 342 two-dimensional MB integrals of $I_2$-type as in Eq.~\eqref{eq:I2}.
These $I_2$-type MB integrals with two-scales are more involved
and require the \texttt{Expand\&Fit} method described in Section~\ref{sec:expfit2scale}.
Since solving these MB integrals is more difficult and computationally much more expensive compared to the $I_1$-type integrals,
it is important to simplify the MB integrals with the \texttt{GammaSimplifyList} command before using the \texttt{Expand\&Fit} method.

For performance reasons, instead of using the automated command \texttt{AIExpandI2},
we use the lower-level command \texttt{AIExpand2DMB} 
to expand the $I_2$-type integrals  in the $x= T/S \to 0$ limit to more than a hundred expansion terms.
At this stage, two expansion series are generated
\bea
I^{(1)} &=&
- \sum_{k=0}^{k_\max} \sum_{n=0}^{N} \, x^{k} \, \log\lb x\rb^{n} \,  \int \limits_{- \ri \infty}^{+ \ri \infty}  \frac{d z_2}{2 \pi \ri} \,  y^{z_2} \, \hat{f}_{(k,n)}^{(1)} \Big( \Gamma, \psi^{(i)}; k_1, z_2 \Big)  \,, \\
I^{(2)} &=& 
\sum_{k_1= 0}^{k_\max} \sum_{k_2= 0}^{k'_{\max}} \sum_{n_1=0}^{N_1} \sum_{n_2=0}^{N_2}  \,  c_{(k_1,n_1,k_2,n_2)}^{(2)} \, x^{k_1} \log\lb x \rb^{n_1} \, y^{k_2} \log\lb y \rb^{n_2}\,.
\eea
where $x=T/S$ and $y=U/S$,
and $k'_{\max}$ in the second series are automatically generated by the algorithm.
The first series still contains a large number of MB integrals involving $z_2$ integrations with $y$-dependence.
These MB integrals with $z_2$ integrations are calculated analytically  by the summation method with the command \texttt{AISum1DMB},
yielding results that contain the exact $y$-dependence expressed in terms of HPLs.
Then we can analytically continue these two series $I^{(1)}$ and $I^{(2)}$ to the physical region by rewriting $y =  (-S-T)/S$ 
and transforming $S \to s$ by using the command \texttt{AIEucl2Phys}.

Once the power-log series in the physical region are obtained in terms of only $s$ and $T$,
we can rewrite the expressions in terms of $x' = T/s$ and set $s= 1$.
We combine $I^{(1)} $ and $I^{(2)} $, and re-expand the series in the $x' \to 0$ limit to obtain
\bea
\label{eq:expnpl1}
I^{\rm(exp)} &=& \sum_{k=0}^{k_\max} \sum_{n=0}^{N} \,  c'_{(k,n)} \, x'{}^{k} \, \log(x')^{n} \,.
\eea
Note that computing the $z_2$-integration in the $I^{(1)}$ series to obtain higher-order $c'(k,n) $ terms can be computationally very expensive.
It can require a few hundred gigabytes of memory to compute $c'(k,n) $ terms for $k>100$ at $\ord(\eps^1)$ and $\ord(m^2)$.
Hence it is important to keep the ansatz size in the next fitting procedure as small as possible.

Now we can perform the fitting procedure with the following basis functions
\bea
f_{\rm RF} &=& \Big\{ \frac{1}{x'}, \, \frac{1}{1-x'} , \,  \frac{1}{x'{}^{2}}, \, \frac{1}{(1-x')^2}  \Big\}  \,,\nonumber \\[4pt]
f_{\rm HPL} &=& \Big\{
H_1(x'),H_{0,1}(x'),H_{0,0,1}(x'),H_{0,1,1}(x'),H_{0,0,0,1}(x'),H_{0,0,1,1}(x'),H_{0,1,1,1}(x'), \nonumber \\
&& \hspace{-1cm} H_{0,0,0,0,1}(x'),H_{0,0,0,1,1}(x'),H_{0,0,1,0,1}(x'),H_{0,0,1,1,1}(x'),H_{0,1,0,1,1}(x'),H_{0,1,1,1,1}(x')
\Big\} \,, 
\eea
where the $\log(x') = H_0(x')$ is again excluded from the HPL basis.
Now the ansatz can be constructed to transcendental weight-5 with 128 undetermined coefficients
\bea
\label{eq:ansatznpl1}
I^{\rm (ans)} &=& \sum_{i,j} \, a'_{(i,j)} \,  f^{(i)}_{\rm RF} (T)  \, f^{(j)}_{\rm HPL} (T) \,.
\eea
To achieve such a minimal ansatz and avoid spurious rational functions in $f_{\rm RF}$,
it is necessary to combine the power-log series obtained from $I_2$-type integrals with all other simpler MB integrals that have been calculated prior to this step.
In this context, we compute the remaining 11308 MB integrals and denote their analytic solution in the physical region as $I^{\rm (known)}$.
It is also worth noting again that higher-order terms in $m$ increase the basis of rational functions $f_{\rm RF}$, while the higher-order terms in $\eps$ increase the basis of HPL functions $f_{\rm HPL}$.
For example, at $\ord(m^0)$ only $\{1/x', 1/(1-x')\}$ are needed for $f_{\rm RF}$, while at $\ord(\ep^0)$ only HPLs up to weight-4 are needed for $f_{\rm HPL}$.
Therefore, up to $\ord(m^2)$ and $\ord(\eps^1)$, at least 128 unknown coefficients $a'_{(i,j)}$ are needed.
By using the \texttt{FitAnsatz} command, we can require
\bea
\lim_{x'\to 0} I^{\rm (ans)} &=& \lim_{x'\to 0} \lb I^{\rm (exp)} + I^{\rm (known)} \rb
\eea
to $\ord(x'{}^{150})$, and determine the coefficients $a'_{(i,j)}$ order-by-order in $\eps$ and $m$.
Once the fitting procedure is complete in the physical region, the $s$-dependence can be reconstructed straightforwardly.
%

Here, we provide a general estimation of the efficiency of \texttt{AsyInt} for solving this $\rm{NPL_{1}}$ integral.
The large number of integrals for  $I^{\rm (known)}$ can be computed in parallel and  do not represent the bottleneck of the computation.
The most computationally expensive part is obtaining $I^{\rm (exp)}$ to higher orders in $x'$.
At $\ord(\eps^1)$, we need to compute at least 65 expansion terms in $x'$ for the fitting procedure at $\ord(m^0)$, 
and 130 expansion terms in $x'$ at $\ord(m^2)$.
While the computations for each order in $x'$ can be parallelised, 
performance is limited by solving the highest-order term due to the presence of complicated intermediate expressions.
To obtain higher-order terms within a reasonable amount of time and memory usage,
\texttt{AsyInt} has been optimised in its symbolic manipulations for these expressions.
With this optimisation, we can solve the 65th expansion term in $x'$ at $\ord(m^0)$ in about ten-hour CPU time,
and the 130th expansion term  in $x'$ at $\ord(m^2)$ in about one-week CPU time.

Finally, in combination with the massless contributions in the hard region $R_1$,
the results of this integral to $\ord(\eps^1)$ and $\ord(m^2)$ are expressed as
\bea
\calI_{\rm NPL_1} [1,1,1,1,1,1,1] & = &
\lb \frac{\mu^2}{s} \rb^{2 \epsilon} \sum_{i=-2}^{1} \sum_{j = -1}^{2} \,  \eps^{i} \, m^{j}  \, f_{(i,j)}\big(s,T,U,m \big)\,,
\eea
with coefficient functions
\bea
f_{(-2,-1)} 
&=& -\frac{i \pi ^2 }{s^2\sqrt{T U}} \,, \\[4pt]
f_{(-2,0)}
&=&\frac{1}{s^2 T U} \bigg[
        2 i \pi  s
        -\frac{5 \pi ^2 s}{3}
        +(2 T
        +i \pi  (T
        -U
        )
        ) H_0\big(
                \hat{T}\big)
        +\big(
                3 i \pi  s
                +(T
                -U
                ) H_0\big(
                        \hat{T}\big)
                \nonumber\\
                &&+(T
                -U
                ) H_1\big(
                        \hat{T}\big)
        \big) H_0\big(
                \hat{m}^2\big)
        +\frac{3}{2} s H_0\big(
                \hat{m}^2\big)^2
+(-2 U
        +i \pi  (T
        -U
        )
        ) H_1\big(
                \hat{T}\big)
\bigg]\,,
 \\[4pt]
f_{(-1,-1)} 
&=&\frac{i}{s^2 \sqrt{T U}} \, \Big[  2  \pi ^2 H_0\big(
        \hat{m}^2\big)
+ 8  \pi ^2 \log (2) \Big] \,,
 \\[4pt]
f_{(-1,0)}
&=&\frac{1}{s^2 T U} \bigg[
        -i \pi  \big(
                16+3 \pi ^2\big) s
        +\big(
                -16 T
                -8 i \pi  U
                +\pi ^2 (4 s
                -6 T
                )
                +2 (4+i \pi ) s H_1\big(
                        \hat{T}\big)
                \nonumber\\
                &&+(-T
                +U
                ) H_1\big(
                        \hat{T}\big)^2
        \big) H_0\big(
                \hat{T}\big)
        +\big(
                -i \pi  s
                +(-T
                +U
                ) H_1\big(
                        \hat{T}\big)
        \big) H_0\big(
                \hat{T}\big)^2
        \nonumber\\
        &&+\frac{1}{3} (-T
        +U
        ) H_0\big(
                \hat{T}\big)^3
        +\big(
                4 i \pi  s
                -\frac{\pi ^2 s}{3}
                +\big(
                        4 T
                        +2 s H_1\big(
                                \hat{T}\big)
                \big) H_0\big(
                        \hat{T}\big)
                \nonumber\\
                &&-s H_0\big(
                        \hat{T}\big)^2
                -4 U H_1\big(
                        \hat{T}\big)
                -s H_1\big(
                        \hat{T}\big)^2
        \big) H_0\big(
                \hat{m}^2\big)
        +\big(
                -4 i \pi  s
                \nonumber\\
                &&+(3 s
                -2 T
                ) H_0\big(
                        \hat{T}\big)
                +(-s
                -2 T
                ) H_1\big(
                        \hat{T}\big)
        \big) H_0\big(
                \hat{m}^2\big)^2
        -\frac{11}{3} s H_0\big(
                \hat{m}^2\big)^3
        \nonumber\\
        &&+\big(
                8 i \pi  T
                +2 \big(
                        8 U
                        +\pi ^2 (s
                        -3 T
                        )
                \big)
        \big) H_1\big(
                \hat{T}\big)
        -i \pi  s H_1\big(
                \hat{T}\big)^2
        +\frac{1}{3} (-T
        +U
        ) H_1\big(
                \hat{T}\big)^3
\bigg] \,, \\[4pt]
f_{(0,-1)}
&=&\frac{-i}{s^2\sqrt{T U}} \bigg[ 2 \pi ^2 H_0\big(
        \hat{m}^2\big)^2
+16 \pi ^2 H_0\big(
        \hat{m}^2\big) \log (2)
+\frac{\pi ^2 \big(
        7 \pi ^2+192 \log ^2(2)\big)}{6} \bigg] \,,
\\[4pt]
 f_{(0,0)} 
&=&
\frac{1}{s^2 T U} \bigg[
        i \big(
                -\frac{\pi ^3}{3}  (3 s
                +10 T
                )
                +12 \pi  T \zeta (3)
                -10 \pi  s (-8+7 \zeta (3))
        \big)
        +\frac{\pi ^4}{60}  (49 s
        +134 T
        )
                \nonumber\\
                &&+\big( i \big(
                        40 \pi  U
                        +\frac{1}{6} \pi ^3 (45 s
                        -38 T
                        )
                \big)
                +\big(
                        20 i \pi  U
                        -10 \big(
                                4 s
                                +\pi ^2 U
                        \big)
                        \nonumber\\
                        &&+10 (T
                        -U
                        ) H_{0,1}\big(
                                \hat{T}\big)
                \big) H_1\big(
                        \hat{T}\big)
                +(10 s
                +2 i \pi  T
                ) H_1\big(
                        \hat{T}\big)^2
                +\frac{2}{3} (s
                -4 T
                ) H_1\big(
                        \hat{T}\big)^3
                \nonumber\\
                &&+(20 (-2 s
                +T
                )
                +4 i \pi  (2 s
                +3 T
                )
                ) H_{0,1}\big(
                        \hat{T}\big)
                +(-48 s
                +40 T
                ) H_{0,0,1}\big(
                        \hat{T}\big)
                \nonumber\\
                &&+26 (-T
                +U
                ) H_{0,1,1}\big(
                        \hat{T}\big)
                +T \big(
                        80-11 \pi ^2-36 \zeta (3)\big)
                -4 s \zeta (3)
        \big) H_0\big(
                \hat{T}\big)
        +\big(
                10 i \pi  U
                \nonumber\\
                &&+\frac{5}{2} \pi ^2 (T
                -U
                )
                +(10 U
                +i \pi  (s
                -4 T
                )
                ) H_1\big(
                        \hat{T}\big)
                +\big(
                        -\frac{s}{2}
                        -T
                \big) H_1\big(
                        \hat{T}\big)^2
                \nonumber\\
                &&+2 (6 s
                -5 T
                ) H_{0,1}\big(
                        \hat{T}\big)
        \big) H_0\big(
                \hat{T}\big)^2
        +\big(
                -\frac{10 T}{3}
                +\frac{4}{3} i \pi  (2 s
                -T
                )
                \nonumber\\
                &&+\big(
                        -2 s
                        +\frac{8 T}{3}
                \big) H_1\big(
                        \hat{T}\big)
        \big) H_0\big(
                \hat{T}\big)^3
        +
        \frac{1}{12} (s
        +8 T
        ) H_0\big(
                \hat{T}\big)^4
        +\big(
                -\frac{1}{6} i \pi  \big(
                        48+11 \pi ^2\big) s
                \nonumber\\
                &&+\big(
                        -8 T
                        -4 i \pi  U
                        +\pi ^2 \big(
                                -\frac{5 s}{2}
                                +T
                        \big)
                        +4 s H_1\big(
                                \hat{T}\big)
                        +s H_1\big(
                                \hat{T}\big)^2
                \big) H_0\big(
                        \hat{T}\big)
                \nonumber\\
                &&+\big(
                        -i \pi  s
                        -s H_1\big(
                                \hat{T}\big)
                \big) H_0\big(
                        \hat{T}\big)^2
                -\frac{1}{3} s H_0\big(
                        \hat{T}\big)^3
                \nonumber\\
                &&+\big(
                        4 i \pi  T
                        +8 U
                        +\pi ^2 \big(
                                \frac{3 s}{2}
                                +T
                        \big)
                \big) H_1\big(
                        \hat{T}\big)
                -i \pi  s H_1\big(
                        \hat{T}\big)^2
                +\frac{1}{3} s H_1\big(
                        \hat{T}\big)^3
                \nonumber\\
                &&+8 s \zeta (3)
        \big) H_0\big(
                \hat{m}^2\big)
        +\big(
                -2 i \pi  s
                +\frac{5 \pi ^2 s}{12}
                -2 T H_0\big(
                        \hat{T}\big)
                +\frac{1}{2} s H_0\big(
                        \hat{T}\big)^2
                \nonumber\\
                &&+2 U H_1\big(
                        \hat{T}\big)
                +\frac{1}{2} s H_1\big(
                        \hat{T}\big)^2
        \big) H_0\big(
                \hat{m}^2\big)^2
        +\big(
                5 i \pi  s
                +\frac{1}{3} (-7 s
                +4 T
                ) H_0\big(
                        \hat{T}\big)
                \nonumber\\
                &&+\big(
                        s
                        +\frac{4 T}{3}
                \big) H_1\big(
                        \hat{T}\big)
        \big) H_0\big(
                \hat{m}^2\big)^3
        +\frac{47}{12} s H_0\big(
                \hat{m}^2\big)^4
        +\big(
                -80 U
                +11 \pi ^2 U
                \nonumber\\
                &&+i \big(
                        -40 \pi  T
                        +\frac{1}{6} \pi ^3 (7 s
                        -38 T
                        )
                \big)
                -14 i \pi  s H_{0,1}\big(
                        \hat{T}\big)
                +10 (-T
                +U
                ) H_{0,0,1}\big(
                        \hat{T}\big)
                \nonumber\\
                &&+4 (s
                +5 T
                ) H_{0,1,1}\big(
                        \hat{T}\big)
                +(26 s
                -36 T
                ) \zeta (3)
        \big) H_1\big(
                \hat{T}\big)
        +\big(
                10 i \pi  T
                +\frac{5}{2} \pi ^2 (
                U - T
                )
        \big) H_1\big(
                \hat{T}\big)^2
        \nonumber\\
        &&+\big(
                \frac{10 U}{3}
                -\frac{4}{3} i \pi  (s
                +T
                )
        \big) H_1\big(
                \hat{T}\big)^3
        +\big(
                \frac{3 s}{4}
                -\frac{2 T}{3}
        \big) H_1\big(
                \hat{T}\big)^4
        +\big(
                20 i \pi  (T
                -U
                )
                \nonumber\\
                &&+10 \pi ^2 (-T
                +U
                )
        \big) H_{0,1}\big(
                \hat{T}\big)
        +(40 s
        -20 T
        -2 i \pi  (11 s
        +6 T
        )
        ) H_{0,0,1}\big(
                \hat{T}\big)
        \nonumber\\
        &&+( 2 i \pi  (17 s
        -6 T
        )
        -20 (s
        +T
        )
        ) H_{0,1,1}\big(
                \hat{T}\big)
        +(72 s
        -60 T
        ) H_{0,0,0,1}\big(
                \hat{T}\big)
        \nonumber\\
        &&+36 (T
        -U
        ) H_{0,0,1,1}\big(
                \hat{T}\big)
        -12 (s
        +5 T
        ) H_{0,1,1,1}\big(
                \hat{T}\big)
        +20 T \zeta (3)
\bigg] \,,
\eea
where $\hat{T} = T/s$, $\hat{m}^2 = m^2/s$ and  $U = s - T$.
The higher-order terms up to $\ord(\eps^1)$ and $\ord(m^2)$ can be found in the ancillary file~\cite{ttplink}.
Note that for applications to non-trivial processes, the calculation of this integral at $\ord(\eps^1)$ is necessary,
and it is much more involved compared to the computations at  $\ord(\eps^0)$.

The analytic results are cross checked against numerical evaluations with \texttt{AMFlow} on various phase space points in the high-energy limit.\footnote{
The $\rm NPL_1$ integral has also been analytically calculated in~\cite{Fiore:2023myh} to $\ord(\eps^0)$, but without publicly available analytic expressions.
Hence, a comparison cannot be performed.
}
For example, with $\sqrt{s} = 2$~TeV, $p_T = \sqrt{u\,t / s} = 400$~GeV and $m=80$~GeV,
the results to  $\ord(m^2)$ agree with \texttt{AMFlow} at a permil level for the real parts and at a few permil level for the imaginary parts.
The high-energy expansion to $\ord(m^{30})$ agrees with \texttt{AMFlow} to 18 digits for the real parts and to 24 digits for the imaginary parts.
%

\subsection{Non-planar integral $\mathrm{NPL_2}$}
As the last two-loop example, we consider the fully-massive non-planar integral $\mathrm{NPL_2}$.
This integral is of great phenomenological interest and also mathematically intriguing, as it hints at new types of beyond-elliptic functions, as discussed in Refs.~\cite{Marzucca:2023gto,Huang:2013kh}.
However, no analytic solution beyond the large-mass limit is known so far.\footnote{
The non-planar fully massive integrals involving three different masses from top quark, Higgs and $Z$/$W$ bosons have been analytically computed in the large top-mass expansion in~\cite{Davies:2023npk}.
}
Here, we present the analytic results of this integral in the high-energy limit to $\ord(\eps^0)$ and $\ord(m^0)$,  i.e., two expansion terms in $m$.\footnote{
The higher-order terms in $\eps$ and $m$ of this integral are usually needed for non-trivial processes.
However, computing these terms requires solving a large number of irreducible two-scale three-dimensional MB integrals,
which are more involved.
Therefore, we defer this calculation to the future.
}
We show that up to these orders, the results can be expressed in terms of HPLs.
\begin{figure}[tb]
  \centering
    \includegraphics[width=.28\textwidth]{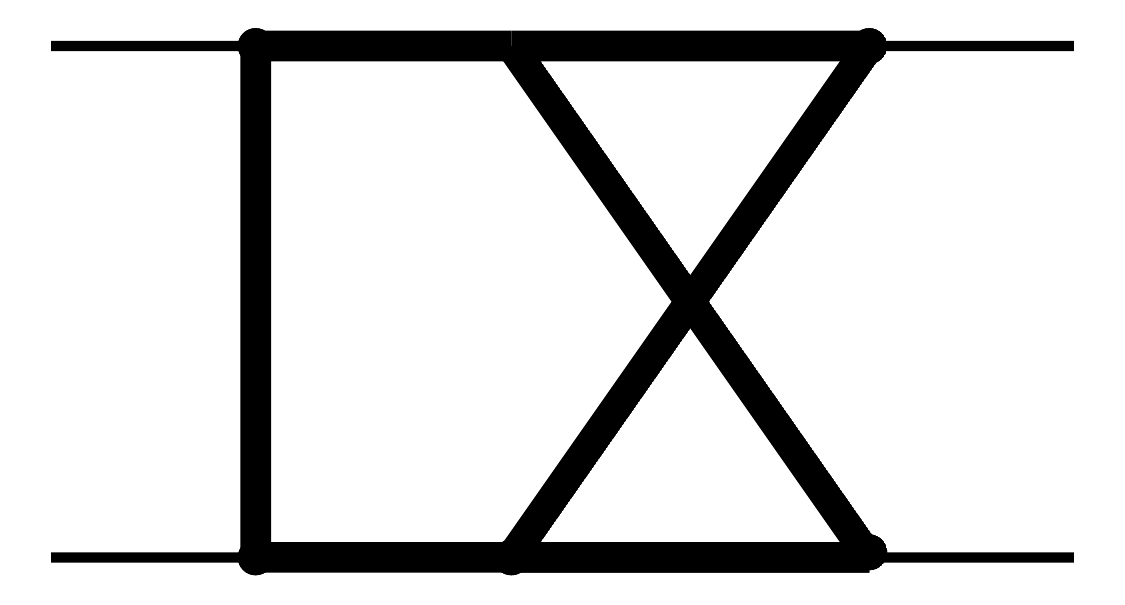} 
  \caption{\label{fig:npl2}
   Two-loop $\mathrm{NPL_2}$ diagram.
    }
\end{figure}

The $\rm NPL_2$ integral has the following propagators
\bea
&\Big\{
m^2-l_1^2,m^2-\left(l_1+q_3\right){}^2,m^2-\left(l_1+l_2+q_2+q_3\right){}^2,m^2-\left(l_1+l_2-q_1\right){}^2, \nonumber \\
& m^2-\left(l_2-q_1\right){}^2,  m^2-l_2^2,m^2-\left(l_2+q_2\right){}^2
\Big\}
\eea
where the kinematics are still the same as Eq.~\eqref{eq:kin} and the numerators are not shown.
The seven-line top-sector integral $\calI_{\rm NPL_2} [1,1,1,1,1,1,1]$ has the Symanzik polynomials
\bea
\calU &=& \left(\alpha _3+\alpha _4\right) \left(\alpha _5+\alpha _6+\alpha _7\right)+\alpha _1 \left(\alpha _3+\alpha _4+\alpha _5+\alpha _6+\alpha _7\right)\nonumber \\
&&+\alpha _2 \left(\alpha _3+\alpha _4+\alpha _5+\alpha _6+\alpha _7\right) \,, \\[4pt]
\calF &=& 
\left(\alpha _1+\alpha _2+\alpha _3+\alpha _4+\alpha _5+\alpha _6+\alpha _7\right) \Big[\left(\alpha _3+\alpha _4\right) \left(\alpha _5+\alpha _6+\alpha _7\right)+\alpha _1 \big(\alpha _3+\alpha _4+\alpha _5 \nonumber \\
&& +\alpha _6+\alpha _7\big)+\alpha _2 \left(\alpha _3+\alpha _4+\alpha _5+\alpha _6+\alpha _7\right)\Big] \, m^2+ \Big[  \left( \alpha _1 \alpha _4+\left(\alpha _1+\alpha _2+\alpha _3+\alpha _4\right) \alpha _5\right) \alpha _7 \nonumber \\
&& + \alpha _2 \alpha _3 \alpha _5 \Big] \, S+ \big( \alpha _2 \alpha _4 \alpha _6 \big) \, T+ \big(\alpha _1 \alpha _3 \alpha _6 \big) \, U \,.
\eea
By imposing the scaling $S,T,U \sim 1$ and $m^2 \sim \rho$, the resulting scalings of alpha parameters in all regions are listed in Table~\ref{tab:npl2region}.
\begin{table}[tb]
\centering
\begin{tabular}{cccccccc}
\text{Region} & $\alpha_1$ & $\alpha_2$ & $\alpha_3$ & $\alpha_4$ & $\alpha_5$ & $\alpha_6$ & $\alpha_7$ \\ \hline
$R_1$ & 0 & 0 & 0 & 0 & 0 & 0 & 0 \\
$R_2$ & 0 & 0 & 0 & 0 & 1 & 1 & 1 \\
$R_3$ & 0 & 0 & 0 & 1 & 1 & 1 & 0 \\
$R_4$ & 0 & 0 & 1 & 0 & 0 & 1 & 1 \\
$R_5$ & 0 & 0 & 1 & 1 & 0 & 0 & 1 \\
$R_6$ & 0 & 0 & 1 & 1 & 1 & 0 & 0 \\
$R_7$ & 0 & 0 & 1 & 1 & 1 & 1 & 1 \\
$R_8$ & 0 & 1/2 & 1/2 & 0 & 0 & 1/2 & 1 \\
$R_9$ & 0 & 1 & 0 & 0 & 0 & 1 & 1 \\
$R_{10}$ & 0 & 1 & 1 & 0 & 0 & 0 & 1 \\
$R_{11}$ & 1/2 & 0 & 0 & 1/2 & 1 & 1/2 & 0 \\
$R_{12}$ & 1 & 0 & 0 & 0 & 1 & 1 & 0 \\
$R_{13}$ & 1 & 0 & 0 & 1 & 1 & 0 & 0 \\
$R_{14}$ & 1 & 1 & 0 & 0 & 0 & 0 & 1 \\
$R_{15}$ & 1 & 1 & 0 & 0 & 1 & 0 & 0 \\
$R_{16}$ & 1 & 1 & 0 & 0 & 1 & 1 & 1 \\
$R_{17}$ & 1 & 1 & 1 & 1 & 0 & 0 & 1 \\
$R_{18}$ & 1 & 1 & 1 & 1 & 1 & 0 & 0 \\
\end{tabular}
\caption{Scalings of alpha parameters for the $\rm NPL_2$ integral, i.e. $\alpha_i \sim \rho^p$ where $p$ is the power in the table. 
$R_1$ is the hard region and $R_2, \dots, R_{18}$ are the asymptotic regions.}
\label{tab:npl2region}
\end{table}
The massless non-planar integral in the hard region $R_1$ can be obtained in the same way as before,
and the asymptotic-region integrals are calculated by \texttt{AsyInt}.
By combining contributions in all regions, 
the final results of this integral to $\ord(\eps^0)$ and $\ord(m^0)$ are obtained as
\bea
\calI_{\rm NPL_2} [1,1,1,1,1,1,1] & = &
\lb \frac{\mu^2}{s} \rb^{2 \epsilon}  \sum_{j=-1}^{0}  \, m^{j}  \, f_{(j)}\big(s,T,U,m^2 \big)\,,
\eea
with coefficient functions
\bea
f_{(-1)} 
&=&-\frac{i \, c_Z \, \pi^2 }{s^2 \sqrt{T U}}
 \,, \\[4pt]
 f_{(0)} 
 &=&
\frac{1}{s^2 T U} \bigg[
        \frac{ \pi ^4}{180} (15 s
        +202 T
        )
        -2 i \pi  \big(
                T \big(
                        \pi ^2-2 \zeta (3)\big)
                +s (-24+23 \zeta (3))
        \big)
        +\big(
                48 T
                -6 \pi ^2 T
                \nonumber\\
                &&+i \big(
                        24 \pi  U
                        +\frac{2}{3} \pi ^3 (7 s
                        -5 T
                        )
                \big)
                +\big(
                        12 i \pi  U
                        +\frac{1}{3} \big(
                                -\big(
                                        \big(
                                                72+13 \pi ^2\big) s\big)
                                +14 \pi ^2 T
                        \big)
                        \nonumber\\
                        &&+6 (T
                        -U
                        ) H_{0,1}\big(
                                \hat{T}\big)
                \big) H_1\big(
                        \hat{T}\big)
                +(6+i \pi ) s H_1\big(
                        \hat{T}\big)^2
                +\frac{1}{3} (s
                -4 T
                ) H_1\big(
                        \hat{T}\big)^3
                \nonumber\\
                &&+(12 (T-2 s
                )
                +4 i \pi  (2 s
                +T
                )
                ) H_{0,1}\big(
                        \hat{T}\big)
                +(24 T -32 s
                ) H_{0,0,1}\big(
                        \hat{T}\big)
                +14 (U-T
                ) H_{0,1,1}\big(
                        \hat{T}\big)
                \nonumber\\
                &&+\frac{2}{3} (s
                -34 T
                ) \zeta (3)
        \big) H_0\big(
                \hat{T}\big)
        +\big(
                6 i \pi  U
                +\frac{1}{3} \pi ^2 (-3 s
                +7 T
                )
                +(-2 i \pi  T
                +6 U
                ) H_1\big(
                        \hat{T}\big)
                \nonumber\\
                &&-T H_1\big(
                        \hat{T}\big)^2
                +(8 s
                -6 T
                ) H_{0,1}\big(
                        \hat{T}\big)
        \big) H_0\big(
                \hat{T}\big)^2
        +\big(
                -2 T
                +\frac{1}{3} i \pi  (5 s
                -2 T
                )
                \nonumber\\
                &&+\big(
                        -s
                        +\frac{4 T}{3}
                \big) H_1\big(
                        \hat{T}
                \big)
        \big) H_0\big(
                \hat{T}\big)^3
        +
        \frac{1}{6} (s
        +2 T
        ) H_0\big(
                \hat{T}\big)^4
        +\big(
                -\frac{2}{3} i \pi  \big(
                        36+7 \pi ^2\big) s
                        \nonumber\\
                        &&+\big( -12 i \pi  U
                        -\frac{8}{3} \big(
                                \big(
                                        9+\pi ^2\big) T
                                -\pi ^2 U
                        \big)
                        +2 (6+i \pi ) s H_1\big(
                                \hat{T}\big)
                        +2 U H_1\big(
                                \hat{T}\big)^2
                \big) H_0\big(
                        \hat{T}\big)
                \nonumber\\
                &&+\big(
                        -2 i \pi  s
                        -2 T H_1\big(
                                \hat{T}\big)
                \big) H_0\big(
                        \hat{T}\big)^2
                -\frac{2}{3} T H_0\big(
                        \hat{T}\big)^3
                \nonumber\\
                &&+\big(
                        12 i \pi  T
                        +24 U
                        +\frac{8}{3} \pi ^2 (-T
                        +U
                        )
                \big) H_1\big(
                        \hat{T}\big)
                -2 i \pi  s H_1\big(
                        \hat{T}\big)^2
                +\frac{2}{3} U H_1\big(
                        \hat{T}\big)^3
                \nonumber\\
                &&-\frac{8 s \zeta (3)}{3}
        \big) H_0\big(
                \hat{m}^2\big)
        +\big(
                6 i \pi  s
                -\frac{5 \pi ^2 s}{3}
                +\big(
                        6 T
                        +i \pi  (T
                        -U
                        )
                        +3 s H_1\big(
                                \hat{T}\big)
                \big) H_0\big(
                        \hat{T}\big)
                \nonumber\\
                &&-s H_0\big(
                        \hat{T}\big)^2
                +(-6 U
                +i \pi  (T
                -U
                )
                ) H_1\big(
                        \hat{T}\big)
                -s H_1\big(
                        \hat{T}\big)^2
        \big) H_0\big(
                \hat{m}^2\big)^2
                \nonumber\\
                && +\big( \frac{4 i \pi  s}{3}
                +\frac{2}{3} (2 s
                +T
                ) H_0\big(
                        \hat{T}\big)
                +\big(
                        -2 s
                        +\frac{2 T}{3}
                \big) H_1\big(
                        \hat{T}
                \big)
        \big) H_0\big(
                \hat{m}^2\big)^3
        -
        \frac{1}{2} s H_0\big(
                \hat{m}^2\big)^4
        \nonumber\\
        &&+\big(
                6 \big(
                        \pi ^2 -8 \big) U
                +i \big(
                        -24 \pi  T
                        +\frac{\pi ^3}{3}  (s
                        -10 T
                        )
                \big)
                -10 i \pi  s H_{0,1}\big(
                        \hat{T}\big)
                +6 (U-T
                ) H_{0,0,1}\big(
                        \hat{T}\big)
                \nonumber\\
                &&+4 (s
                +3 T
                ) H_{0,1,1}\big(
                        \hat{T}\big)
                +\frac{4}{3} (9 s
                -17 T
                ) \zeta (3)
        \big) H_1\big(
                \hat{T}\big)
        +\big(
                6 i \pi  T
                \nonumber\\
                &&+\frac{1}{3} \pi ^2 (4 s
                -7 T
                )
        \big) H_1\big(
                \hat{T}\big)^2
        +\big(
                2 U
                -\frac{1}{3} i \pi  (3 s
                +2 T
                )
        \big) H_1\big(
                \hat{T}\big)^3
        +\frac{1}{6} (3 s
        -2 T
        ) H_1\big(
                \hat{T}\big)^4
        \nonumber\\
        &&+\big(
                12 i \pi  (T
                -U
                )
                +\frac{14}{3} \pi ^2 (
                U-T
                )
        \big) H_{0,1}\big(
                \hat{T}\big)
        +(24 s
        -12 T
        -2 i \pi  (9 s
        +2 T
        )
        ) H_{0,0,1}\big(
                \hat{T}\big)
        \nonumber\\
        &&+(i \pi  (22 s
        -4 T
        )
        -12 (s
        +T
        )
        ) H_{0,1,1}\big(
                \hat{T}\big)
        +(48 s
        -36 T
        ) H_{0,0,0,1}\big(
                \hat{T}\big)
        \nonumber\\
        &&+20 (T
        -U
        ) H_{0,0,1,1}\big(
                \hat{T}\big)
        -12 (s
        +3 T
        ) H_{0,1,1,1}\big(
                \hat{T}\big)
        +\frac{4}{9} \pi ^2 s \psi ^{(1)}\big(
                \frac{1}{3}\big)
        -\frac{1}{3} s \psi ^{(1)}\big(
                \frac{1}{3}\big)^2
        \nonumber\\
        &&+12 T \zeta (3)
\bigg] \,,
\eea
where $\hat{T} = T/s$, $\hat{m}^2 = m^2/s$ and  $U = s - T$.
Note that the constant $c_Z$ is defined in Eq.~\eqref{eq:cz}, and its numerical value 
\bea
c_Z
 &=& 17.695031908454309764234228747255048751062059438637  \dots 
\eea 
can be computed to arbitrary numerical precision.
These results are cross checked against numerical evaluations with \texttt{pySecDec}~\cite{Borowka:2017idc} on several phase space points in the high-energy limit.
For example, with the phase space point $\sqrt{s} = 2$~TeV, $p_T = 400$~GeV and $m=80$~GeV,
the results to $\ord(m^0)$ agree with \texttt{pySecDec}  at a few percent level for the real part and at a percent level for the imaginary part, 
given the default percent-level accuracy of the \texttt{pySecDec}  result.

\section{Conclusions}

The high-energy behaviour of massive two-loop four-point Feynman integrals
is of both phenomenological and mathematical interest.
For example,
as an effective probe of new physics effects beyond the Standard Model,
studies of multi-Higgs-boson and associated-Higgs-boson production  with large transverse momenta at the LHC and future high-energy colliders
require precise theoretical predictions in the high-energy region.
In this region, both higher-order EW and QCD radiative corrections are relevant.
In particular,  the massive particles such as top quark, Higgs and vector bosons are resolved in the virtual loops.
Therefore, the calculation of massive two-loop four-point Feynman integrals is of paramount importance.

In this paper, we have presented analytic techniques and the \texttt{Mathematica} toolbox \texttt{AsyInt} 
for calculating massive two-loop four-point Feynman integrals 
in the high-energy region.
By treating particle masses as small expansion parameters,
direct integrations of these integrals can be achieved in the parametric space with the asymptotic expansion and the MB approach.
In particular, we have presented the analytic \texttt{Expand\&Fit} method to systematically solve two types of complicated irreducible MB integrals:  one-scale two-dimensional MB integrals with nested non-vanishing arc contributions,
and two-scale two-dimensional MB integrals.
With this method, analytic results to higher orders in the small-mass expansion parameter and the dimensional regulator $\eps$
can be obtained using \texttt{AsyInt} in an algorithmic fashion.
This kind of analytic calculation
is the bottleneck of the deep high-energy expansion
approach,
which can provide precise predictions to $2\to 2$ scattering processes 
across a vast range of interesting phase space regions.
These results may further serve as boundary conditions to the differential equations approach in the high-energy limit.
Currently, \texttt{AsyInt} is  limited by its capability for reducing the dimensionality  of MB integrals,
which will be improved in future work.

Finally, we have discussed three representative massive two-loop four-point Feynman integrals, including both planar and non-planar examples.
Their analytic results are also provided in the ancillary file~\cite{ttplink}.

\section*{Acknowledgements}
H.Z. thanks Kay Sch\"onwald for close collaboration on master integral calculations that are valuable to the development of \texttt{AsyInt}. H.Z. thanks Joshua Davies, Go Mishima and Matthias Steinhauser for various helpful discussions during the collaborations.
H.Z. also thanks J.D., K.S. and M.S. for carefully reading the manuscript and their comments.
This research is supported by the 
Deutsche Forschungsgemeinschaft (DFG, German Research Foundation) under the grant 396021762 -- TRR~257 “Particle Physics Phenomenology after the Higgs Discovery”.
The Feynman diagrams in this paper are drawn using \texttt{FeynGame}~\cite{Harlander:2020cyh}.

\appendix
\section{Commands of \texttt{AsyInt}}
\label{app:func}

\definecolor{lightgray}{gray}{0.95}

\lstdefinestyle{overallstyle}{
basicstyle=\ttfamily\small,
backgroundcolor=\color{lightgray}
}

The  commands of the public version \texttt{AsyInt 1.0} are listed in this Appendix.
%
The global parameters in \texttt{AsyInt} are \texttt{ep} and \texttt{D}, \texttt{rho}, \texttt{im},
where $\texttt{D} = 4- 2 \texttt{ep}$, \texttt{rho} is the scaling power-counting parameter used internally, and \texttt{im} denotes the imaginary $\rm i$.
In the following, we use \texttt{=>} to denote the output of commands.

\subsection{Toolkit I: generate integrals}
\begin{lstlisting}[frame=single, style=overallstyle]
GenerateInput[loops, props, kinematics, SmallInv, ScalePara, xlist,
  EuclInv:{SS,TT,UU}, EuclInvSum:0, UserDefineRelation:{}]
  => {UFpoly, Region, Scaling}
\end{lstlisting}
\begin{lstlisting}[frame=single,style=overallstyle]
GenerateInputNum[loops, props, numidx, kinematics, SmallInv, ScalePara, 
  xlistNum, EuclInv:{SS,TT,UU}, EuclInvSum:0, UserDefineRelation:{}]
  => {UFpoly, Region, Scaling, UFpolyNum}
\end{lstlisting}
The inputs are
\begin{itemize}
\item \texttt{loops}: a list of loop momenta, e.g. \texttt{\{l1, l2\}}.
\item \texttt{props}: a list of propagators, e.g. \texttt{\{m^2-(l1+q1)^2, \dots\}}.
\item \texttt{kinematics}: rules of external kinematics, e.g. \texttt{\{q1*q2->-SS/2, q1*q3->-TT/2, \dots\}}.
\item \texttt{EuclInv}: Euclidean kinematic invariants (\texttt{SS=-s, TT=-t, UU=-u} by default).
\item \texttt{EuclInvSum}: sum of Euclidean invariants (\texttt{SS+TT+UU=0} by default).
\item \texttt{SmallInv}: a small expansion parameter, e.g. \texttt{m^2}.
\item \texttt{ScalePara}: a scaling parameter, e.g. $\texttt{rhos} = \texttt{rho^2} \sim \texttt{m^2/SS}$.
\item \texttt{xlist}: a list of alpha parameters of a Feynman diagram, e.g. \texttt{\{x[1], x[2], \dots\}}.
\item \texttt{xlistNum}: a list of alpha parameters of an extended Feynman integral with numerators .
\item \texttt{numidx}: a list of numbers that indicates the positions of numerators in the \texttt{props} list.
\item \texttt{UserDefineRelation}: a replacement rule provided by the user, e.g. \texttt{\{UU->-SS-TT\}}.
\end{itemize}
The outputs are
\begin{itemize}
\item \texttt{UFpoly}: Symanzik polynomials of a Feynman diagram.
\item \texttt{UFpolyNum}:  Symanzik polynomials of an extended Feynman integrals with numerators.
\item \texttt{Region}: a list of regions from the asymptotic expansion. The first region is the hard region be default.
\item \texttt{Scaling}: the scaling of alpha parameters in terms of \texttt{rho} power countings.
\end{itemize}
\vspace{0.2cm}
\begin{lstlisting}[frame=single,style=overallstyle]
AlphaRepForTempInt[UFpoly, Scaling, dlist, xlist, ScalePara]
  => {AlphaRepRegion}
\end{lstlisting}
The additional input is
\begin{itemize}
\item \texttt{dlist}: a list of delta regulators of a Feynman diagram, e.g. \texttt{\{d1, d2, \dots\}}. They also serve as symbolic propagator-power shifts for template integrals.
\end{itemize}
The output is
\begin{itemize}
\item \texttt{AlphaRepRegion}: a list of alpha representations of asymptotic regions.
\end{itemize}
\vspace{0.2cm}
\begin{lstlisting}[frame=single,style=overallstyle]
AsyExp2MB[UFpoly, Scaling, dlist, xlist, ExpLowOrd, ExpMaxOrd, 
  SmallInv, ScalePara, TempIntList, HardIntList, Zrule, epOrd, 
  DotShift: {}, AddExpOrd: 0]
  => {MBexp}
\end{lstlisting}
\begin{lstlisting}[frame=single,style=overallstyle]
AsyExpNum2MB[UFpolyNum, numidx, UFpoly, Scaling, dlistNum, xlistNum,
  ExpLowOrd, ExpMaxOrd, SmallInv, ScalePara, TempIntList,
  HardIntList, Zrule, epOrd, AddExpOrd:0]
  => {MBexp}
\end{lstlisting}
The additional inputs are
\begin{itemize}
\item \texttt{dlistNum}: a list of delta regulators of an extended Feynman integral with numerators.
\item \texttt{ExpLowOrd}: the lowest power of the expansion parameter \texttt{SmallInv}, where the high-energy expansion begins. Typically, it is \texttt{0} for planar integrals, but \texttt{-1/2} for non-planar integrals.
\item \texttt{ExpMaxOrd}: the highest power of the expansion parameter  \texttt{SmallInv}, where the high-energy expansion terminates.
\item \texttt{AddExpOrd}: an additional expansion depth in the square root of the expansion parameter  \texttt{Sqrt[SmallInv]}. Typically, it is \texttt{0} for integrals without numerators, but \texttt{1} or \texttt{2} for integrals with numerators.
Note that \texttt{AsyInt} performs internal  safety checks during the expansion. Therefore, the command will abort if potential missing expansion terms are detected, and prompt users to increase the value of \texttt{AddExpOrd}.
\item \texttt{epOrd}: the highest power of dimensional regularisation parameter \texttt{ep}, where the $\ep$-series terminates.
\item \texttt{HardIntList}: a list with only one hard region (the first region), e.g. \texttt{\{R1tmp\}}. Note that the \texttt{R1tmp} does not need to be the \texttt{Head} of a function; it serves as a placeholder to ensure that \texttt{Length[HardIntList]=1}. 
For simple integrals, users can also calculate the template integral for the hard region and set \texttt{HardIntList=\{\}} to handle the hard region integral on the same footing as the other asymptotic region integrals.
\item \texttt{TempIntList}: a list of template integrals in asymptotic regions, e.g. \texttt{\{R2tmp, R3tmp, \dots\}}. 
They must be the \texttt{Head} of template-integral functions that return the MB integrand representation. 
The convention of template-integral functions is demonstrated by the example 
\[\texttt{R2tmp[dlist,ep]:= Gamma[d1+ep+Z1]*Gamma[d2-ep-Z2]*\dots} \]
where \texttt{Z1,Z2} are integration variables in the complex plane.
\item \texttt{Zrule}: a rule fixing the real parts of straight integration lines parallel to the imaginary axis. The default choice is \texttt{\{Z1->-1/7, Z2->-1/11, Z3->-1/17, Z4->-1/19\}}.
\item \texttt{DotShift}: a list indicating which propagator is raised to higher powers. For example, if the second propagator contains \texttt{n} dots, i.e. raised to power \texttt{(n+1)}, users should set \texttt{\{d2,n\}}.
\end{itemize}
The output is
\begin{itemize}
\item \texttt{MBexp}: expressions containing Mellin-Barnes integrands.
\end{itemize}
The option is:
\begin{itemize}
\item  \texttt{ExpSafetyCheck:} an option to switch on or off the safety check on the expansion depth. 
It is \texttt{True} by default, meaning that \texttt{AsyInt} will expand two more terms in \texttt{rho} to check whether the expansion depth is sufficient to the desired order \texttt{ExpMaxOrd}.
If the safety check fails, the commands will prompt users to increase the value of \texttt{AddExpOrd}.
For performance reasons, users can disable this feature by setting it to \texttt{False}.
\item \texttt{ResolveMBexact0}: an option to switch on or off the internal check and resolution of problematic outputs with exactly vanishing MB integrals. 
It is \texttt{False} by default.
Usually, \texttt{AsyInt} has internal checks to avoid this kind of problematic outputs.
However, in complicated calculations, problematic outputs may still occur.
Users can set this option to \texttt{True} and \texttt{AsyInt} will try to resolve the issue.
\end{itemize}
Note that   warning messages can appear during the evaluation,
which are handled by \texttt{AsyInt} internally.
Users can ignore these massages as long as the program does not abort.

\subsection{Toolkit II: solve integrals}

\begin{lstlisting}[frame=single,style=overallstyle]
AISum1DMB[MBexp, Z1, k1, Zcontour, SmallInv, MBscale:"none", 
  MBscale2:"none", LRpreset:"R", AddShift:1, BSpresent:0]
  => {Result}
\end{lstlisting}
\begin{lstlisting}[frame=single,style=overallstyle]
AISum2DMB[MBexp, Z1, Z2, k1, k2, Zcontour, SmallInv, MBscale:"none",
  MBscale2:"none", LRpreset:{"R","R"}, AddShift:1, split:0]
  => {Result}
\end{lstlisting}
The inputs are
\begin{itemize}
\item \texttt{Z1, Z2}: MB integration variables in the complex plane.
\item \texttt{k1, k2}: positive integers for the infinity residues representations. Note that \texttt{AsyInt} imposes positive-integer assumptions on \texttt{k1, k2}.
If users choose other parameters, additional assumptions are required.
\item \texttt{Zcontour}: same as \texttt{Zrule}.
\item \texttt{MBscale}: a scale associated with the MB integrals, i.e. \texttt{MBscale^Z1} is present. Note that \texttt{MBscale="none"} implies scaleless MB integrals.
\item \texttt{MBscale2}: a second scale that is not associated with the MB integrals, i.e. no \texttt{Z1} or \texttt{Z2} dependence. It is only relevant for non-planar integrals.
\item \texttt{LRpreset}: variables determining which side of  semi-circle to close with. For example, \texttt{\{"R","L"\}} will close the \texttt{Z1} integration contour to the right semi-circle and the \texttt{Z2} integration contour to the left semi-circle. Note that the \texttt{Z1}-dependent \texttt{Z2} integration contour will be decided by \texttt{AISum1DMB} and \texttt{AISum2DMB} automatically, and user inputs have no effect on it.
\item \texttt{AddShift}: a level parameter dealing with the extra residues from left- and right-merging poles, which are separated from infinity series representations expressed in terms of \texttt{k1, k2}. Note that \texttt{AISum1DMB} and \texttt{AISum2DMB} will adjust its value when necessary.
\item \texttt{BSpresent}: an parameter indicating whether binomial sums are handled or not. By default, this feature is disabled with \texttt{BSpresent=0}.
\item \texttt{split}: an parameter indicating whether to split the nested residue calculations for performance reasons. By default, this feature is disabled with \texttt{split=0}.
\end{itemize}
\vspace{0.2cm}
Note that with the public version \texttt{AsyInt 1.0},
the   commands \texttt{AISum1DMB} and \texttt{AISum2DMB} only support computations of MB integrals with integer-valued residues.
\vspace{0.4cm}

\begin{lstlisting}[frame=single,style=overallstyle]
AINumRec1DMB[MBexp, KinList, TransParaList, ConstList, Zcontour, 
  PrecList, MaxPower:200, KnownRes:0]
  => {Result}
\end{lstlisting}
The additional inputs are
\begin{itemize}
\item \texttt{KinList}: a list of kinematic variables appearing in scaleless MB integrals, e.g. \texttt{\{TT, UU\}}.
\item \texttt{TransParaList}: a list of transcendental parameters used for dividing MB integrals into smaller blocks, e.g. \texttt{\{Log[m^2], Zeta[3], Pi\}}. This list is necessary for handling large expressions of MB integrals.
\item \texttt{ConstList}: a list of constants used for numerical reconstruction with \texttt{PSLQ} algorithm.
\item \texttt{PrecList}: a list of two targeted precisions for numerical evaluations, specified in terms of digits. For example, \texttt{\{900, 1000\}} indicates numerical evaluations aiming at 900- and 1000-digit precision. Note that \texttt{AISum2DMB} requires identical final result from both evaluations for a successful numerical reconstruction.  Therefore, users should ensure a sufficient difference between these two precisions.
\item \texttt{MaxPower}: the maximal truncating powers of parameters in \texttt{KinList} used during the evaluation. This is typically relevant for the \texttt{Expand\&Fit} method.
\item \texttt{KnownRes}: previously calculated analytic results used to combined with MB integrals for the numerical reconstruction. This can reduce the size of \texttt{ConstList} by exploiting cancellations of spurious terms.
\end{itemize}
The output \texttt{Result} is the analytic solutions to the MB integrals.

The options are:
\begin{itemize}
\item \texttt{NoGammaSimplify}: an option to switch on or off the usage of \texttt{GammaSimplifyList} inside the command \texttt{AINumRec1DMB}. It is \texttt{False} by default. Users can enable this feature by setting it to \texttt{True} for some specific types of integrals.
\item \texttt{MBintOneByOne}: an option to switch on or off  the one-by-one MB integral numerical evaluation. It is \texttt{False} by default.
For complicated integrals involving Hypergeometric functions like \texttt{HypergeometricPFQ}, users may enable this feature by setting it to \texttt{True}.
\end{itemize}
\vspace{0.4cm}

\begin{lstlisting}[frame=single,style=overallstyle]
AIEucl2Phys[exp, SmallInv, TT, UU, SS, s, mus, IntDef, LoopOrd, epOrd,
  NoEpExp:False, EuclKinRelation:{UU->-TT-SS}]
  => {PhysResult}
\end{lstlisting}
The additional inputs are
\begin{itemize}
\item \texttt{SS,TT,UU}: the positive invariants in the Euclidean region.
\item \texttt{s,TT,UU}: the positive invariant in the physical region.
\item \texttt{mus}: regularisation scale squared $\mu^2$
\item \texttt{IntDef}: a list of integral definition, e.g. a seven-line integral is \texttt{\{1,1,1,1,1,1,0,0\}}.
\item \texttt{LoopOrd}: the loop order.
\item \texttt{NoEpExp}: if \texttt{NoEpExp =!= False}, no \texttt{(mus/SS)^ep} or \texttt{(mus/s)^ep} prefactor will be attached, and no $\ep$-expansion will be performed. This feature is disabled by default.
\item \texttt{EuclKinRelation}: the kinematic relation in Euclidean region used in the analytic continuation.
\end{itemize}
The output is
\begin{itemize}
\item \texttt{PhysResult}: results in the physical region.
\end{itemize}
Note that \texttt{AIEucl2Phys} does not contain a complete list of analytic continuation rules.
For more crossing and analytic continuation rules, please refer to Ref.~\cite{Davies:2022ram}.
\vspace{0.4cm}

\begin{lstlisting}[frame=single,style=overallstyle]
AC2Phys[exp, SmallInv, TT, UU, SS, s, EuclKinRelation:{UU->-TT-SS}]
  => {PhysResult}
\end{lstlisting}
This is a lower-level command version of \texttt{AIEucl2Phys}, but without  \texttt{s}-dependence and regularisation scale reconstructions.
It is suitable for manipulating individual functions, rather than the full expression.

\subsubsection{\texttt{Expand\&Fit} module}

\begin{lstlisting}[frame=single,style=overallstyle]
AIExpandI1[MBexp, Z1, k1, Z2, k2, Zcontour, Z1ExpOrd, KinList, 
  TransParaList, ConstList, PrecList, LRpreset:{"R","R"}, AddShift:2]
  => {ResExp}
\end{lstlisting}

\begin{lstlisting}[frame=single,style=overallstyle]
AIExpandI2[MBexp, Z1, k1, Z2, k2, Zcontour, Z1ExpOrd, MBscale,
  MBscale2, SmallInv, LRpreset:{"R","R"}, AddShift:2]
  => {ResExp}
\end{lstlisting}
The additional input is
\begin{itemize}
\item \texttt{Z1ExpOrd}: the highest  power of \texttt{Z1} in the expansion of \texttt{MBscale^Z1} in the $\texttt{MBscale} \to 0$ limit.
\end{itemize}
The output is
\begin{itemize}
\item \texttt{ResExp}: expanded results in the $\texttt{MBscale} \to 0$ limit.
\end{itemize}
Note that for the \texttt{AIExpandI2} command, the second scale \texttt{MBscale2} is also associated with MB integrals, i.e. \texttt{MBscale2^Z2} is present.
\vspace{0.4cm}

\begin{lstlisting}[frame=single,style=overallstyle]
AIExpand2DMB[MBexp, Z1, k1, Z2, k2, Zcontour, Z1ExpOrd, 
  LRpreset:{"R","R"}, AddShift:2, IgnoreBDconst:0, ToSum:0]
  => {MBexp, ExtraResidue}
\end{lstlisting}

\vspace{0.2cm}
Note that with the public version \texttt{AsyInt 1.0},
the  commands \texttt{AIExpand2DMB}, \texttt{AIExpandI1} and \texttt{AIExpandI2} only support computations of MB integrals with integer-valued residues.
\vspace{0.4cm}

\begin{lstlisting}[frame=single,style=overallstyle]
FitAnsatz[Ansatz, ResExp, MBscale, SmallInv, CoeffSymbol, ExpOrd,
  AnsatzExpRule, TransParaList:{Pi}, InitGuess:{}]
  => {Result}
\end{lstlisting}
The additional inputs are
\begin{itemize}
\item \texttt{Ansatz}: an ansatz constructed from transcendental functions and rational functions with unknown coefficients.
\item \texttt{CoeffSymbol}: the symbolic \texttt{Head} of unknown coefficients in the ansatz. For example, \texttt{CoeffSymbol=a} for coefficients \texttt{a[i,j]}.
\item \texttt{AnsatzExpRule}: expansion rules for transcendental functions appearing in the ansatz in the $\texttt{MBscale} \to 0$ limit.
\item \texttt{ExpOrd}: expansion order used for  fitting the ansatz, with its maximal value being \texttt{Z1ExpOrd}.
\item \texttt{InitGuess}: prior knowledge or initial guesses that may reduce the size of the ansatz. 
\end{itemize}

\subsection{Other commands}
\label{sec:utilfunc}

\begin{lstlisting}[frame=single,style=overallstyle]
IntTypeA[integrand, var, varResList:{}] 
\end{lstlisting}
\begin{lstlisting}[frame=single,style=overallstyle]
IntTypeB[integrand, var, varResList:{}]
\end{lstlisting}
\begin{lstlisting}[frame=single,style=overallstyle]
MBsplit[integrand, X, Y, Zvar]
\end{lstlisting}
The command \texttt{IntTypeA} is the integration routine of Eq.~\eqref{eq:inttype1},
the command \texttt{IntTypeB} is the integration routine of Eq.~\eqref{eq:inttype2},
and the command \texttt{MBsplit} is the Mellin transformation routine of Eq.~\eqref{eq:mellin}.
The inputs are
\begin{itemize}
\item \texttt{integrand}: the integrand in alpha representation.
\item \texttt{var}: the alpha variable to be integrated over.
\item \texttt{varResList}: a list of alpha variables to be rescaled. For example, if \texttt{varResList = \{x2, x3\} } and \texttt{var = x1},
these routines will rescale \texttt{x2->x2*x1, x3->x3*x1}, and then integrate over \texttt{x1}. The Jacobian determinants are taken into account.
\item \texttt{X,Y}: expressions of the $X$ and $Y$ as in Eq.~\eqref{eq:mellin}.
\item \texttt{Zvar}: the MB integration variable $z$ introduced in Eq.~\eqref{eq:mellin}.
\end{itemize}
The outputs are the integrated or Mellin transformed results.
\vspace{0.4cm}

\begin{lstlisting}[frame=single,style=overallstyle]
GammaSimplifyList[exp]
\end{lstlisting}
This  is an efficient and momery-economic recursive simplification command designed to simplify large expressions of Gamma and PolyGamma functions through their recursions.
\vspace{0.4cm}

\begin{lstlisting}[frame=single,style=overallstyle]
SortPatternList[exp, PatternList]
\end{lstlisting}
This  is a pattern sorting command that outputs a list of expressions. 
For example, if \texttt{PatternList = \{Z1, Z2\}}, then the output is a list of non-vanishing expressions in the form
\{no \texttt{Z1} or \texttt{Z2}, only \texttt{Z1}, only \texttt{Z2}, both \texttt{Z1} and \texttt{Z2}\}.
\vspace{0.4cm}

\begin{lstlisting}[frame=single,style=overallstyle]
KinToX[exp, TT, SS, x, Z]
\end{lstlisting}
This is a command that identifies MB integrals with scales. It translate \texttt{TT^(Z+a)*SS^(-Z+b)} to \texttt{x^Z*TT^a*SS^b}.

%


\providecommand{\href}[2]{#2}\begingroup\raggedright\endgroup

\end{document}